\documentclass[useAMS,usenatbib,usegraphicx]{mn2e}
\newcommand{\Rj}{\hbox{R$_{Jup}$\,}}
\newcommand{\Mj}{\hbox{M$_{Jup}$\,}}
\newcommand{\Rsol}{\hbox{R$_{\odot}$\,}}
\newcommand{\Msol}{\hbox{M$_{\odot}$\,}}
\newcommand{\Ranti}{\hbox{$\Delta \chi^{2}/\Delta \chi_{-}^{2}$}}

\usepackage{subfigure}
\usepackage{lscape}
\usepackage{amssymb}
\usepackage{times}

\title[SuperWASP-N Extra-solar Planet Candidates]{SuperWASP-N Extra-solar Planet 
Candidates Between 18hr $<$ RA $<$ 21hr}
\author[R.A. Street et al.]{
R.A. Street$^{1,12,13}$,
D.J. Christian$^{1}$,
W.I. Clarkson$^{2,8}$,
A. Collier Cameron$^{3}$,
B.~Enoch$^{2}$,
\newauthor
S.R.~Kane$^{3,9}$,
T.A.~Lister$^{3,4,12}$,
R.G.~West$^{7}$,
D.M.~Wilson$^{4}$,
A. Evans$^{4}$,
\newauthor
A. Fitzsimmons$^{1}$,
C.A. Haswell$^{2}$,
C. Hellier$^{4}$,
S.T. Hodgkin$^{5}$, 
K. Horne$^{3}$,
\newauthor
J. Irwin$^{5}$,
F.P. Keenan$^{1}$,
A.J. Norton$^{2}$,
J. Osborne$^{7}$,
D.L. Pollacco$^{1}$,
\newauthor
R. Ryans$^{1}$,
I. Skillen$^{6}$,
P.J. Wheatley$^{10}$,
and J. Barnes$^{3,11}$. \\
$^{1}$Astrophysics Research Centre, Department of Physics and Astronomy, 
Queen's University Belfast, Belfast, BT7 1NN, UK,\\ 
$^{2}$Department of Physics \& Astronomy, The Open University, Milton Keynes, 
MK7 6AA, UK,  \\
$^{3}$School of Physics \& Astronomy, University of St. Andrews, North Haugh, 
St. Andrews, Fife, KY16 9SS, UK,  \\
$^{4}$Astrophysics Group, School of Chemistry \& Physics, Keele University, 
Staffordshire, ST5 5BG, UK,  \\
$^{5}$Institute of Astronomy, University of Cambridge, Madingley Road, 
Cambridge, CB3 0HA, UK, \\ 
$^{6}$Isaac Newton Group of Telescopes, Apartado de correos 321,
E-38700 Santa Cruz de la Palma, Tenerife, Spain,\\ 
$^{7}$Department of Physics \& Astronomy, University of Leicester, Leicester, 
LE1 7RH, UK,\\
$^{8}$Space Telescope Science Institute (STScI), 3700 San Martin Drive,
Baltimore, MD~21218, USA,\\
$^{9}$University of Florida, PO~Box~112005, 211 Bryant Space Science Center,
Gainesville, FL, USA,\\
$^{10}$Dept. of Physics, University of Warwick, Coventry, CV4 7AL, UK.\\
$^{11}$Centre for Astrophysics Research, Science \& Technology Research
Institute, University of Hertfordshire, Hatfield, AL10 9AB. \\
$^{12}$Las Cumbres Observatory, 6740B Cortona Drive, Goleta, CA 93117, USA.\\
$^{13}$Dept. of Physics, Broida Hall, University of California, Santa Barbara, 
CA 93106-9530, USA.\\
} 

\date{Accepted 2006 ?? ??. Received 2006 March ??; in original form 2006 
August ??}

\pagerange{\pageref{firstpage}--\pageref{lastpage}} \pubyear{2006}

\begin{document}

\maketitle

\label{firstpage}

\begin{abstract}

The SuperWASP-I instrument observed 6.7\,million stars between 8 -- 15\,mag from
La Palma during the 2004 May -- September season.  Our transit-hunting algorithm
selected 11,626 objects from the 184,442 stars within the range RA 18\,hr --
21\,hr.  We describe our thorough selection procedure whereby catalogue
information is exploited along with careful study of the SuperWASP data to
filter out, as far as possible, transit mimics.  We have identified
35 candidates which we recommend for follow-up observations.  

\end{abstract}



\section{Introduction}

The $\sim$200 exoplanets found to date have revolutionised our understanding of
how planetary systems form and evolve (\citealt{lin96}, \citealt{burrows00}). 
In particular, the discovery of `hot Jupiters' - Jovian-mass planets in orbits
of period$\leq$5\,days where conditions are too hot for them to have formed -
led to a reevaluation of the theory of orbital migration (\citealt{ipatov93},
\citealt{lin96}).  This class of planets have a comparatively high ($\sim$10\%)
probability of transiting across the face of their parent star.  Transiting
exoplanets are highly sought-after as an exceptional range of information can be
derived from them; to date 19\footnote{The Exoplanet Encyclopedia, exoplanet.eu} systems have been discovered.  Unambiguous
measurements of their physical and orbital parameters can be made, thereby
providing quantitative data against which to test evolutionary models (e.g.
\citealt{chabrier04}).  Research into the brightest transiting systems has, among
other ground-breaking advances, detected components of exoplanetary atmospheres
\citep{charbonneau02} and trailing exosphere (\citealt{vidalmadjar03},
\citealt{vidalmadjar04}), and placed limits on the existence of moons
\citep{brown01} and other planets in the same system \citep{steffen05}.  For a
comprehensive review of this exciting field, see \cite{charbonneau07}.  

In Section~\ref{sec:instobs} we introduce the SuperWASP project\footnote{www.superwasp.org} \citep{pollacco06}, a wide-angle photometric survey
searching for bright transiting planets.  Inevitably, all surveys looking for
low-amplitude, periodic eclipses will find those caused by stellar as well as
planetary objects.  \citet{brown03} and \citet{odonovan06} discuss several
astrophysical systems which can masquerade as transiting exoplanets.  The fact
that photometric data alone cannot identify transiting planets conclusively was
demonstrated by the OGLE project (e.g. \citealt{udalski04}), who have found to
date 177 eclipsing candidates, of which 5 have been confirmed as planetary.

We therefore need an effective filtering strategy to eliminate `false positives'
wherever possible in advance of time-consuming follow-up observations. 
Section~\ref{sec:candsel} describes our system of evaluating candidates to
select high-priority objects for follow-up.   We discuss the transit candidates
discovered within the RA range 18\,hr -- 21\,hr during SW-N's 2004 observing
season in Sections~\ref{sec:results} -- \ref{sec:concs}.

\section{Observations \& Data Reduction} 
\protect\label{sec:instobs} 

SuperWASP-North at the Isaac Newton Group of observatories, La Palma, Canary Islands
(hereafter SW-N), is a dedicated ultra-wide field photometric survey instrument
observing northern field stars of V$\sim$8--15\,mag.  Our science goals are
designed to explore long baseline (months--years) time domain astronomy, in
particular the search for transiting exoplanets.  The station supported five
cameras in 2004, each with a field of view of 7.8$^{\circ}\times$7.8$^{\circ}$. 
The instrumentation, observing strategy and data reduction pipeline are
described in detail in \citet{pollacco06}.  

The fields monitored were carefully selected to avoid the Galactic plane, in
contrast to some other transit surveys. The ecliptic plane was also avoided
wherever possible to minimise the sky background due to the Moon and to exclude
(Solar System) planets.  During the 2004 season we acquired lightcurves for some
6.7\,million objects.  

A custom-written, fully automated data reduction pipeline, developed by our
Consortium, has been applied to the 2004 data (see \citealt{pollacco06} \&
\citealt{cameron06}).  The photometric output is stored in, and exploited from, the
SuperWASP Data Archive held at the University of Leicester.  The pipeline routinely
achieves a photometric precision of $\sim$5\,millimag for stars with V$\sim$9.5,
rising to $\sim$0.02\,mag at V$\sim$13. This gives us a sample of
$\sim$1.2\,million stars with which to search for transits from SW-N's first season
(see \citealt{christian06} \& \citealt{lister07}). 

\subsection{RA range 18\,hr -- 21\,hr} 

The {\sc huntsman} algorithm \citep{cameron06} was applied to search for
transits in the lightcurves of stars with an RMS of $\lesssim$0.02\,mag or in
practice, those brighter than 13\,mag.  We note that transits can be detected
around late type stars of fainter magnitudes; these will be the subject of a
follow-up paper owing to the computational demands of searching much larger
numbers of stars.  We further constrain our searches to those stars for which we
have at least 500 photometric measurements, spanning a period of
$\geq$10\,nights.  In total, 184,442 stars met these conditions within the RA
range 18\,hr -- 21\,hr, and their distribution is summarised in
Table~\ref{tab:fields}.  

Our ability to detect transiting planets in these data depends on several
factors: the spectral types of monitored stars and the numbers for which we
achieve adequately precise photometry, the degree of crowding in the fields, our
observing window function and length of the dataset, and not least, the
frequency of hot Jovian exoplanets and the distributions of their periods and
other physical parameters.  

\cite{brown03} presents a thorough discussion of the transit recovery rates
expected for wide-field transit surveys, emphasising that it is a strong
function of planetary period for single-site observations such as ours.  He also
found that the rate of transit recovery depends on the distribution of spectral
types surveyed.  Early ground-based surveys (e.g. STARE, Vulcan) concentrated on
Galactic Plane fields in order to maximize the numbers of stars monitored. 
While large numbers of stars are crucial to any such survey, the larger
populations of early-type main sequence and giant stars in Galactic Plane fields
only serve to exacerbate the blending.  These stars do not contribute
significantly to the detection statistics since transit amplitude is inversely
proportional to the stellar radius, making planetary companions
difficult to detect.  

\begin{figure*} \def\subfigtopskip{4pt} \def\subfigbottomskip{8pt}
\def\subfigcapskip{4pt} \centering \begin{tabular}{cc}

\subfigure[]{\label{fig:starcensus}
\includegraphics[angle=270,width=8cm]{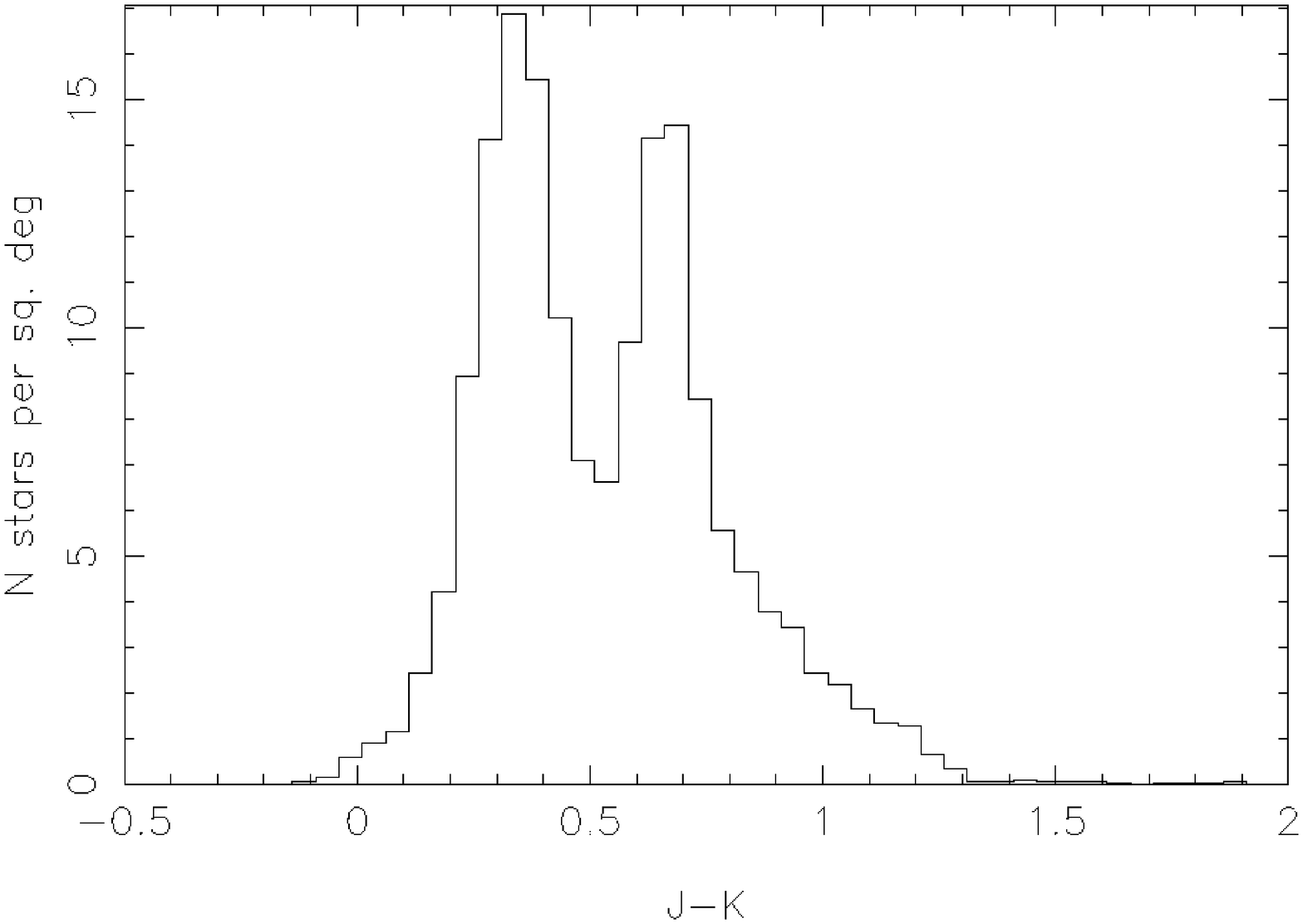}}
& 
\subfigure[]{\label{fig:ccdiagram}
\includegraphics[angle=270,width=8cm]{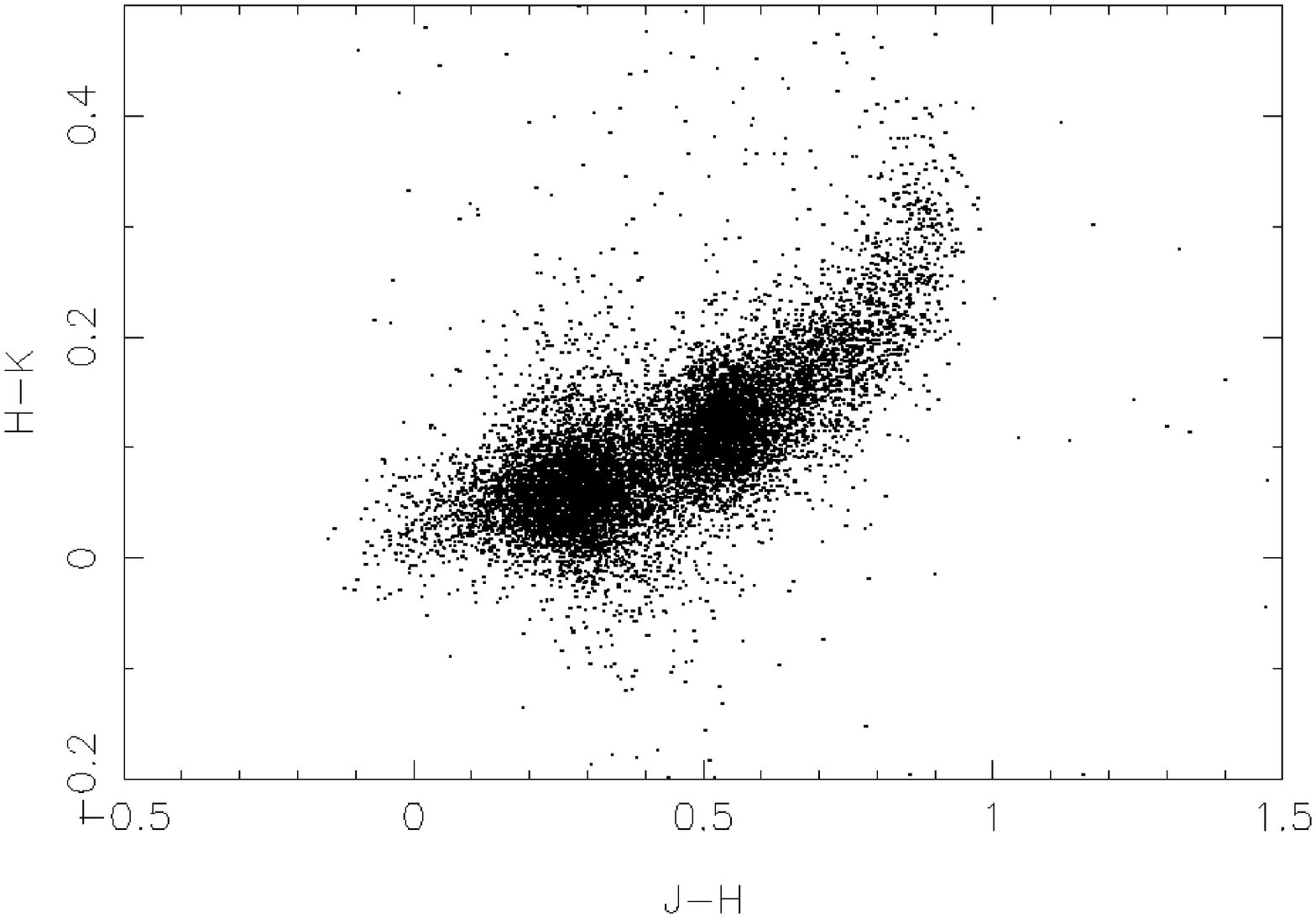}} \\

\end{tabular}
\caption{A census of the population of stars monitored in RA=18\,hr -- 21\,hr. 
The colour information is derived from the 2MASS catalogue.}
\protect\label{fig:starcensushist}
\end{figure*}

For this reason, SW-N has deliberately avoided the crowded Galactic Plane
fields, relying instead on our ultra-wide field of view to gather sufficient
numbers of stars.  Figure~\ref{fig:starcensus} provides a census of the spectral
types covered by our data from a representative field (SW2045+1628), deriving
colour information for each star from the 2MASS catalogue.  Main sequence stars
make up the dominant peak ($J-K$$<$0.5) in the SW-N sample.  To complement this,
Figure~\ref{fig:ccdiagram} presents the colour-colour diagram for the same data,
extending from $\sim$late A/early F stars down to approximately early-M type and
showing a cluster of points around the solar values of $J-H$$\sim$0.3,
$H-K$$\sim$0.1.  

\cite{pont06} highlighted the detrimental effect of residual systematic noise in the
photometry of this type of survey.  While we have gone to great lengths to minimise
these systematics (see Section~\ref{sec:huntsman}), the noise in our data is `red'
rather than `white'.  This has the effect of raising the signal-to-noise (S/N)
required to detect transiting systems (\cite{smith06} investigates the implications
for our survey characteristics in detail). In practical terms, an observer must
obtain longer baseline data including larger numbers of transits to boost the S/N.  

\begin{figure*}
\def\subfigtopskip{4pt}
\def\subfigbottomskip{8pt}
\def\subfigcapskip{4pt}
\centering
\begin{tabular}{cc}

\subfigure[SW1817+2326 129 nights of data]{\label{fig:comp1817+2326}
\includegraphics[angle=270,width=7.5cm]{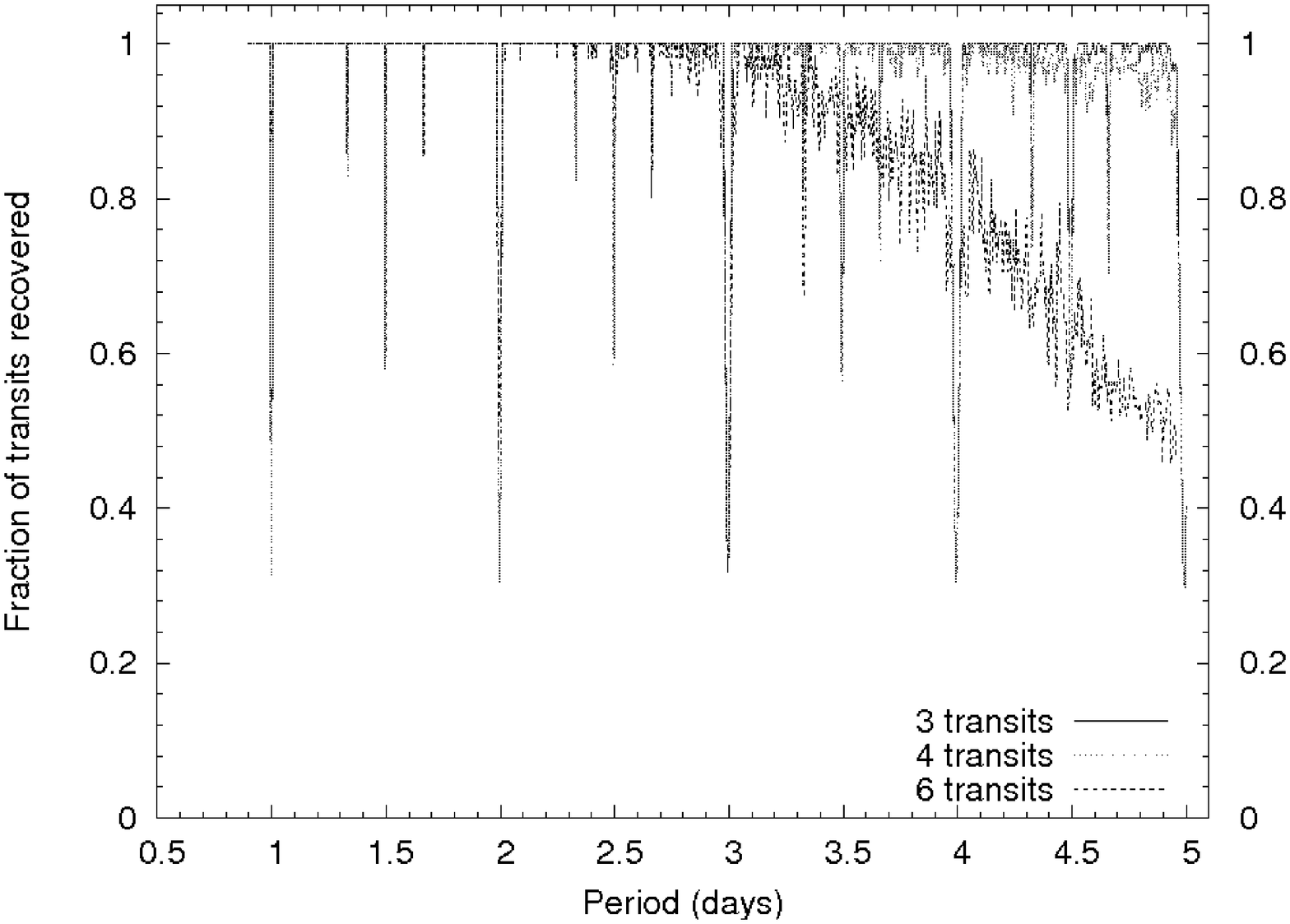}} 
&
\subfigure[SW1820+4723 116 nights of data]{\label{fig:comp1820+4723}
\includegraphics[angle=270,width=7.5cm]{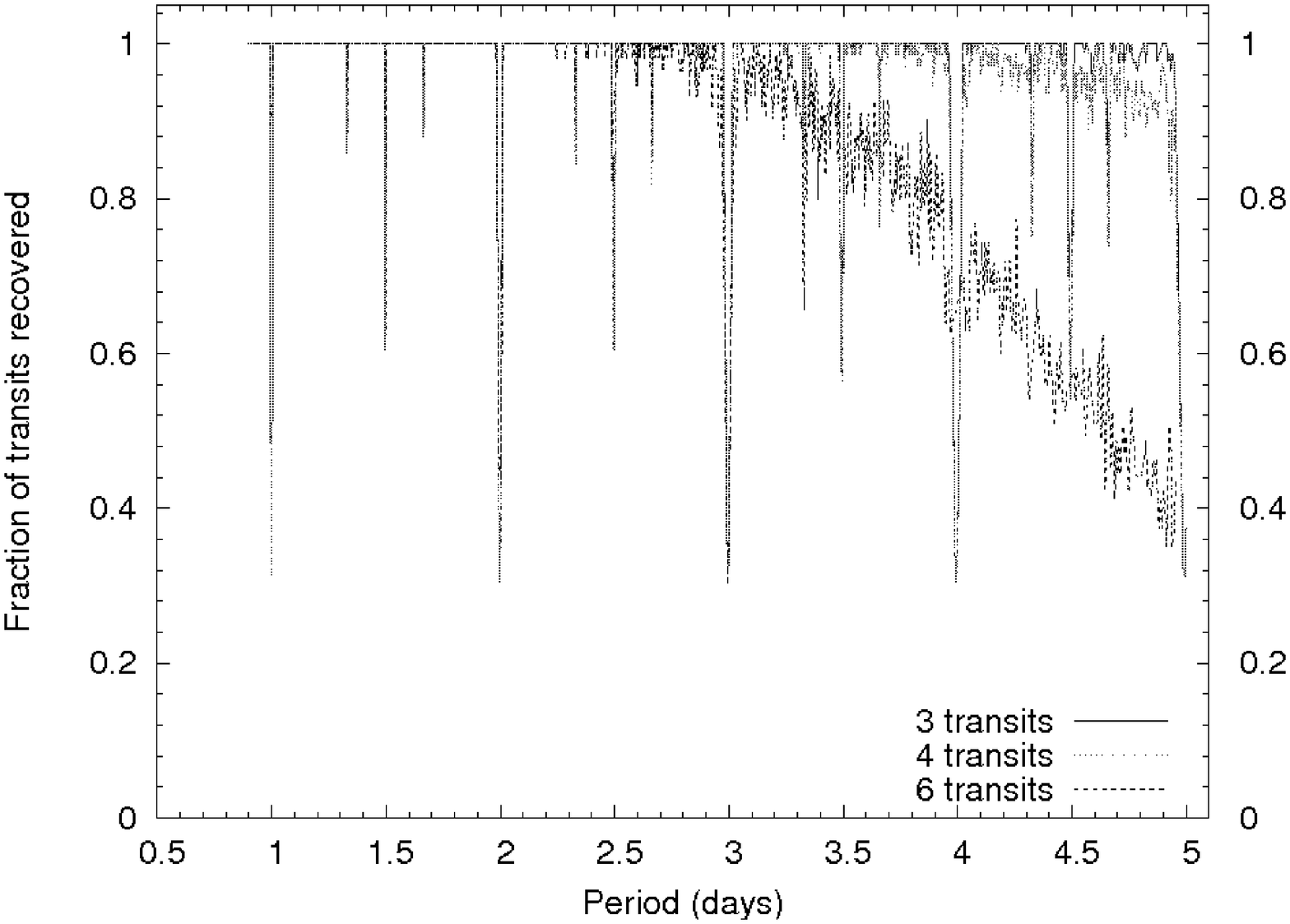}} \\

\subfigure[SW2045+0928 97 nights of data]{\label{fig:comp2045+0928}
\includegraphics[angle=270,width=7.5cm]{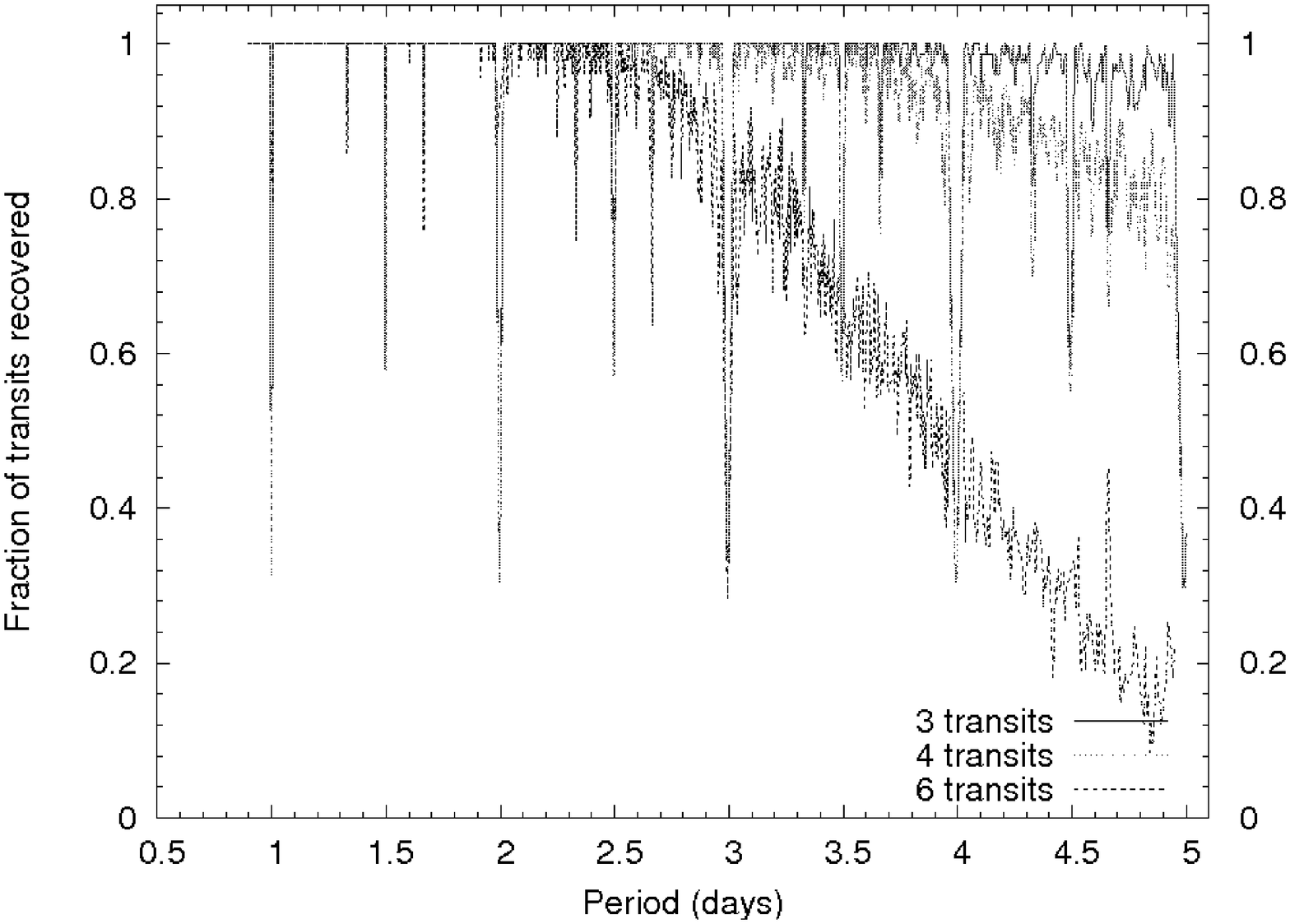}} \\

\end{tabular}
\caption{Probabilities of observing more than $N_{t}$ transits from the 2004 SW-N 
data for fields within the range RA=18\,hr -- 21\,hr, as a function of the 
planetary orbital period.}
\protect\label{fig:completeness}
\end{figure*}

To illustrate this, Figure~\ref{fig:completeness} demonstrates the probability
of detecting $N_{t}$ or more transits as a function of orbital period, $P$, from
the data obtained for several fields illustrating the range of observation
intervals spanned in this dataset.  A transit is counted as `observed' if data
were obtained within the phase range of $\phi$$<$0.1 $w/P$ or
$\phi$$>$$1-0.1w/P$, where $w$ is the expected transit duration, estimated from
$w \sim \frac{P R_{*}}{\pi a}$, where the separation, $a$ is calculated from
Kepler's third law.   All cases assumed the host star to be a dwarf star of mass
0.9\Msol and radius $R_{*}$ = 0.9\Rsol.  

SW produces well-sampled data of acceptable quality most nights and  generally
$\gtrsim$40 per cent of a given transit is observed during a detectable event. 
Setting a detection threshold of only three transits, our data returns 100\% of
all transiting systems for almost all orbital periods up to $\sim$5\,days.  As
our observations contain daytime gaps, the probability of identifying systems
with periods close to an integer multiples of 1\,day or 1.5\,days is only
$\sim$35\%.  The recovery rate also drops for $P\geq$4\,days, implying a longer
timebase of observations is required.  This is particularly noticeable in the
SW2045+0928 field, which has the shortest timebase.  When the required number of
transits is increased to 6, the detectable planets are confined to shorter
periods ($\leq$3\,days).  Two fields in the RA range, SW2115+0828 \&
SW2116+1527, have significantly less data than the others: 5 nights in total
(spread over $>$10\,nights).  They were included in the search automatically as
they pass the data criteria, but produced understandably fewer candidates.  

\begin{table*} 
\centering
\caption{J2000.0 coordinates of field centres surveyed in this work, giving for
each field the number of targets searched by the transit-hunting algorithm, 
and the number of stars selected by it.} 
\label{tab:fields} 
\begin{tabular}{llcclc} 
\hline 
 RA   	  & Dec        & No. nights & No. targets  & No. stars    & DAS \\
      	  &   	       &      	  &   	       & extracted  & \\
\hline
18 16 00  & +31 26 00  & 127   	  & 19,810     & 1396	    & 3 \\
18 17 00  & +23 26 00  & 129  	  & 24,220     & 1737	    & 4 \\
18 20 00  & +39 23 00  & 118  	  & 16,429     & 850	    & 4 \\
18 20 00  & +47 23 00  & 116  	  & 14,085     & 1011	    & 3 \\
20 45 00  & +09 28 00  & 97   	  & 21,390     & 1090	    & 1 \\
20 45 00  & +16 28 00  & 5	  & 2,259      & 90	    & 1 \\
20 45 00  & +16 28 00  & 116  	  & 25,971     & 1226	    & 2 \\
20 46 00  & +24 45 00  & 104  	  & 26,873     & 1669	    & 5 \\
21 14 00  & +16 28 00  & 116  	  & 17,747     & 1220	    & 3 \\
21 15 00  & +08 28 00  & 116  	  & 14,225     & 1200	    & 4 \\
21 15 00  & +23 51 00  & 5    	  & 689        &   55	    & 3 \\
21 16 00  & +15 27 00  & 5    	  & 744        &   82	    & 4 \\
\hline
Total 	  &   	       &      	  & 184,442    & 11,626     & \\
\hline
\end{tabular}
\end{table*}

\section{The Candidate Selection Procedure}
\protect\label{sec:candsel}

\subsection{Stage 1: The {\sc huntsman} Transit Finding Package}  
\protect\label{sec:huntsman}

\citet{cameron06} presents a detailed discussion of the corrections applied to
the SW-N photometry and the nature of the adapted-Box-fitting Least Squares
transit-hunting algorithm employed here.  It produces a `periodogram' of the
difference in the goodness-of-fit statistic $\Delta \chi^{2}$ between each model
relative to the no-transit case, plotted against transit frequency.  

{\sc huntsman} rejects obviously variable stars with $\chi^{2} > 3.5N$ ($N$=number of
datapoints), those less than 2 transits, and those solutions which have phase gaps in
the folded lightcurve greater than 2.5$\times$ the transit duration.   A candidate's
signal-to-red noise ratio, $S_{red}$, must be greater than 5.0, taking account of the
dominance of systematics in the photometric noise \citep{pont06}.   The strongest
peaks in the $\Delta \chi^{2}$ periodogram corresponding to brightening and dimming
are used to define the ``anti-transit ratio'' \citep{burke06}, \Ranti.  Candidates
must have \Ranti$>$1.5.  The algorithm also estimates the degree of ellipsoidal
variation in the out-of-transit lightcurve by producing a signal-to-noise statistic,
$S/N_{ellip}$.  

{\sc huntsman} selected 11,626 candidates in total from the fields in this dataset,
summarised in Table~\ref{tab:fields}.  In the next section we describe the subsequent
stages of systematic candidate assessment employed to eliminate interlopers.  

\subsection{Stage 2: Visual Assessment of Lightcurves} 
\protect\label{sec:visasses}

A visual inspection was made of each lightcurve in conjunction with the
corresponding periodogram of $\Delta \chi^{2}$ plotted against frequency.  For a
candidate to be selected, it had to display a clear transit with credible
amplitude, width and period and a smoothly sampled folded lightcurve.  Our
finite-length, single-site observations meant that lightcurves folded on
multiples of 1~day were by far the most common transit mimic.  The vast majority
of these cases were rapidly eliminated on sight as they showed no clear transit
signal.  Many classes of obvious stellar binaries or variables were also
removed from the candidate list.  

We developed the following 4-digit coding scheme to try to quantify this
subjective inspection process as far as possible.  

\begin{itemize}
\item Digit 1: Shape and visibility of the transit.  
  \begin{enumerate}
  \item Clear transit-shaped signal of credible width and depth.
  \item Shallow/noisy but clearly visible transit signal.
  \item Transit barely visible, either very shallow, lost in noise or ill-shaped.
  \item Partial transit or gaps around phase 0 but still showing clear transit
  morphology.
  \item Signs of a dip at phase 0 but no clear in/egress.
  \end{enumerate}
  
\item Digit 2: Out-of-transit lightcurve.
  \begin{enumerate}
  \item Clean and flat, no other variations.
  \item Noisy but flat.  
  \item Signs of ellipsoidal variation or suspected secondary eclipses (includes
  some candidates which have been folded on twice the period).
  \item Shows low-amplitude sinusoidal variation on short timescales, giving a 
  `knotty' appearance (can indicate that the lightcurve is folded on the wrong
  period).  
  \item Realistic variability of some other form out of transit.  
  \item Multi-level or `jumpy' lightcurves (can indicate the wrong period or
  photometry artifacts).  
  \end{enumerate}
  
\item Digit 3: Distribution of points in the folded lightcurve.
  \begin{enumerate}
  \item Smoothly sampled with a similar density of points throughout.
  \item Some minor regions with slightly lower density of points, retaining a
  clear signal. 
  \item Significant clumpy of data points (can indicate a pathological period).
  \end{enumerate}
  
\item Digit 4: Credibility of determined period.
  \begin{enumerate}
  \item No reason to doubt measured period, clear peak in $\Delta \chi^{2}$
  periodogram.  
  \item Period gives a secure signal visible in the folded lightcurve, but
  peak lies close to a known alias.  Sometimes associated with gaps in the folded
  lightcurve.  
  \item Signal visible in folded lightcurve but period is a known alias or peak
  lies at a commonly-occurring frequency.  
  \item Lightcurve suggests that the measured period is wrong.  
  \end{enumerate}
\end{itemize}

We emphasize that this is designed to guide the manual selection of targets,
rather than to provide a hard `statistic' on which a threshold cut might be
applied.  The code for each star was assessed on a case-by-case basis.  That
said, stars coded `[4,5]nnn', `n[5,6]nn', or `nn[3]n' were almost always
eliminated unless there were very clear signs of a planet-like transit within
the lightcurve despite its shortcomings.  Candidates with `[3]nnn' or `n[3,4]nn'
were assessed with caution.  However, targets with `n[3]nn' and/or `nnn[4]' that
otherwise showed a clear transit signal were retained and alternative periods
were explored.

This process uncovered several exciting, high S/N planetary candidates but 
inevitably also produced a number of cases close to the threshold.   Like all
our candidates, such cases were required to have believable transit-like
lightcurves and credible parameters sufficient to pass our criteria. 
Nevertheless, some stars, while intriguing, only just made the cut.  For
instance, some objects demonstrated a clear, transit-like lightcurve, but had a
period close to an integer multiple of 1\,day.  Others were close to the cut-off
for ellipsoidal variation.  Since objects in this category were potentially
low-mass-star or brown dwarf binaries and therefore of independent interest,
they were retained in the candidate list but not shortlisted after Stage 4.

\subsection{Stage 3: Selection Criteria}
\protect\label{sec:sortncuts}

Surviving candidates were subject to the following requirements:
\begin{itemize}
\item The $S_{red}$ must be at least 8.0.  
\item The period must be $\geq 1.05$\,days. This criterion is implemented in order
to reject candidates folded on one-day aliases.  
\item The number of transits observed must be $\geq 3$.  
\item Anti-transit ratio must be greater than 2.0.  
\item The $S/N_{ellip}$ should be less than about 8.0.  While this threshold was
generally reliable, a number of objects were found which had a value of $S/N_{ellip}$
exceeding this threshold yet the out-of-transit lightcurve appeared flat to
visual inspection.  In cases with exceptionally clear, believable transit-like
lightcurves, a degree of human discretion was afforded.  
\end{itemize}

We elected not to search for transits with periods less than 1.05\,d as early
test runs resulted in unfeasibly large numbers of false alarms folded on periods
that are integer fractions of 1\,d.  It was decided that separate searches would
be run for very short (and long) period planets after the present work had
cultivated experience in false-positive rejection.  

\subsection{Stage 4: Compilation of Catalogue Data}

Objects surviving this cull were submitted to SW's online {\em Variable Star
Investigator} tool \citep{wilson2006}, which performs automated queries on a
number of existing photometric catalogues including 2MASS \citep{2mass}, Tycho-2
\citep{tycho2}, Simbad \citep{simbad} and Hipparcos \citep{hipparcos} among
others.  This provided for each candidate a table of multi-colour photometric
information, lists of other nearby objects falling within SW-N's photometric
aperture of $\sim$48\arcsec and 3\arcmin$\times$3\arcmin\, and findercharts from
DSS \citep{dss} and 2MASS.  The latter information was used to assess the degree
to which each star is blended in the SW-N photometry, a major cause of false
positives.  If a brighter object was found within a candidate's aperture, then
that star was removed from the target list. 

Two separate temperature-colour relationships were employed to estimate the
temperature of each candidate star, assuming it to be main sequence and that the
measured colours were not contaminated by light from the companion (as expected under
the exoplanet hypothesis).  The first relationship uses Tycho-2 $V_{T}$ and 2MASS $K$
with an uncertainty of 91K and the second, 2MASS $J$ \& $H$ (uncertainty 186K):

\begin{eqnarray}	   
T_{eff} &=& 213.19(V_{T}-K)^{2} - 1920.1(V_{T}-K) + 8335.7,\\
T_{eff} &=& -4369.5(J-H) + 7188.2,
\end{eqnarray}

These were derived from the temperature data on 30,000 FGK dwarf stars presented
in \citet{ammons06} for which the precision of the Tycho-2 and 2MASS photometry
is better than 1\%.  The use of the second relation, based on infrared colours,
is more sensitive to the presence of cooler companion bodies.  A significant
discrepancy between the two temperature (and hence radius) estimates can
therefore indicate the presence of a companion (often stellar).  

The colour indices, together with the USNO-B1.0 proper motions ($\mu$) were also used
as an indicator of the luminosity class of the target.  The Reduced Proper Motion
($RPM_{J}$) was computed from:

\begin{equation}
RPM_{J} = J + 5\log_{10}{\mu}.
\end{equation}

Plotted against the $J-H$ index, dwarfs are separated from giants, as they lean 
towards higher values of $RPM_{J}$ and low $J-H$.  A polynomial boundary was set
between the two groups so that {\em VSI} could issue a warning when this
threshold is crossed.  \cite{brown03} demonstrated that $J-K$ colours can also
act as a rough indicator of luminosity based on data from the {\sc stare}
project.  Taking this and \cite{charbonneau04} as a guide, {\em VSI} flags any
star with a $J-K$$>$0.7 as a possible giant.  

The derived $T_{eff}$ values were then used to estimate the spectral type of the
host star based on data from \cite{astroquantities} while the radius and mass
were estimated using data from \cite{gray92}.  For $T_{eff}$$<$7000\,K, the RMS
of the fit of polynomial functions describing $T_{eff}$ .vs. radius and mass
were 0.016\,K in both cases.  

A minimum limit on the radius of the companion, $R_{p}$, was estimated from the
stellar radius, $R_{*}$ and the transit amplitude, $\delta$, using the
relationship derived by \cite{tingley05} for the I-band:

\begin{equation}
R_{c} \approx \sqrt{\frac{\delta}{1.3} R_{*}^{2}},
\end{equation}

Our unfiltered, wide bandpass photometry is dominated by the red sensitivity of
the CCD and the uncertainty introduced by approximating to I-band is smaller
than that of the stellar radius estimate.  

Although electron degeneracy means that there can be little difference in the
radii of objects between 0.5\Msol -- $\sim$1\Mj, we concentrated on objects with
predicted $R_{p}$ of less than $\sim$2\Rj.  To aid selection, we also employed
the $\eta_{p}$ diagnostic derived by \citet{tingley05}, comparing the observed
transit duration $D_{obs}$ with that theoretically predicted ($D_{pred}$) for a
transiting hot Jupiter:

\begin{eqnarray}
\eta_{p} &=& \frac{D_{obs}}{D_{pred}} \\
      	&=& \frac{D_{obs}}{2Z(1+\sqrt{1.3/\delta})} 
\left ( \frac{2\pi G M_{\odot}}{P} \right )^{\frac{1}{3}} R_{c}^{-\frac{7}{12}} 
R_{\odot}^{-\frac{5}{12}} \left ( \frac{1.3}{\delta} \right )^{\frac{5}{24}},
\end{eqnarray}

where $Z$ is a factor representing the effects of the projected orbital
inclination, set equal to 1 (see discussion in \citealt{tingley05}), $\delta$ is the
depth of the transit and $P$ is the period.  Strong exoplanet candidates are
expected to have $\eta_{p} \sim 1$.  However, caution was exercised when using
this criterion to judge our candidates since the value of $R_{p}$ depends heavily
on the value of $R_{*}$, the estimate of which is subject to significant
uncertainty when made from colour indices alone.

Our assessment of characteristics was quantified using three additional indices
from the following coding scheme.  

\begin{itemize}

\item Planetary radius, $R_{p}$.\\ 
  A. $R_{p}$ $<$ 1.6\,\Rj.\\
  B. 1.6 $\leq$ $R_{p}$ $\leq$ 1.75 \Rj. \\
  C. $R_{p}$ $\geq$ 1.75 \Rj.\\

\item Exoplanet diagnostic $\eta_{p}$. \\
  A. 0.5 $\geq$ $\eta_{p}$ $\geq$ 1.5.\\
  B. $\eta_{p}$ $<$ 0.5.\\
  C. $\eta_{p}$ $\geq$ 1.5.\\
  
\item Blending. \\
  A. No other objects within aperture.\\
  B. 1 or 2 other objects less than 5\,mags fainter than target within
  aperture.\\
  C. More than 2 objects less than 5\,mags fainter than target within
  aperture.\\
  D. Brighter objects within the aperture. \\

\end{itemize}  

Each candidate was then assessed in turn, taking into account all available data,
and a final shortlist of high-priority candidates was produced.  In the next
section we summarise the results for stars in the RA range 18\,hr - 21\,hr.  

It can be seen from this discussion that some selection cuts are repeated during
subsequent stages using increasingly stringent thresholds.  For instance, {\sc
huntsman} executes an automatic cut of objects with $S_{red}<5.0$, while at Stage
3, a further cut is made at $S_{red}<8.0$.  In exploring the first large-scale
transit hunting results from SW, we took a cautious approach in order to
investigate the most effective selection criteria.  Not wanting the algorithm to
dismiss interesting objects before human interpretation, the initial thresholds
were set low, systematically rising for successive stages of evaluation. 
Needless to say, lessons learned from this season's work will enable us to
streamline the procedure in future.  

\section{Results}
\protect\label{sec:results}

The {\sc huntsman} algorithm flagged 11,626 objects for attention.  Stage 2
visual inspection concluded that 775 of these were of geniune interest.  The
Stage 3 selection requirements detailed in Section~\ref{sec:sortncuts} sifted
this list down to 77 stars, the details of which are presented in
Table~\ref{tab:stage3list}.  

The visual lightcurve assessment of each star is quantified by a 4-digit code in
column 11.  At this stage, the list contained 19 borderline candidates, many of which
are likely low-mass binaries.  As these objects are of independent interest, we have
included their full parameters in Tables~\ref{tab:stage3list} \&
\ref{tab:stage4list}, marked by $^{\dagger}$, although these objects were not carried
through to the final shortlisting as the present paper deals with planetary
candidates only.  

\begin{table*}
\centering
\caption{Initial list of candidates after Stage 3. Borderline candidates are marked with $^{\dagger}$ and are listed 
for information.  } 
\protect\label{tab:stage3list}
\begin{tabular}{lcccccccccc}
\hline
 Identifier   	      	    & $V_{SW}$  & Period   & Duration  & $\delta$  & $N_{tr}$  & $S_{red}$	& $\Delta \chi^{2}$ & $S/N_{ellip}$  & $\Ranti$ & Code \\ 
1SWASP...	            & (mag)   	& (days)   & (hrs)     & (mag)     &           &        	&     	      	    & 	      	   &  	      	& \\
\hline
$^{\dagger}$J175919.79+353935.1  & 11.824	 & 4.846186 & 4.272  & 0.026  & 6     & 9.264	 & 338.197	     & 0.605	    & 5.327	 & 2223 \\
$^{\dagger}$J180103.13+511557.1  & 9.988	 & 4.785081 & 3.672  & 0.0145 & 8     & 11.215   & 928.888	     & 2.401	    & 3.467	 & 2423 \\
J180304.96+264805.4  & 11.782	 & 2.364723 & 5.136  & 0.0254 & 20    & 13.453   & 1454.973	     & 4.145	    & 9.616	 & 3211 \\
J180726.64+224227.9  & 12.568	 & 2.121623 & 5.256  & 0.0173 & 21    & 9.548	 & 375.908	     & 4.150	    & 3.302	 & 3314 \\
$^{\dagger}$J181129.19+235412.4  & 12.884	 & 4.234895 & 8.568  & 0.0578 & 16    & 11.584   & 2580.622	     & 1.699	    & 9.145	 & 1314 \\
{\bf J181317.03+305356.0}  & {\bf 12.046}  & {\bf 4.498677} & {\bf 1.92}   & {\bf 0.0194} & {\bf 13} & {\bf 14.446} & {\bf 540.914} & {\bf 4.992} & {\bf 6.704} & {\bf 1134} \\
{\bf J181454.99+391146.0}  & {\bf 12.796}  & {\bf 1.102625} & {\bf 1.56}   & {\bf 0.0235} & {\bf 25} & {\bf 13.297} & {\bf 219.564} & {\bf 0.659} & {\bf 5.134} & {\bf 1212} \\
{\bf J181958.25+492329.9}  & {\bf 10.6}    & {\bf 2.368548} & {\bf 2.424}  & {\bf 0.0061} & {\bf 16} & {\bf 10.759} & {\bf 145.924} & {\bf 0.241} & {\bf 2.902} & {\bf 3111} \\
J182127.09+200011.7  & 11.449	 & 2.647752 & 4.248  & 0.0366 & 18    & 16.824   & 2831.871	     & 2.789	    & 15.396	 & 1111 \\
J182131.07+483735.5  & 12.164	 & 1.809191 & 2.832  & 0.0167 & 16    & 9.781	 & 470.931	     & 2.314	    & 4.140	 & 3211 \\
J182333.22+222801.2  & 12.788  	 & 1.821008 & 3.432  & 0.0421 & 18    & 17.315 	 & 983.6475    	     & 8.064   	    & 12.2324  	 & 1211 \\
$^{\dagger}$J182339.64+210805.5  & 12.794	 & 1.585846 & 2.088  & 0.0245 & 22    & 10.374   & 306.613	     & 6.991	    & 2.312	 & 1314 \\
J182346.12+434241.3  & 11.771	 & 2.969366 & 3.384  & 0.0295 & 11    & 19.982   & 444.963	     & 0.895	    & 11.656	 & 1124 \\
{\bf J182620.36+475902.8}  & {\bf 11.584}  & {\bf 3.04365}  & {\bf 4.032}  & {\bf 0.0628} & {\bf 13} & {\bf 24.415} & {\bf 10754.299}& {\bf 4.225}& {\bf 11.474}& {\bf 1112} \\
$^{\dagger}$J182626.38+374954.8  & 11.614	 & 4.698312 & 4.944  & 0.0157 & 8     & 13.104    & 317.828	     & 1.643	    & 6.417	 & 2213 \\
J182916.00+235724.8  & 12.043	 & 4.465326 & 1.752  & 0.0373 & 7     & 12.356    & 578.442	     & 3.163	    & 11.565	 & 2224 \\
{\bf J182924.67+232200.2}  & {\bf 11.331}   & {\bf 3.678186} & {\bf 2.952}  & {\bf 0.0173} & {\bf 10}& {\bf 14.248}  & {\bf 244.980}& {\bf 2.174} & {\bf 2.639} & {\bf 3123} \\
$^{\dagger}$J182927.04+233217.1  & 10.8	 & 4.903747 & 4.704  & 0.0063 & 9     & 8.214	  & 146.459	     & 1.954	    & 2.299	 & 3214 \\
$^{\dagger}$J183043.97+230526.1  & 9.31	 & 3.680977 & 4.296  & 0.0098 & 9     & 11.139    & 628.645	     & 3.278	    & 4.027	 & 2311 \\
{\bf J183104.01+323942.7}  & {\bf 11.027}   & {\bf 2.378781} & {\bf 1.776}  & {\bf 0.0089} & {\bf 15}& {\bf 11.013}  & {\bf 256.230}& {\bf 2.065} & {\bf 4.873} & {\bf 2111} \\
J183104.12+243739.3  & 12.789    & 1.492383 & 1.92   & 0.0197 & 20    & 10.218    & 188.009	    & 4.411        & 2.836	& 1314 \\
{\bf J183431.62+353941.4}  & {\bf 10.485}   & {\bf 1.846796} & {\bf 2.28}   & {\bf 0.0127} & {\bf 17}& {\bf 12.111}  & {\bf 787.959}& {\bf 0.691} & {\bf 3.635} & {\bf 1111} \\
 $^{\dagger}$J183517.51+390316.2  & 9.823    & 4.073428 & 5.16   & 0.012  & 8     & 9.282     & 1320.766	    & 5.377        & 2.225	& 1123 \\
 J183723.62+373721.9  & 11.851   & 3.300887 & 4.32   & 0.0251 & 13    & 13.599 	  & 841.3629   	    & 8.779   	   & 10.1919  	& 1213 \\
{\bf J183805.57+423432.3}  & {\bf 12.641}    & {\bf 3.515957} & {\bf 4.104}  & {\bf 0.0197} & {\bf 9} & {\bf 8.815}  & {\bf 127.097}  & {\bf 0.999} & {\bf 3.693} & {\bf 3131} \\
{\bf J184119.02+403008.4}  & {\bf 12.157}    & {\bf 3.734014} & {\bf 4.224}  & {\bf 0.0148} & {\bf 11}& {\bf 9.449}  & {\bf 198.451}  & {\bf 0.502} & {\bf 2.720} & {\bf 3133} \\
{\bf J184303.62+462656.4}  & {\bf 11.935}    & {\bf 3.338103} & {\bf 4.08}   & {\bf 0.0265} & {\bf 11}& {\bf 12.248} & {\bf 1065.843} & {\bf 1.867} & {\bf 9.098} & {\bf 4124} \\
 J202820.25+094651.0  & 11.108    & 2.146933 & 4.776  & 0.0085 & 16    & 12.533    & 294.491	    & 3.910        & 5.344	& 2111 \\
{\bf J202824.02+192310.2}  & {\bf 12.16}     & {\bf 1.257835} & {\bf 2.424}  & {\bf 0.0222} & {\bf 23} & {\bf 13.111} & {\bf 589.550} & {\bf 3.355} & {\bf 7.095} & {\bf 1111} \\
 $^{\dagger}$J202907.09+171631.7  & 12.786    & 4.117398 & 4.968  & 0.0309 & 11    & 9.996     & 450.143	    & 1.126        & 3.844	& 2223 \\
{\bf J203054.12+062546.4}  & {\bf 11.98}     & {\bf 2.152102} & {\bf 1.296}  & {\bf 0.0168} & {\bf 11}& {\bf 9.463}  & {\bf 217.184}& {\bf 5.522}   & {\bf 3.262} & {\bf 1111} \\
 $^{\dagger}$J203229.10+132820.9  & 12.471    & 4.632829 & 4.608  & 0.047  & 9     & 12.773    & 1385.902	    & 2.670        & 11.318	& 2213 \\
 J203247.55+182805.3  & 12.157    & 2.522688 & 7.776  & 0.0118 & 22    & 11.579    & 308.408	    & 0.875        & 5.324	& 3113 \\
{\bf J203314.77+092823.4}  & {\bf 11.78}     & {\bf 1.753056} & {\bf 3.048}  & {\bf 0.0316} & {\bf 18}& {\bf 14.221} & {\bf 2154.619}& {\bf 7.012}  & {\bf 8.927} & {\bf 1111} \\
{\bf J203315.84+092854.2}  & {\bf 11.943}    & {\bf 1.752371} & {\bf 2.784} & {\bf 0.0413} & {\bf 16} & {\bf 13.545} & {\bf 2796.5991} & {\bf 9.663} & {\bf 11.4699}& {\bf 1211} \\
 J203543.98+072641.1  & 10.094    & 1.85463  & 2.76   & 0.0195 & 13    & 16.884    & 3354.689	    & 1.083        & 10.542	& 1112 \\
{\bf J203704.92+191525.1}  & {\bf 11.301}    & {\bf 1.68011}  & {\bf 1.416}  & {\bf 0.0095} & {\bf 16}& {\bf 9.344}  & {\bf 245.231}& {\bf 3.226}   & {\bf 2.826} & {\bf 3111} \\
 J203717.02+114253.5  & 11.327    & 3.118049 & 2.496  & 0.0274 & 8     & 12.11     & 2792.375	    & 3.870        & 21.267	& 1111 \\
{\bf J203906.39+171345.9}  & {\bf 9.716} & {\bf 1.348858} & {\bf 1.968} & {\bf 0.0173} & {\bf 18} & {\bf 17.059}	 & {\bf 2934.2539} & {\bf 8.365} & {\bf 47.1445} & {\bf 1124} \\
$^{\dagger}$J203932.30+162451.1  & 10.904    & 1.520504 & 8.976  & 0.02   & 39    & 14.359    & 10012.064	    & 0.966        & 2.936	& 2311 \\
{\bf J204125.28+163911.8}  & {\bf 11.243}    & {\bf 1.221506} & {\bf 2.88}   & {\bf 0.008}  & {\bf 28}& {\bf 11.48}  & {\bf 518.131} & {\bf 2.703} & {\bf 3.151}  & {\bf 3111} \\

\hline  									         
\end{tabular}							        		      
\end{table*}

\begin{table*}
\centering
\contcaption{Initial list of candidates after Stage 3. Borderline candidates are marked with $^{\dagger}$ and are listed 
for information.  Parenthesis around an object indicates that spectroscopic data are discussed in Section~\ref{sec:specdata}.} 
\protect\label{tab:initlist2}
\begin{tabular}{lcccccccccc}
\hline
 Identifier   	      	    & $V_{SW}$  & Period   & Duration  & $\delta$  & $N_{tr}$  & $S_{red}$  & $\Delta \chi^{2}$ & $S/N_{ellip}$  & $\Ranti$ & Code \\
 1SWASP...        	    & (mag)   	& (days)   & (hrs)     & (mag)  &           &  	      	 &		    &		   &		& \\
\hline

 $^{\dagger}$J204142.31+052007.5  & 12.422    & 3.216912 & 4.776  & 0.0279 & 8     & 10.462   & 317.078	    & 0.533        & 7.574      & 2232 \\
{\bf J204142.49+075051.5}  & {\bf 12.082}   & {\bf 1.381342} & {\bf 1.968}  & {\bf 0.0096} & {\bf 19}& {\bf 11.739} & {\bf 165.756} & {\bf 1.413}& {\bf 7.403} & {\bf 3114} \\
 J204211.19+240145.1  & 11.588    & 1.792911 & 2.424  & 0.0518 & 10    & 14.079   & 1074.758	    & 6.535        & 2.917      & 4134 \\
{\bf J204323.83+263818.7}  & {\bf 11.561}   & {\bf 1.419959} & {\bf 1.2}& {\bf 0.0369} & {\bf 10}    & {\bf 18.496}   & {\bf 179.712}   & {\bf 0.971} & {\bf 2.440} & {\bf 1224} \\
 $^{\dagger}$J204328.95+054823.1  & 12.616    & 3.939179 & 2.328  & 0.0617 & 10    & 16.96    & 1989.211	    & 5.293        & 17.029     & 1322 \\
 (J204456.57+182136.0  & 12.596    & 2.71611  & 4.584  & 0.0202 & 16    & 12.164   & 525.040	    & 1.287        & 14.612     & 3214) \\
{\bf J204617.02+085412.0}  & {\bf 12.28}    & {\bf 1.947141} & {\bf 2.184}  & {\bf 0.0095} & {\bf 14}& {\bf 9.436}  & {\bf 92.943} & {\bf 0.647} & {\bf 2.163} & {\bf 3112} \\
{\bf J204712.42+202544.5}  & {\bf 12.386}   & {\bf 2.61264}  & {\bf 2.064}  & {\bf 0.0275} & {\bf 10}& {\bf 13.103} & {\bf 355.276} & {\bf 3.327} & {\bf 6.693} & {\bf 2211} \\
{\bf J204745.08+103347.9}  & {\bf 11.648}   & {\bf 3.235407} & {\bf 3.648}  & {\bf 0.0289} & {\bf 8} & {\bf 16.376} & {\bf 1336.114} & {\bf 5.186} & {\bf 16.348} & {\bf 1112} \\
 $^{\dagger}$J204905.55+110000.4  & 12.891    & 1.371571 & 1.584  & 0.023  & 20    & 12.8     & 244.376	    & 4.343        & 4.619      & 1311 \\
{\bf J205027.33+064022.9}  & {\bf 10.164}   & {\bf 1.229345} & {\bf 3.192}  & {\bf 0.0096} & {\bf 20}& {\bf 13.641} & {\bf 1198.006} & {\bf 6.691} & {\bf 5.830} & {\bf 3111} \\
 $^{\dagger}$J205218.75+182330.0  & 11.991    & 2.197814 & 3.48   & 0.0441 & 16    & 19.038   & 3378.642	    & 3.256        & 22.912     & 1131 \\
 J205223.03+151046.8  & 11.493    & 1.454887 & 2.4    & 0.0301 & 23    & 19.47    & 3389.000	    & 2.060        & 21.470     & 1114 \\
 J205302.40+201748.3  & 10.853    & 4.931719 & 8.88   & 0.0084 & 9     & 8.327    & 360.930         & 0.093   	   & 2.553    	& 3123 \\
{\bf J205308.03+192152.7}  & {\bf 11.13}    & {\bf 1.676449} & {\bf 2.736}  & {\bf 0.0068} & {\bf 23}& {\bf 10.406} & {\bf 213.332} & {\bf 0.668} & {\bf 3.508} & {\bf 2111} \\
 J205438.05+105040.7  & 11.428    & 2.623442 & 2.664  & 0.0405 & 11    & 16.117   & 3278.368	    & 4.645        & 8.251      & 1114 \\
{\bf J210009.75+193107.1}  & {\bf 10.422}   & {\bf 3.054875} & {\bf 2.424}  & {\bf 0.0082} & {\bf 9} & {\bf 8.877}  & {\bf 303.455} & {\bf 1.646} & {\bf 2.612} & {\bf 3113} \\
 $^{\dagger}$J210130.24+190021.7  & 12.14     & 2.683587 & 1.584  & 0.0697 & 12    & 23.253	& 1860.082	    & 5.557	   & 31.460   	& 1311 \\
{\bf J210151.43+072326.7}  & {\bf 12.476}   & {\bf 2.220785} & {\bf 2.472}  & {\bf 0.0138} & {\bf 15}& {\bf 8.764}  & {\bf 108.956}& {\bf 0.948} & {\bf 2.396} & {\bf 3213} \\
 $^{\dagger}$J210231.79+101014.5  & 12.635    & 1.506187 & 1.608  & 0.0296 & 16    & 14.97    & 258.766	    & 6.760        & 2.971   	& 1332 \\
{\bf J210318.01+080117.8}  & {\bf 11.909}   & {\bf 1.223824} & {\bf 1.92}   & {\bf 0.0167} & {\bf 24}& {\bf 12.784} & {\bf 466.284} & {\bf 0.248} & {\bf 4.999} & {\bf 1111} \\
 $^{\dagger}$J210335.82+125637.6  & 12.387    & 1.447543 & 2.856  & 0.0146 & 24    & 9.082	& 268.208	    & 1.208	   & 4.420    	& 2213 \\
{\bf J210352.56+083258.9}  & {\bf 11.636}   & {\bf 3.89368}  & {\bf 3.504}  & {\bf 0.0227} & {\bf 11}& {\bf 13.38}  & {\bf 953.011}& {\bf 7.066} & {\bf 11.909}  & {\bf 1112} \\
{\bf J210909.05+184950.9}  & {\bf 9.912}    & {\bf 2.91879}  & {\bf 2.664}  & {\bf 0.0083} & {\bf 13}& {\bf 9.718}  & {\bf 801.126}& {\bf 0.121} & {\bf 3.041} & {\bf 3112} \\
{\bf J210912.02+073843.3}  & {\bf 11.262}   & {\bf 1.36983}  & {\bf 2.28}   & {\bf 0.0213} & {\bf 22}& {\bf 16.035} & {\bf 1594.4681}& {\bf 12.508} & {\bf 20.6406} & {\bf 1111} \\
 J211127.41+182653.3  & 12.291    & 4.216933 & 3.168  & 0.0464 & 8     & 20.186   & 1043.324        & 0.775	   & 25.743	& 2211 \\
 J211417.15+112741.0  & 11.246    & 2.519934 & 2.784  & 0.0336 & 11    & 10.555   & 2902.904        & 1.290	   & 3.334	& 3214 \\
 J211448.98+203557.1  & 12.453    & 4.864666 & 4.632  & 0.0525 & 8     & 13.794	& 1939.578	    & 4.542	   & 16.558   	& 1212 \\
{\bf J211608.42+163220.3}  & {\bf 11.308}   & {\bf 3.468244} & {\bf 1.992}  & {\bf 0.0131} & {\bf 10}& {\bf 13.461} & {\bf 228.680} & {\bf 0.781} & {\bf 5.584} & {\bf 1111} \\
{\bf J211645.22+192136.8}  & {\bf 9.432}    & {\bf 1.466001} & {\bf 1.68}   & {\bf 0.012}  & {\bf 16}& {\bf 12.273} & {\bf 1379.556} & {\bf 2.033} & {\bf 3.516} & {\bf 2124} \\
 J211817.92+182659.9  & 12.395    & 4.419854 & 3.36   & 0.0274 & 9     & 12.149   & 716.481	    & 1.194        & 9.733      & 3214 \\
{\bf J212532.55+082904.4}  & {\bf 11.343}   & {\bf 3.125014} & {\bf 2.688}  & {\bf 0.0267} & {\bf 9} & {\bf 14.313} & {\bf 1013.935} & {\bf 1.980} & {\bf 7.591} & {\bf 1212} \\
 J212749.35+190246.0  & 12.317    & 4.870738 & 3.408  & 0.0438 & 10    & 10.158   & 1332.879	    & 1.191        & 2.215      & 2224 \\
 $^{\dagger}$J212815.28+082933.7  & 10.165    & 4.91815  & 5.592  & 0.0083 & 9     & 8.493    & 374.959	    & 0.644        & 2.249      & 3414 \\
{\bf J212843.62+160806.2}  & {\bf 11.453}    & {\bf 1.375647} & {\bf 2.64}   & {\bf 0.0159} & {\bf 25} & {\bf 15.572} & {\bf 1288.665} & {\bf 8.841} & {\bf 9.5499} & {\bf 1111} \\
{\bf J212855.03+075753.5}  & {\bf 12.241}    & {\bf 4.688048} & {\bf 1.92}   & {\bf 0.0297} & {\bf 5}& {\bf 9.54}   & {\bf 188.137} & {\bf 0.953} & {\bf 2.503} & {\bf 3213} \\

\hline  			
\end{tabular}										   
\end{table*}
											   
The remaining 58 objects surviving to Stage 4 could be grouped into three broad
classes.  Twenty-four stars received the best grades ( between `1111' and
`2222'), indicating a clear, credible transit signal in a flat, well sampled
lightcurve.  Seventeen objects were flagged as displaying a credible transit
signal, but on a period not correctly identified.  A further 17 candidates were
found to show plausible transits signals and were only downgraded on the grounds
of low S/N.  

At this stage we attempted to eliminate astrophysical false positives by
considering the catalogue information available, estimating the companion radius
and corresponding value of $\eta_{p}$ and assessing the degree of blending in
the field.  

Table~\ref{tab:stage4list} gives the full set of parameters for these
candidates.  Each candidate was then evaluated on its merits, including a visual
examination of both folded and unfolded lightcurves.  Where relevant, target 
lightcurves were re-folded on the periods of the alternative peaks from the
periodogram.  In a small number of cases, this showed that the true period fell
outside {\sc huntsman}'s search range of 0.9 -- 5\,days.  We then applied the
algorithm developed by \citet{schwarzenberg89},\citet{schwarzenberg99} (referred
to as S-C) to determine the correct period.  

\begin{figure*}
\def\subfigtopskip{4pt}
\def\subfigbottomskip{8pt}
\def\subfigcapskip{4pt}
\centering
\begin{tabular}{cc}

\subfigure[1SWASP J181317.03+305356.0]{\label{fig:J181317.03+305356.0}
\includegraphics[angle=270,width=6.5cm]{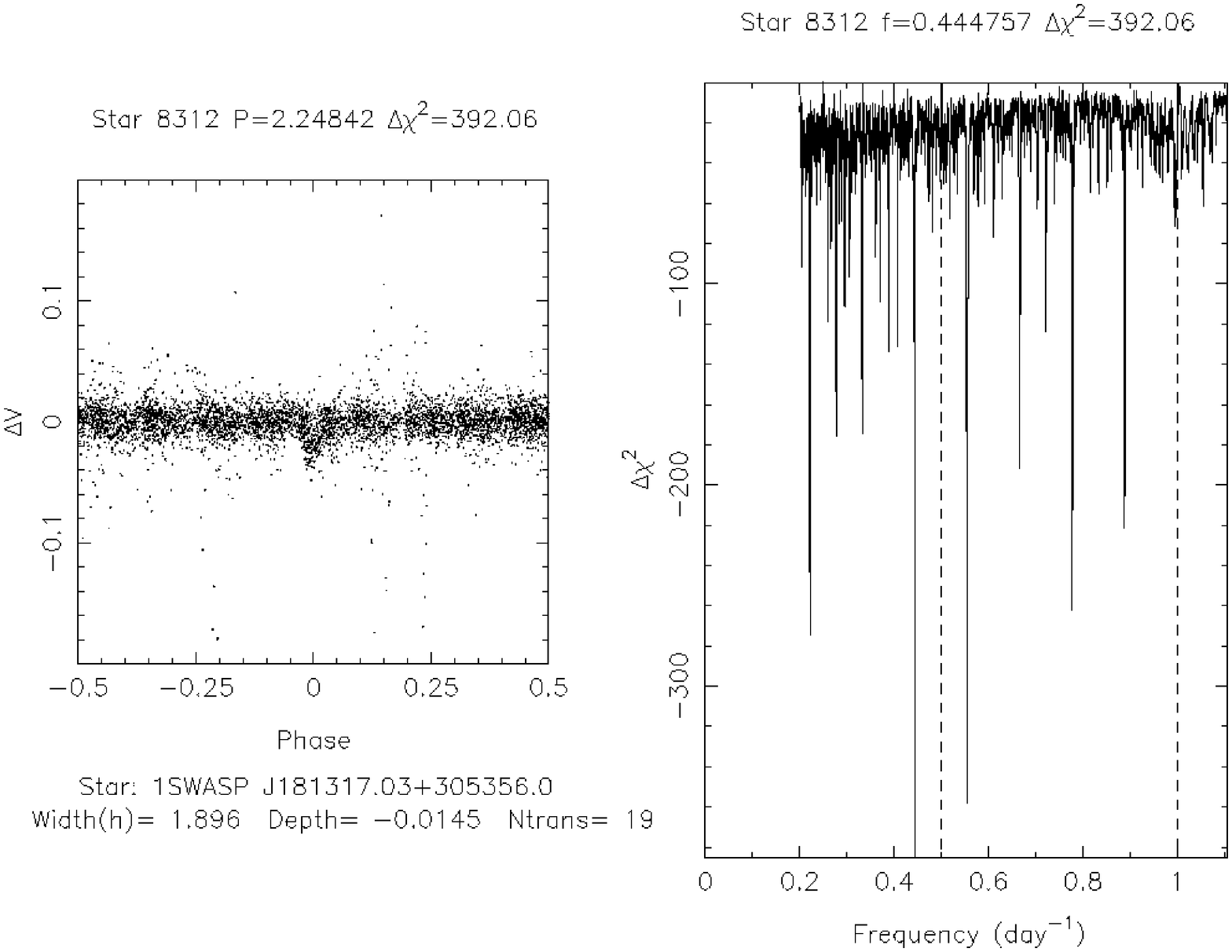}} 
&
\subfigure[1SWASP J181454.99+391146.0]{\label{fig:J181454.99+391146.0}
\includegraphics[angle=270,width=6.5cm]{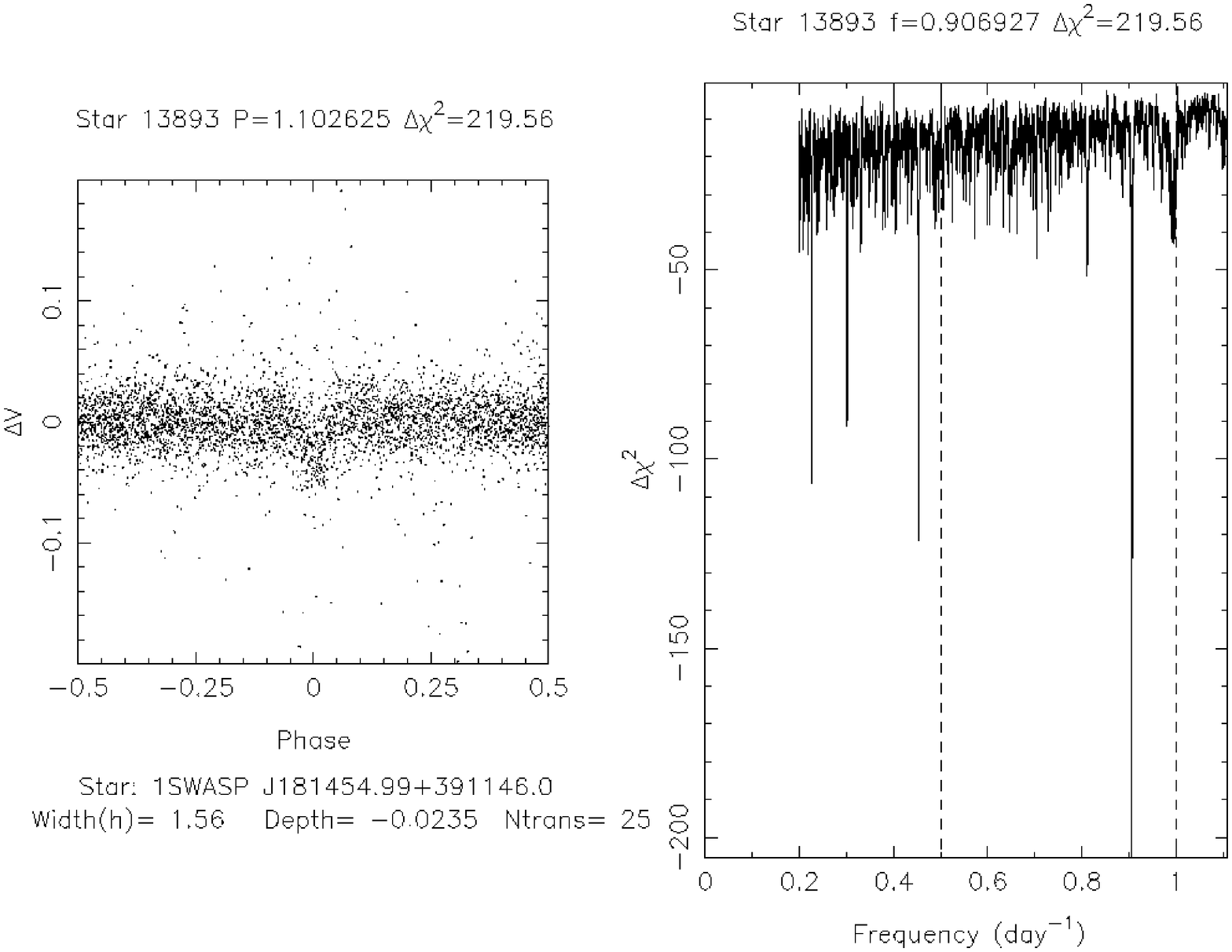}} \\

\subfigure[1SWASP J181958.25+492329.9]{\label{fig:J181958.25+492329.9}
\includegraphics[angle=270,width=6.5cm]{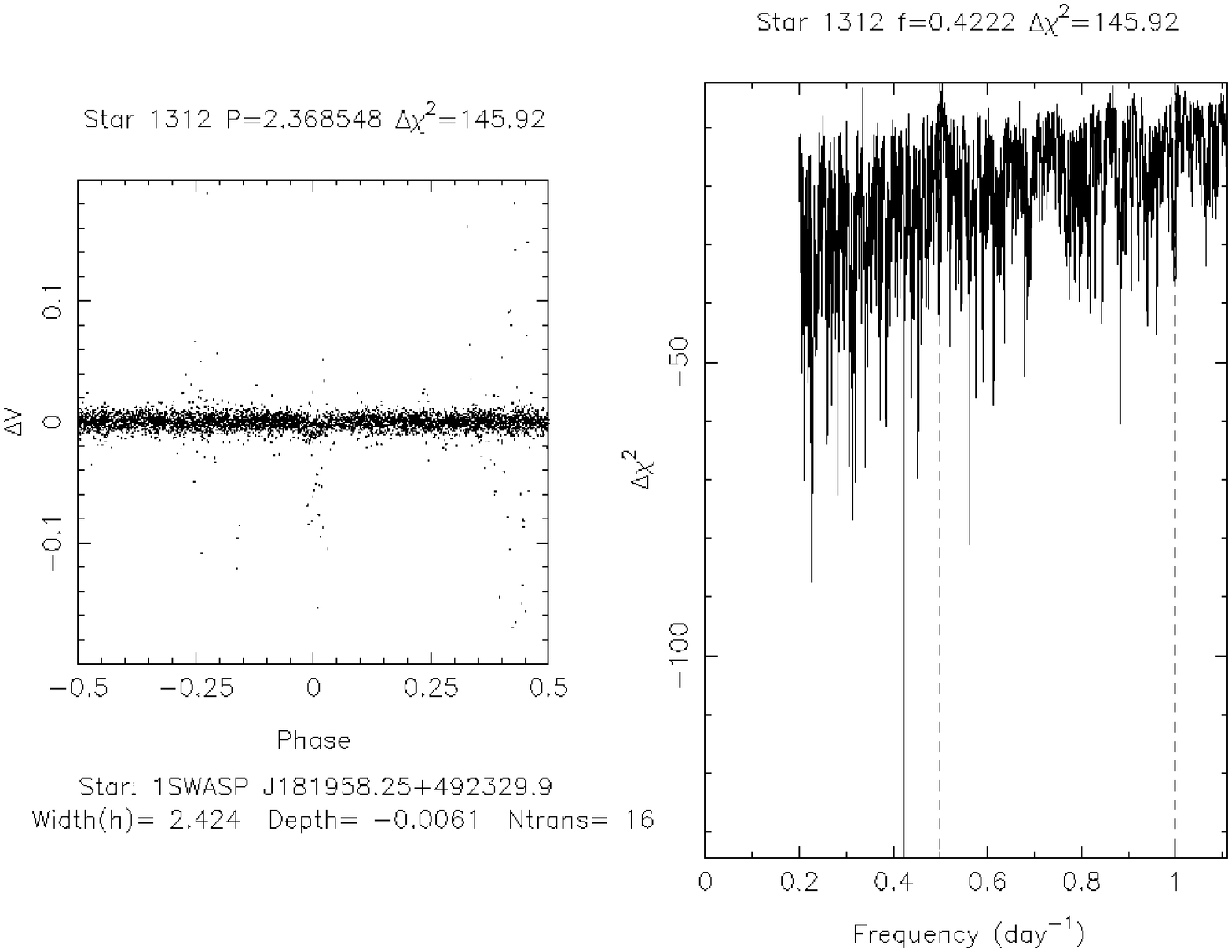}} 
&
\subfigure[1SWASP J182620.36+475902.8]{\label{fig:J182620.36+475902.8}
\includegraphics[angle=270,width=6.5cm]{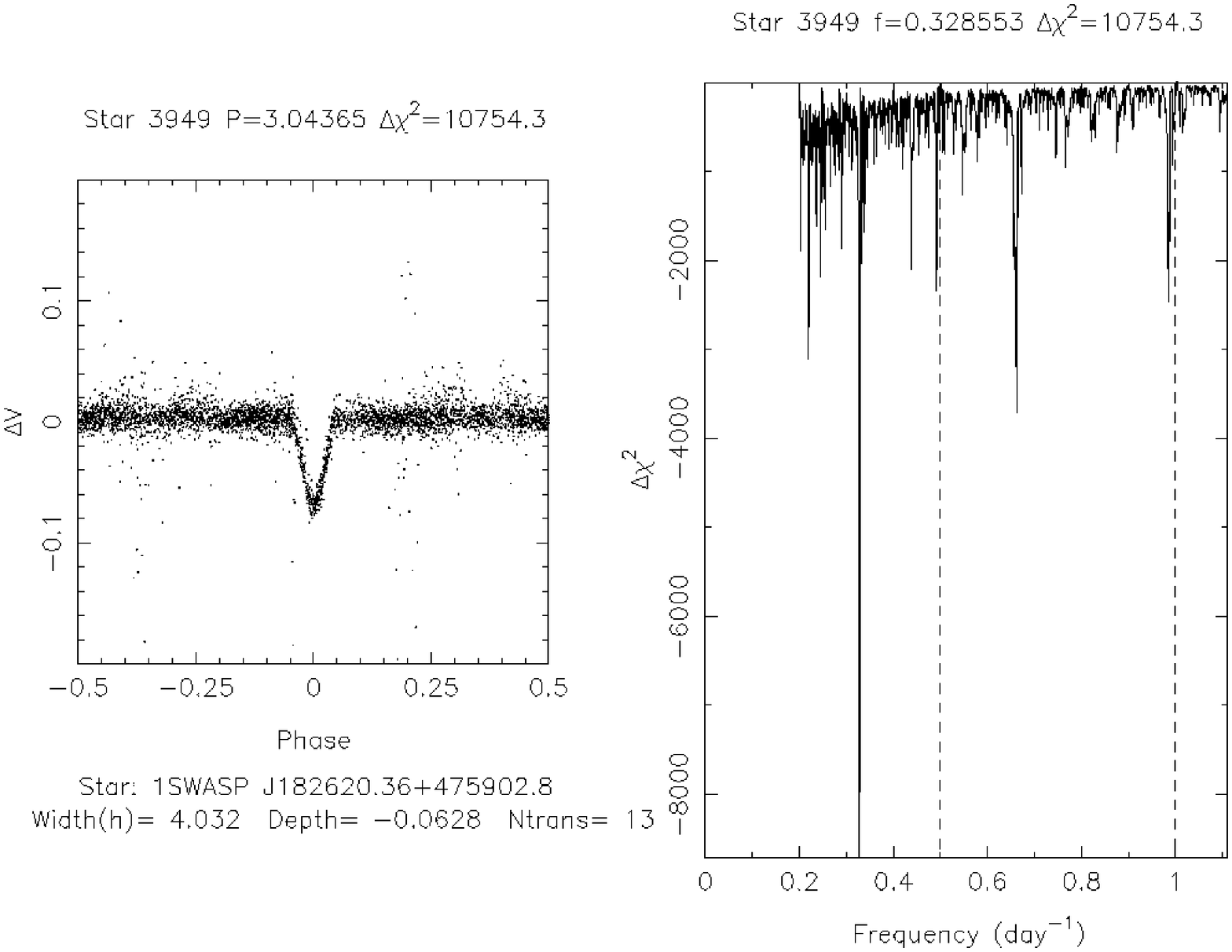}} \\

\subfigure[1SWASP J182924.67+232200.2]{\label{fig:J182924.67+232200.2}
\includegraphics[angle=270,width=6.5cm]{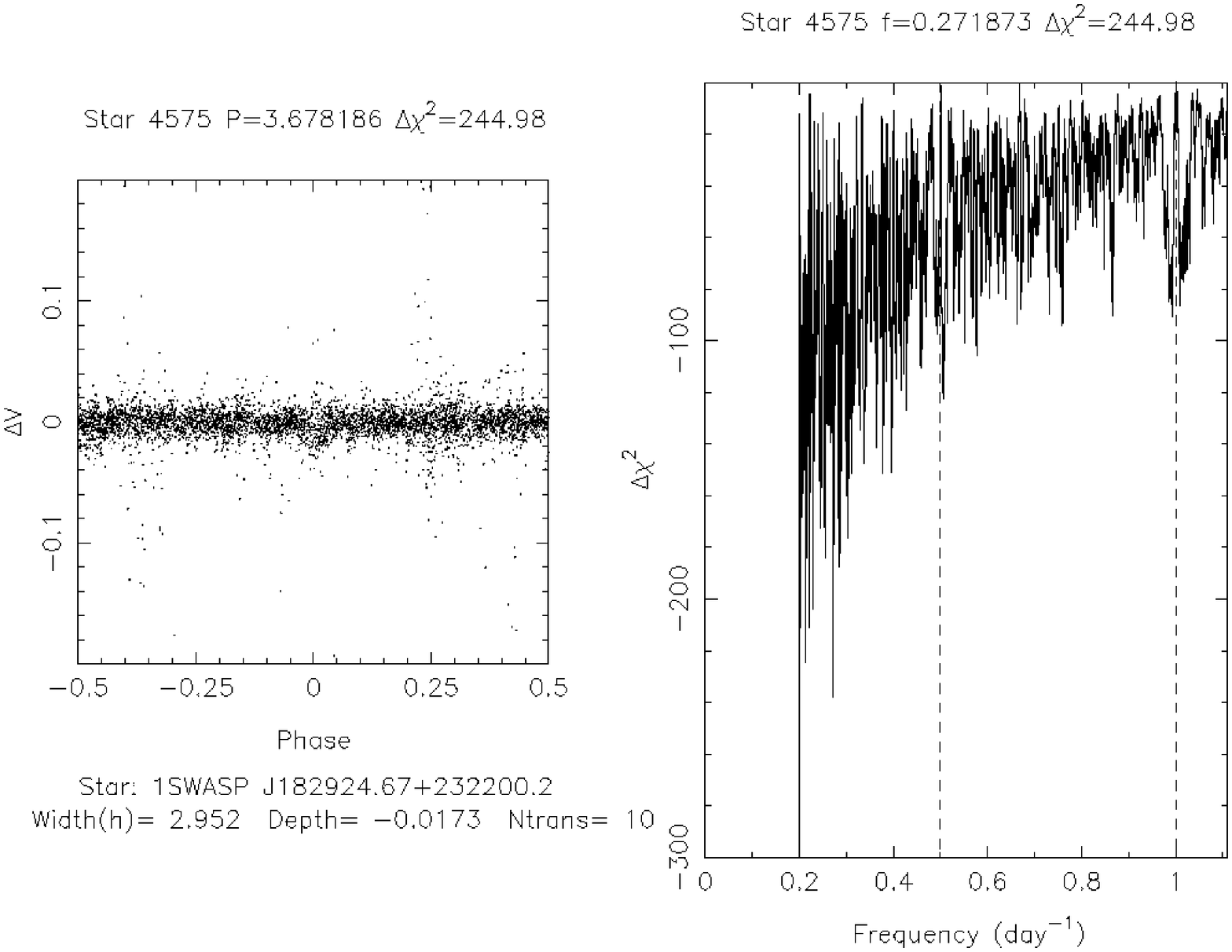}} 
&
\subfigure[1SWASP J183104.01+323942.7]{\label{fig:J183104.01+323942.7}
\includegraphics[angle=270,width=6.5cm]{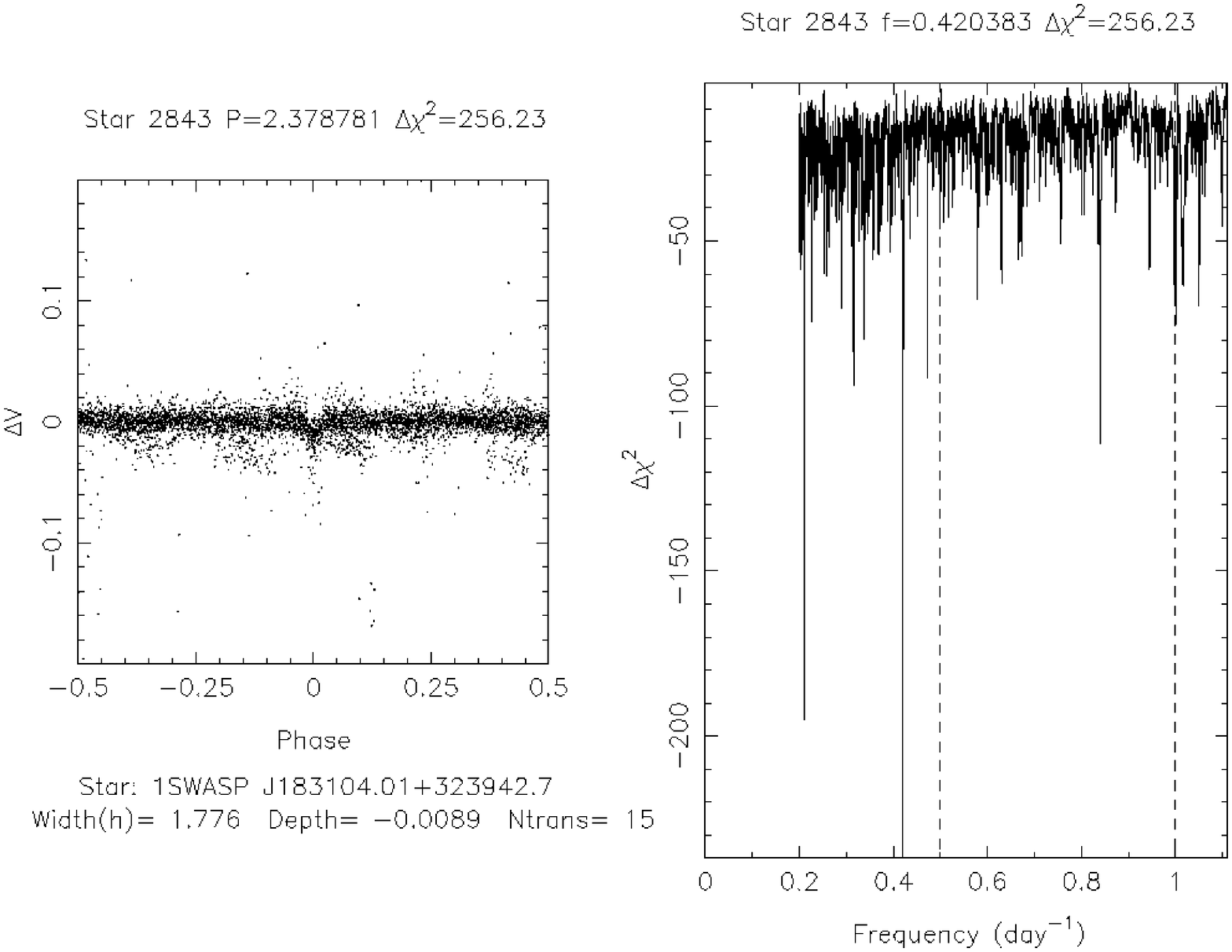}} \\

\subfigure[1SWASP J183431.62+353941.4]{\label{fig:J183431.62+353941.4}
\includegraphics[angle=270,width=6.5cm]{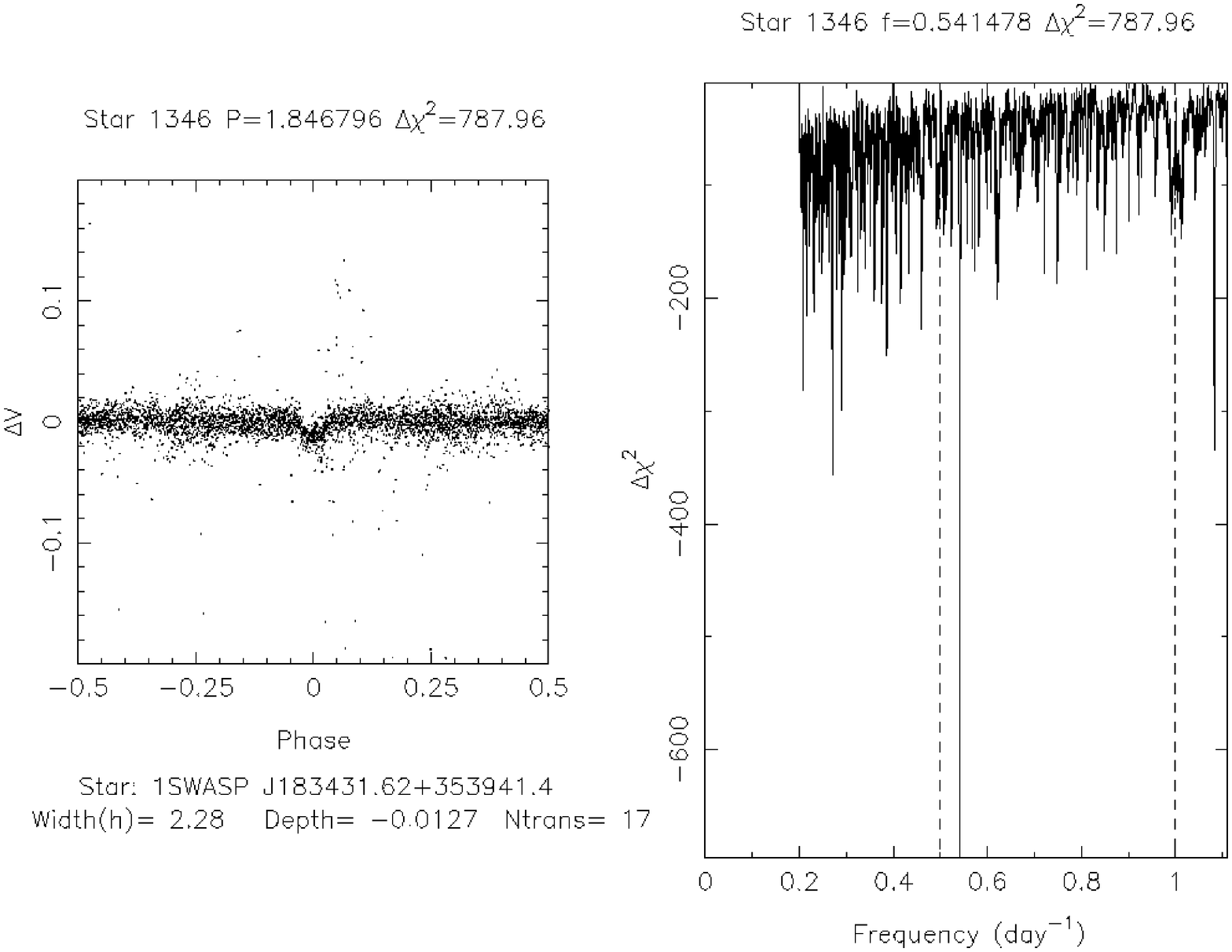}} 
&
\subfigure[1SWASP J183805.57+423432.3]{\label{fig:J183805.57+423432.3}
\includegraphics[angle=270,width=6.5cm]{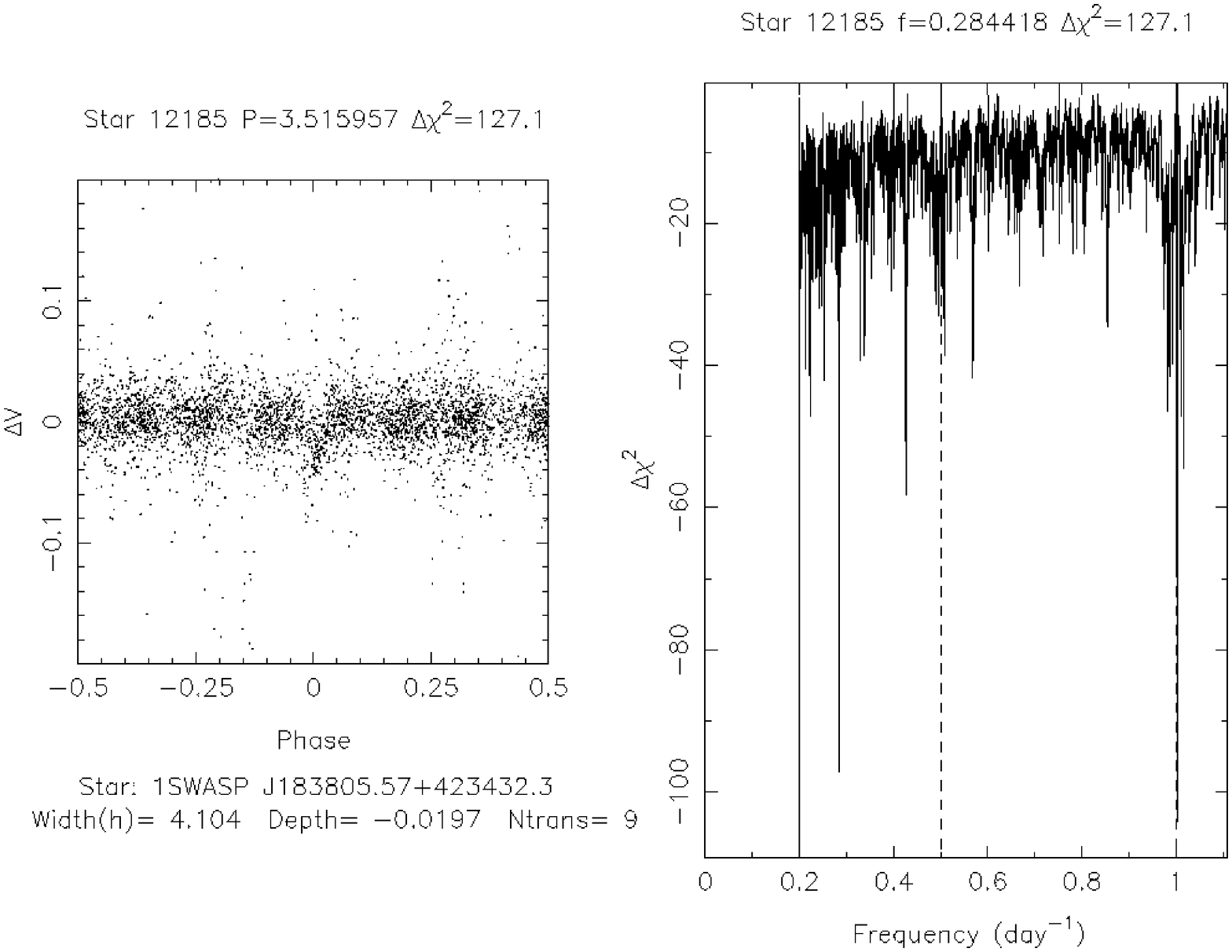}} \\

\end{tabular}
\caption{The lightcurves of the selected transit candidates, folded on the measured
period.  }
\protect\label{fig:candlcs1}
\end{figure*}

\begin{figure*}
\def\subfigtopskip{4pt}
\def\subfigbottomskip{8pt}
\def\subfigcapskip{4pt}
\centering
\begin{tabular}{cc}

\subfigure[1SWASP J204125.28+163911.8]{\label{fig:J204125.28+163911.8}
\includegraphics[angle=270,width=6.5cm]{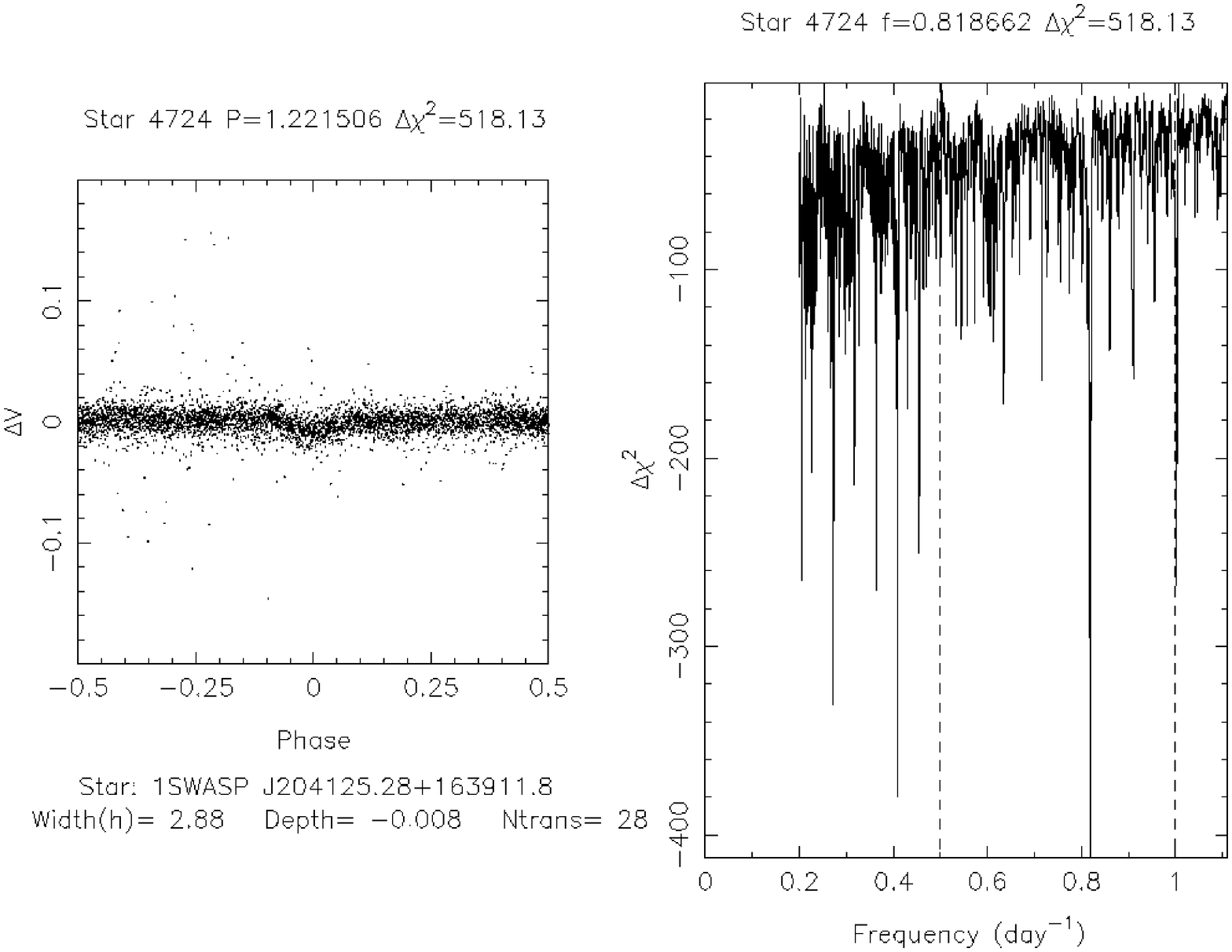}}
&
\subfigure[1SWASP J204142.49+075051.5]{\label{fig:J204142.49+075051.5}
\includegraphics[angle=270,width=6.5cm]{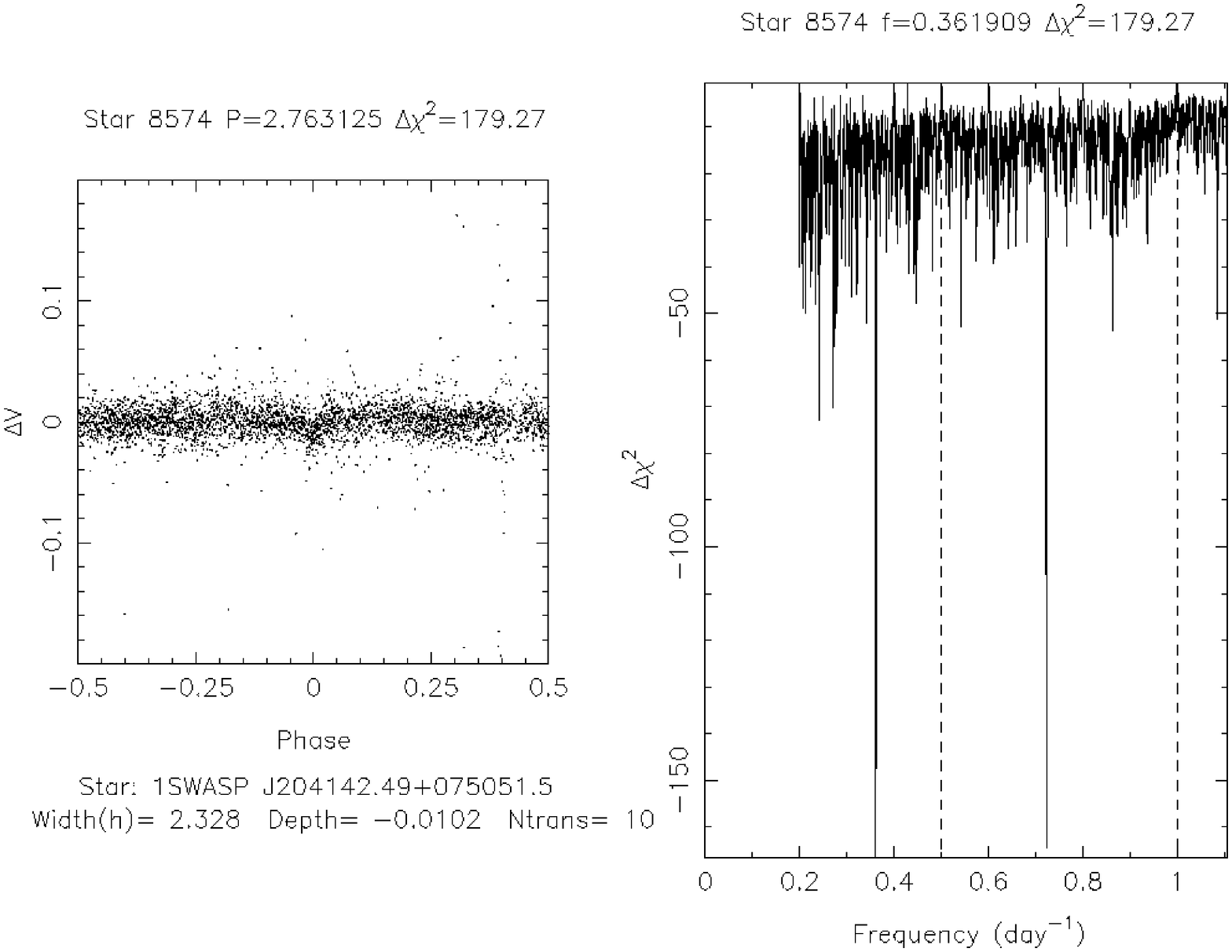}} \\

\subfigure[1SWASP J204323.83+263818.7]{\label{fig:J204323.83+263818.7}
\includegraphics[angle=270,width=6.5cm]{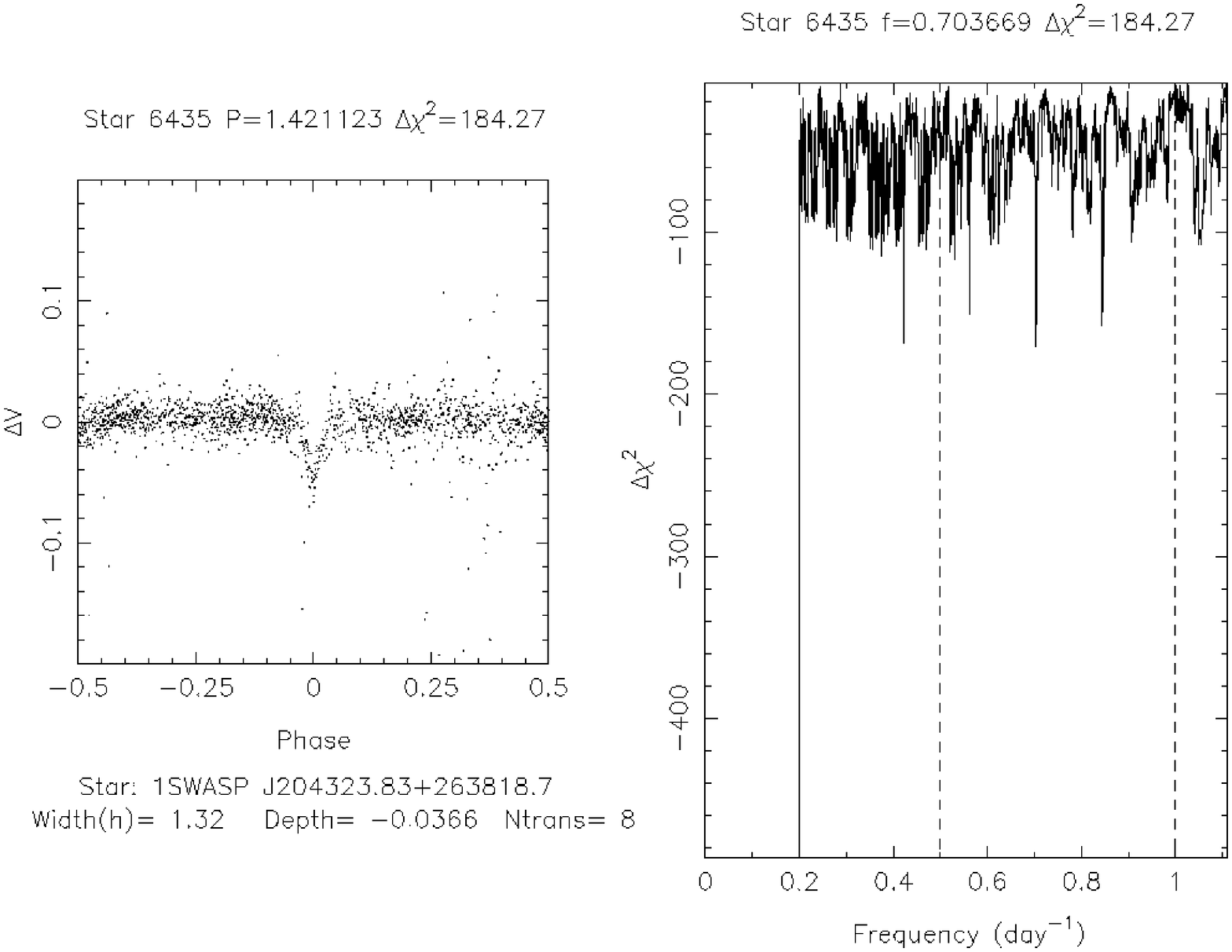}} 
&
\subfigure[1SWASP J204617.02+085412.0]{\label{fig:J204617.02+085412.0}
\includegraphics[angle=270,width=6.5cm]{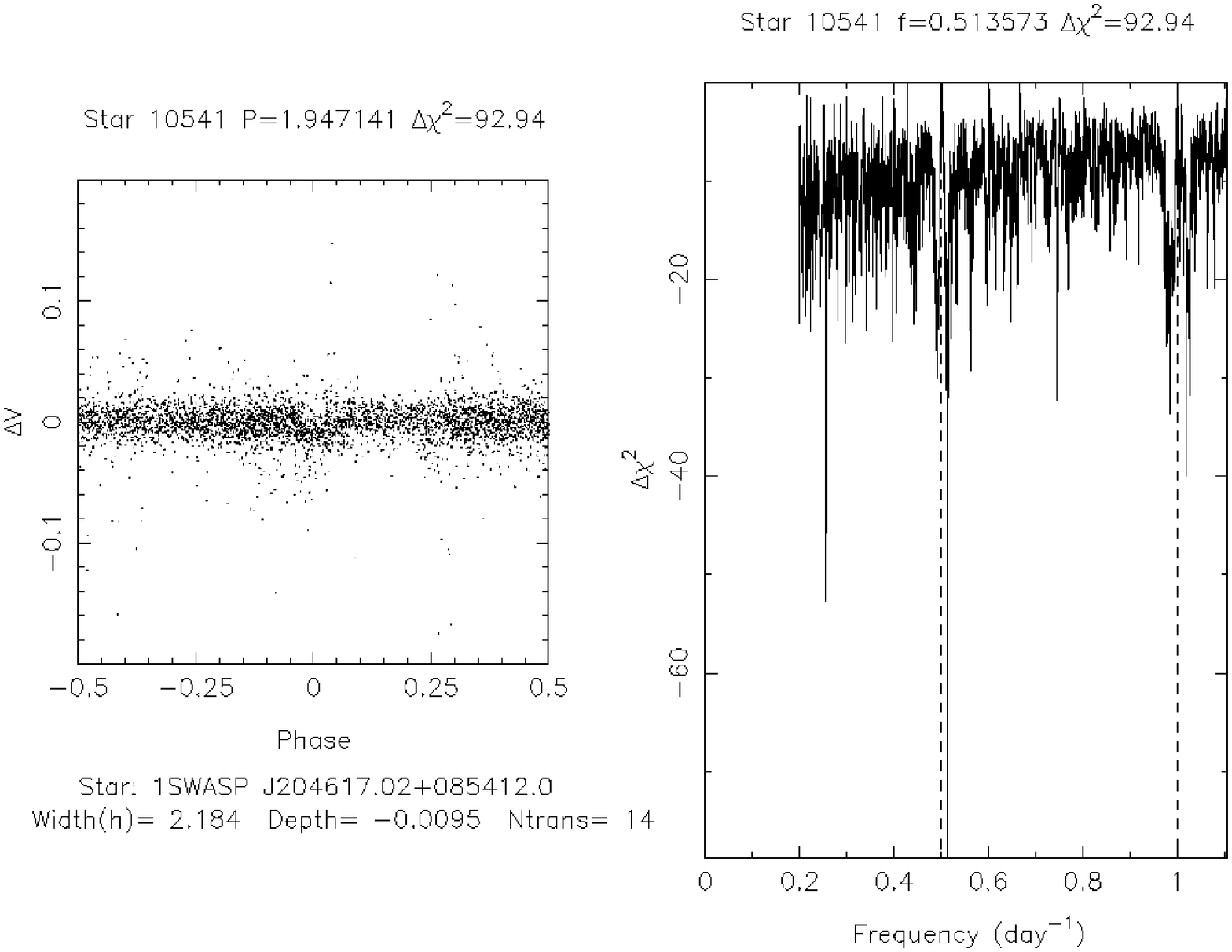}} \\

\subfigure[1SWASP J204712.42+202544.5]{\label{fig:J204712.42+202544.5}
\includegraphics[angle=270,width=6.5cm]{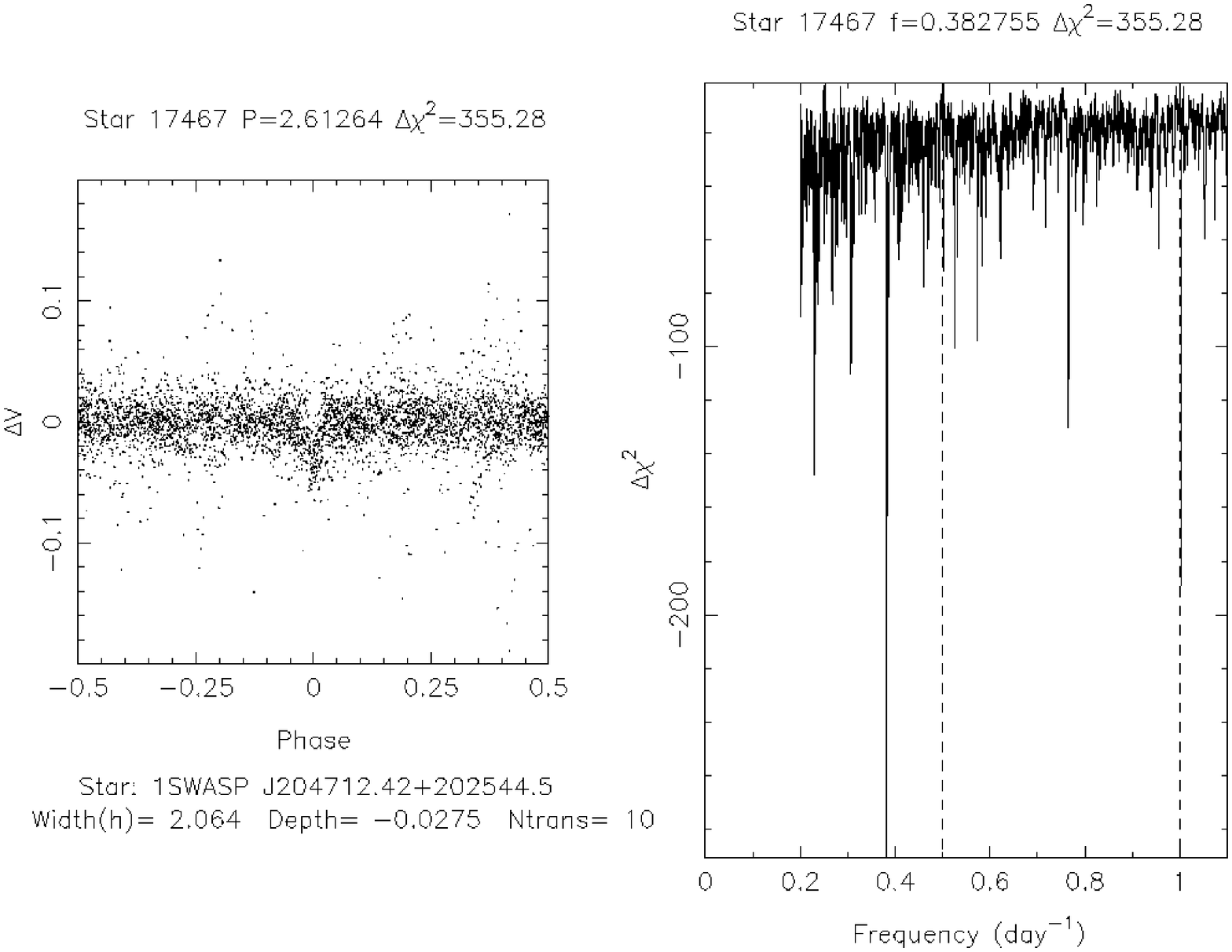}} 
&
\subfigure[1SWASP J204745.08+103347.9]{\label{fig:J204745.08+103347.9}
\includegraphics[angle=270,width=6.5cm]{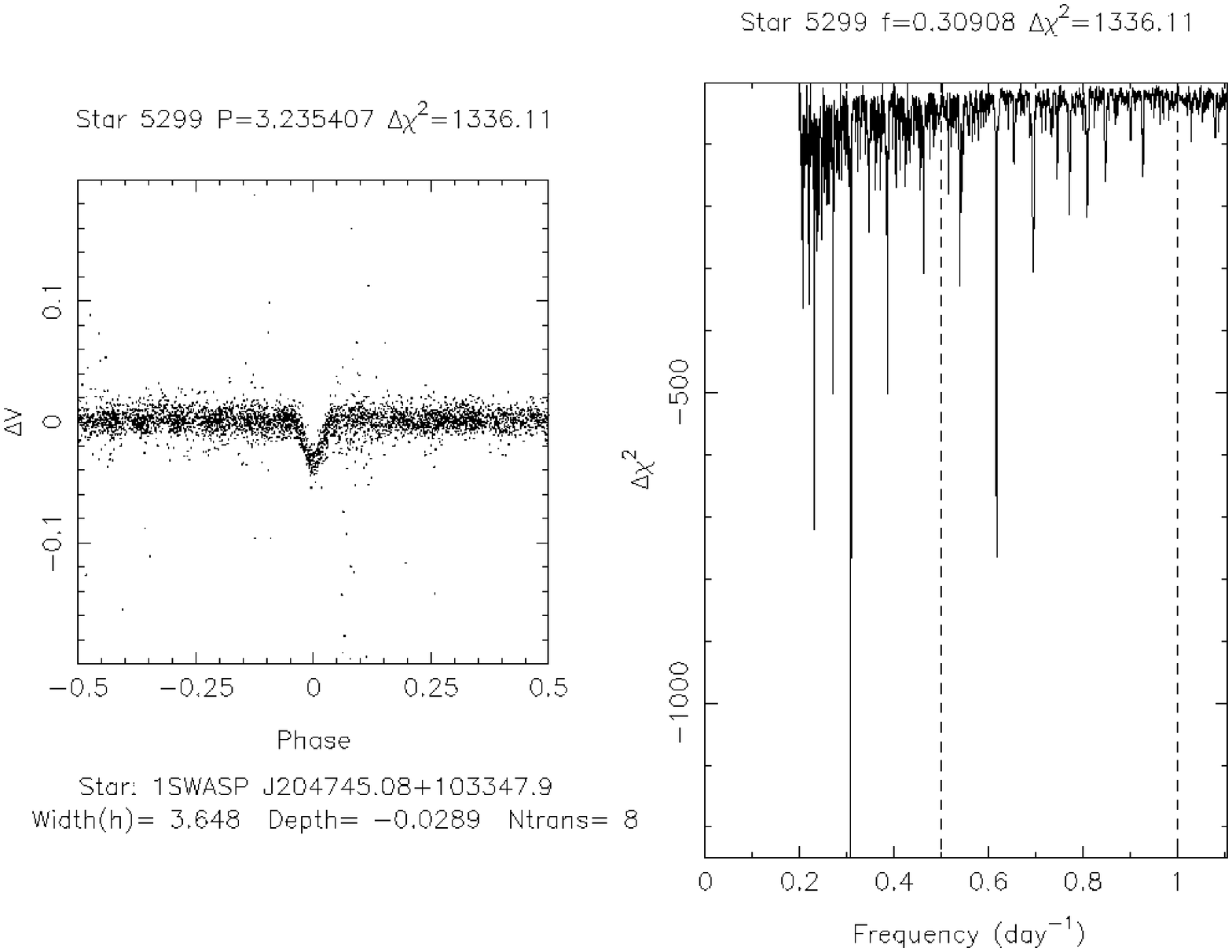}} \\

\subfigure[1SWASP J205027.33+064022.9]{\label{fig:J205027.33+064022.9}
\includegraphics[angle=270,width=6.5cm]{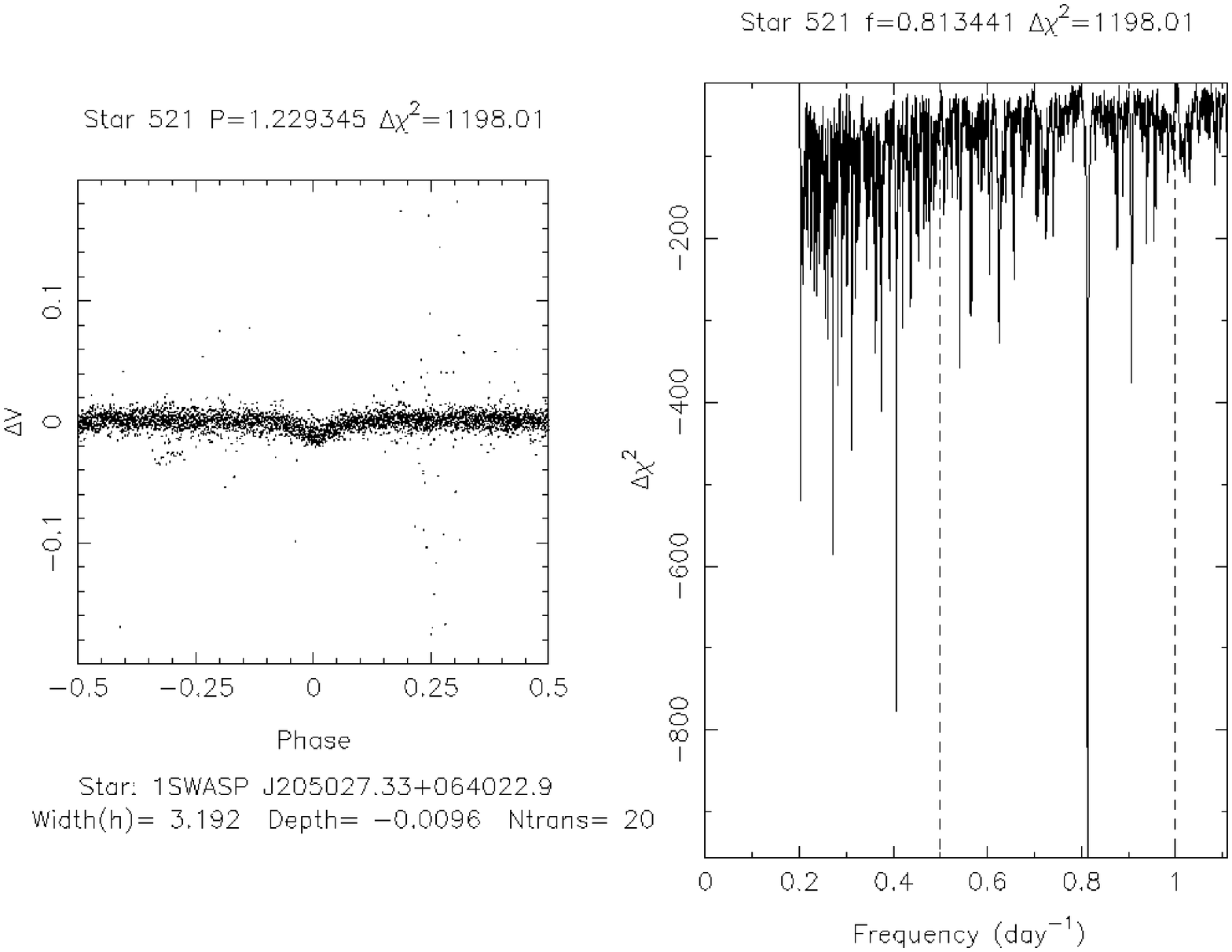}} 
&
\subfigure[1SWASP J205308.03+192152.7]{\label{fig:J205308.03+192152.7}
\includegraphics[angle=270,width=6.5cm]{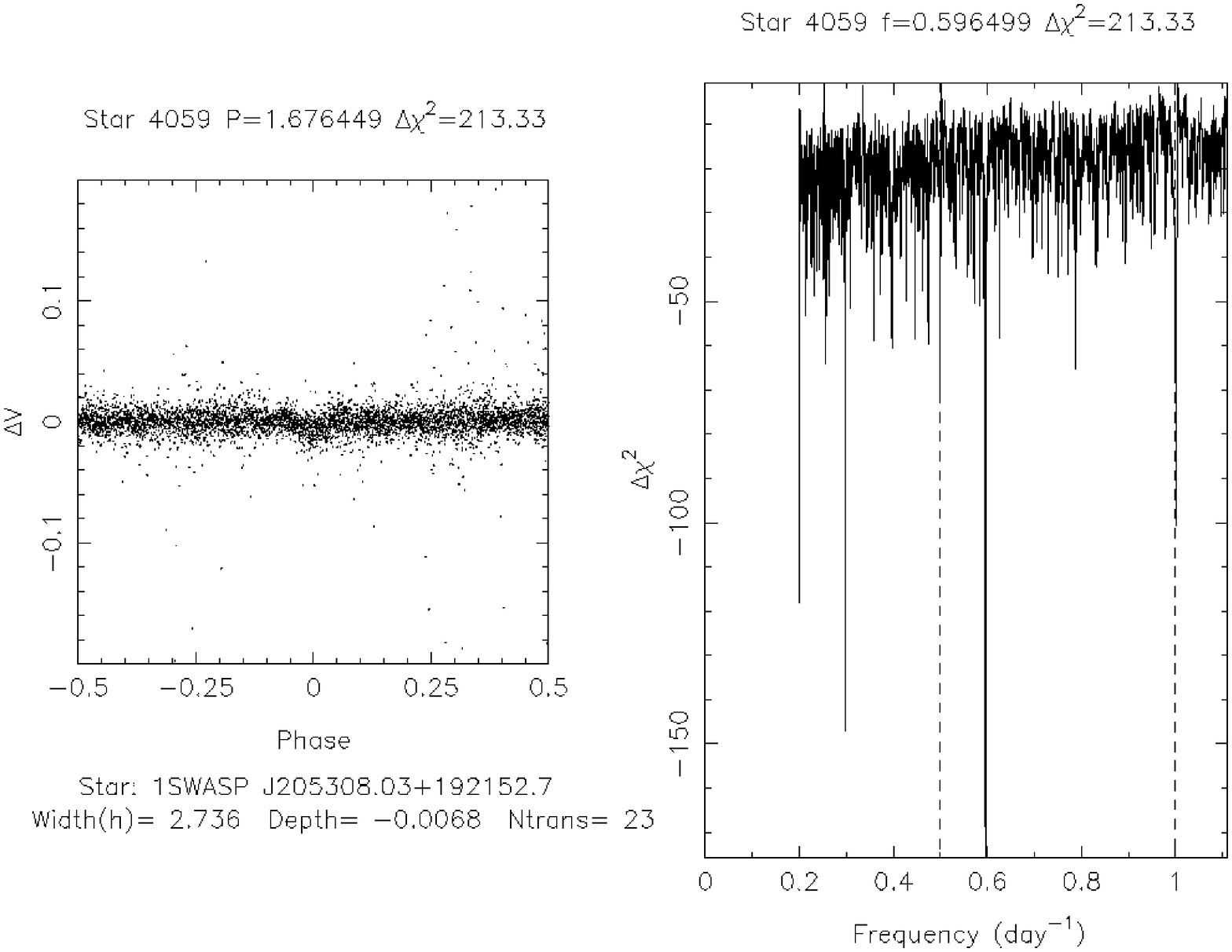}} \\

\end{tabular}
\caption{The lightcurves of the selected transit candidates, folded on the measured
period.  }
\protect\label{fig:candlcs3}
\end{figure*}

\begin{figure*}
\def\subfigtopskip{4pt}
\def\subfigbottomskip{8pt}
\def\subfigcapskip{4pt}
\centering
\begin{tabular}{cc}

\subfigure[1SWASP J210009.75+193107.1]{\label{fig:J210009.75+193107.1}
\includegraphics[angle=270,width=6.5cm]{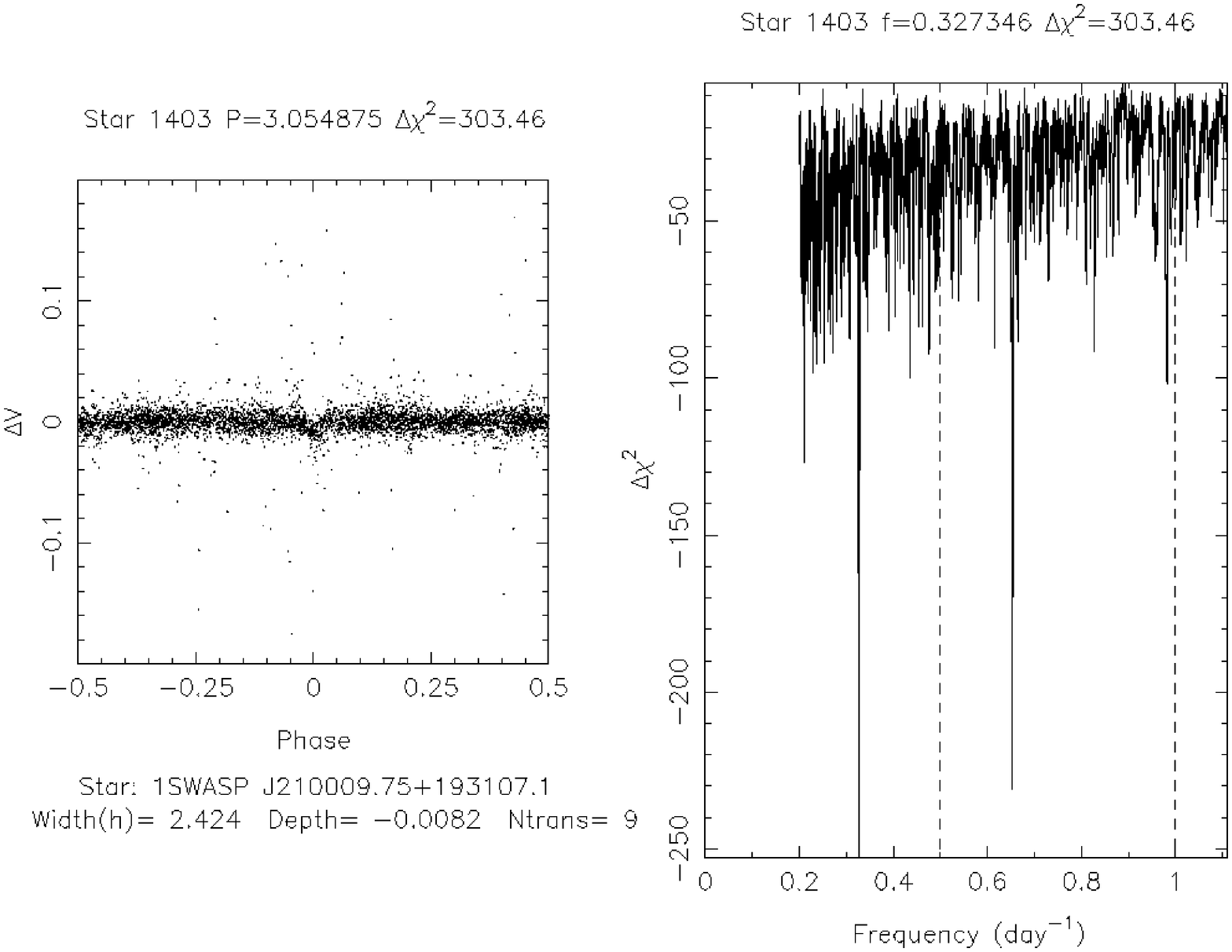}} 
&
\subfigure[1SWASP J210151.43+072326.7]{\label{fig:J210151.43+072326.7}
\includegraphics[angle=270,width=6.5cm]{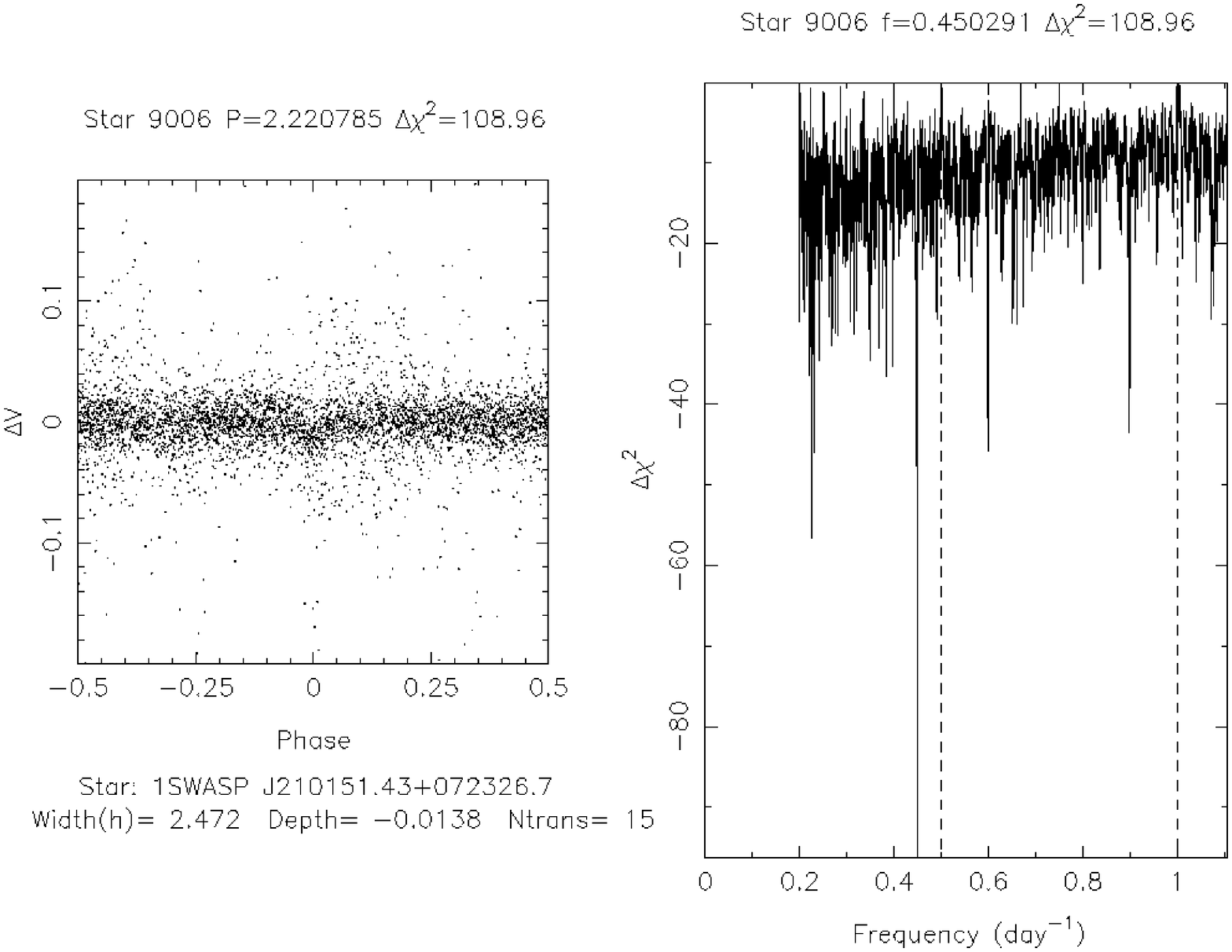}} \\

\subfigure[1SWASP J210318.01+080117.8]{\label{fig:J210318.01+080117.8}
\includegraphics[angle=270,width=6.5cm]{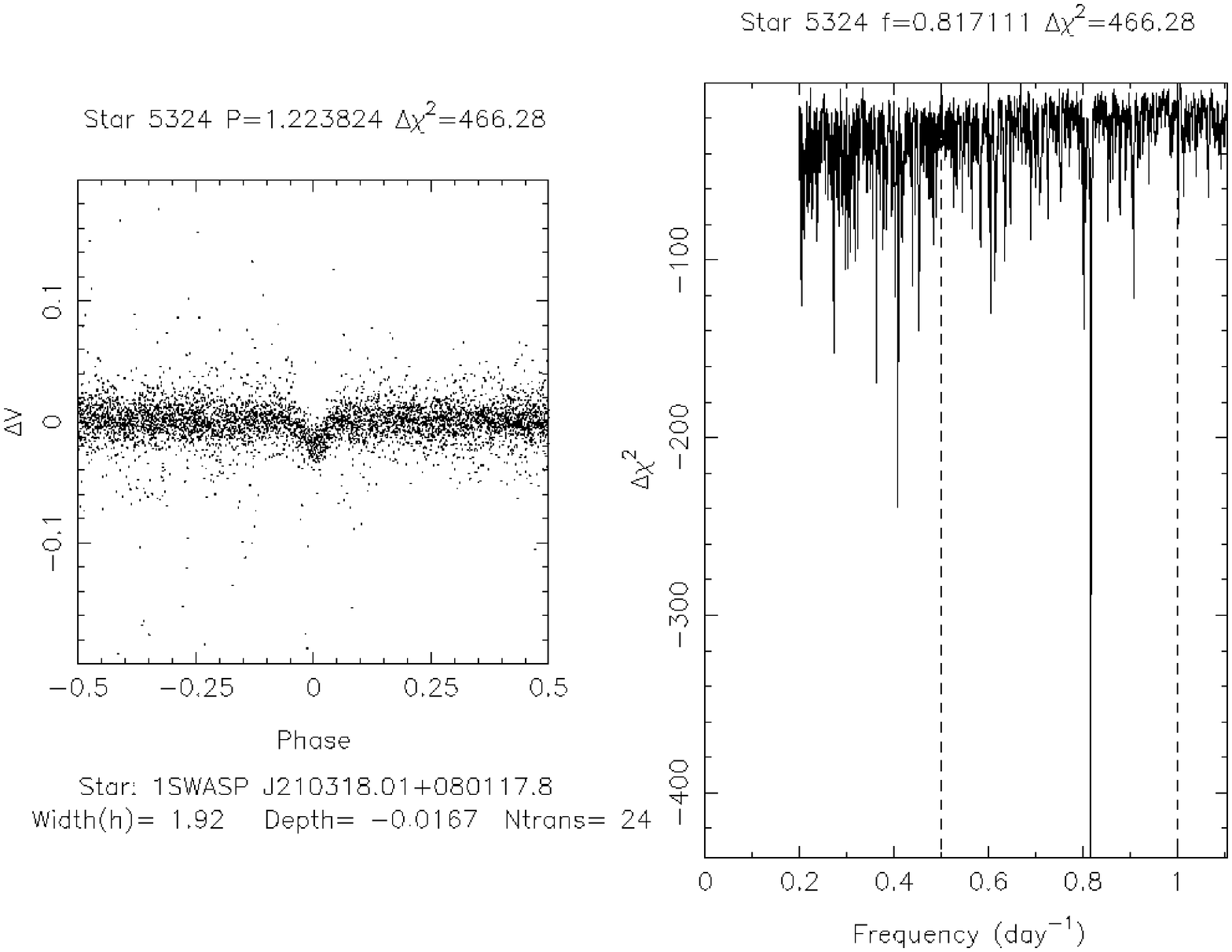}} 
&
\subfigure[1SWASP J210352.56+083258.9]{\label{fig:J210352.56+083258.9}
\includegraphics[angle=270,width=6.5cm]{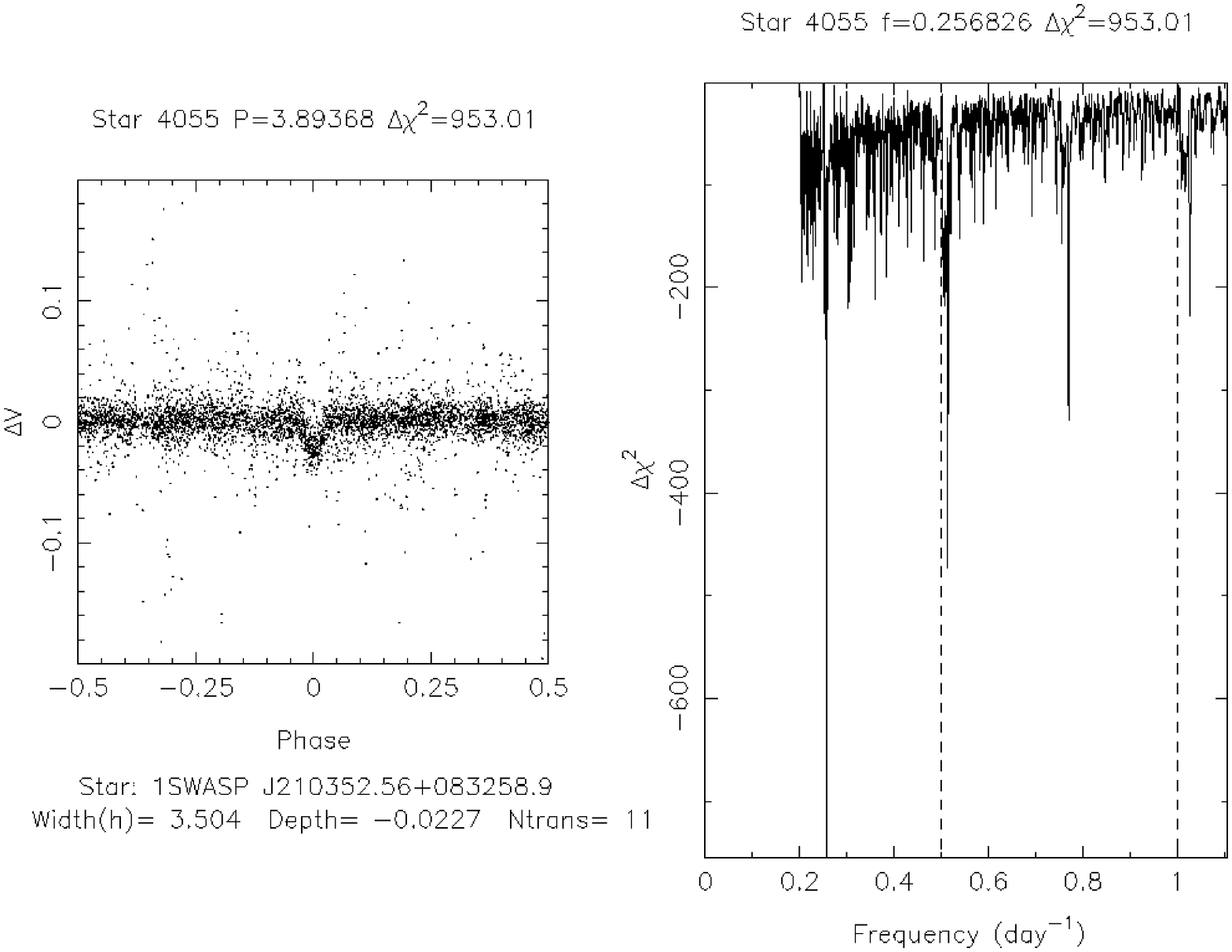}} \\

\subfigure[1SWASP J210909.05+184950.9]{\label{fig:J210909.05+184950.9}
\includegraphics[angle=270,width=6.5cm]{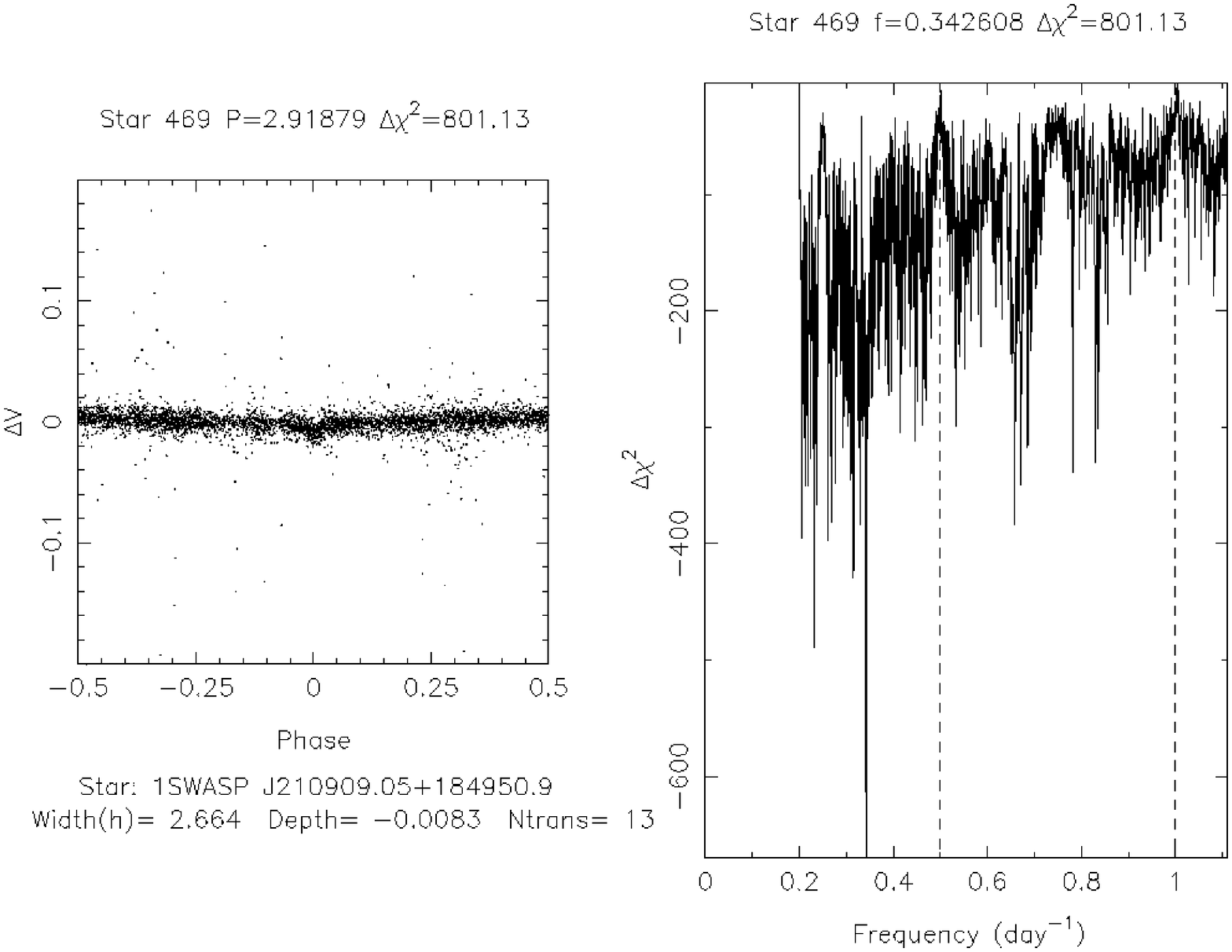}} 
&
\subfigure[1SWASP J210912.02+073843.3]{\label{fig:J210912.02+073843.3}
\includegraphics[angle=270,width=6.5cm]{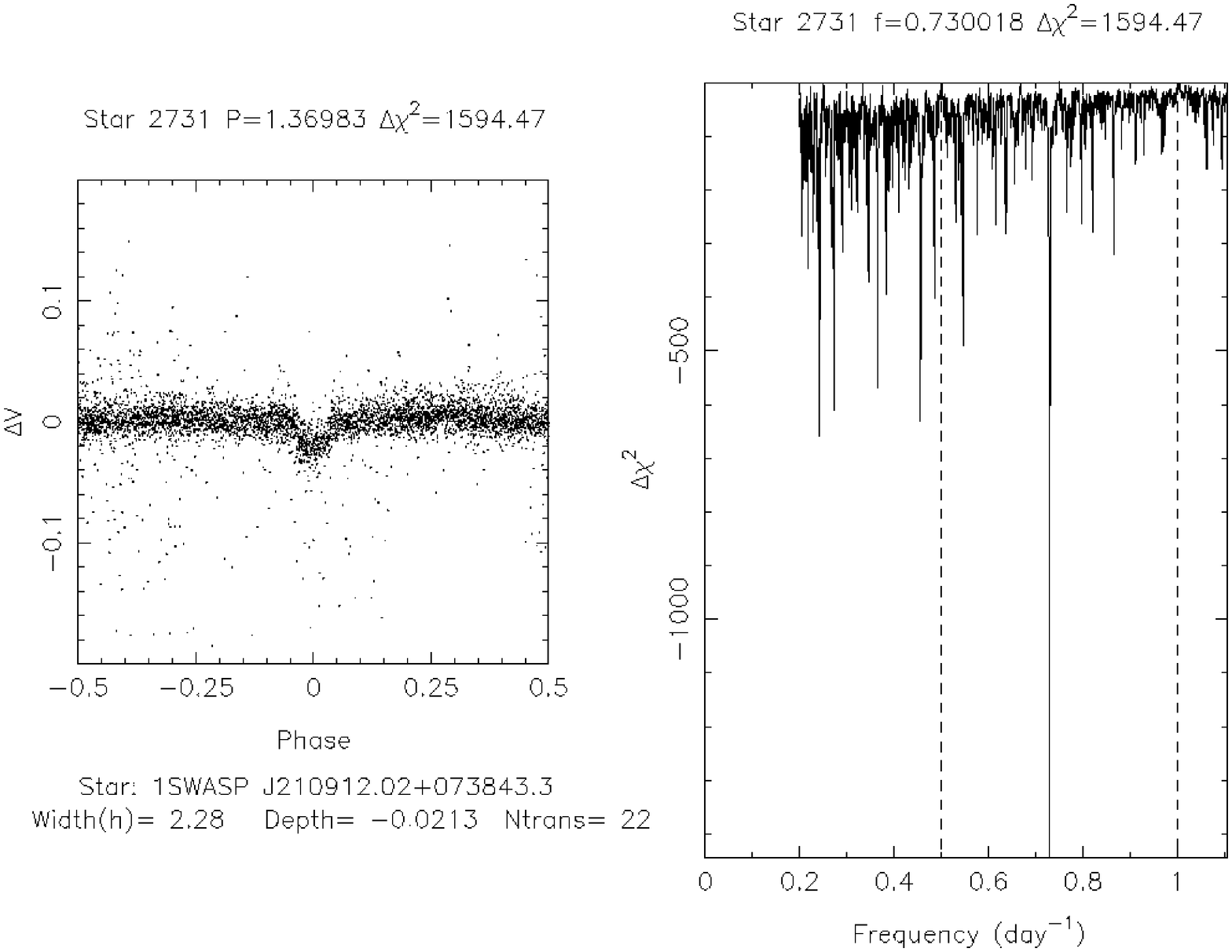}} \\

\subfigure[1SWASP J211608.42+163220.3]{\label{fig:J211608.42+163220.3}
\includegraphics[angle=270,width=6.5cm]{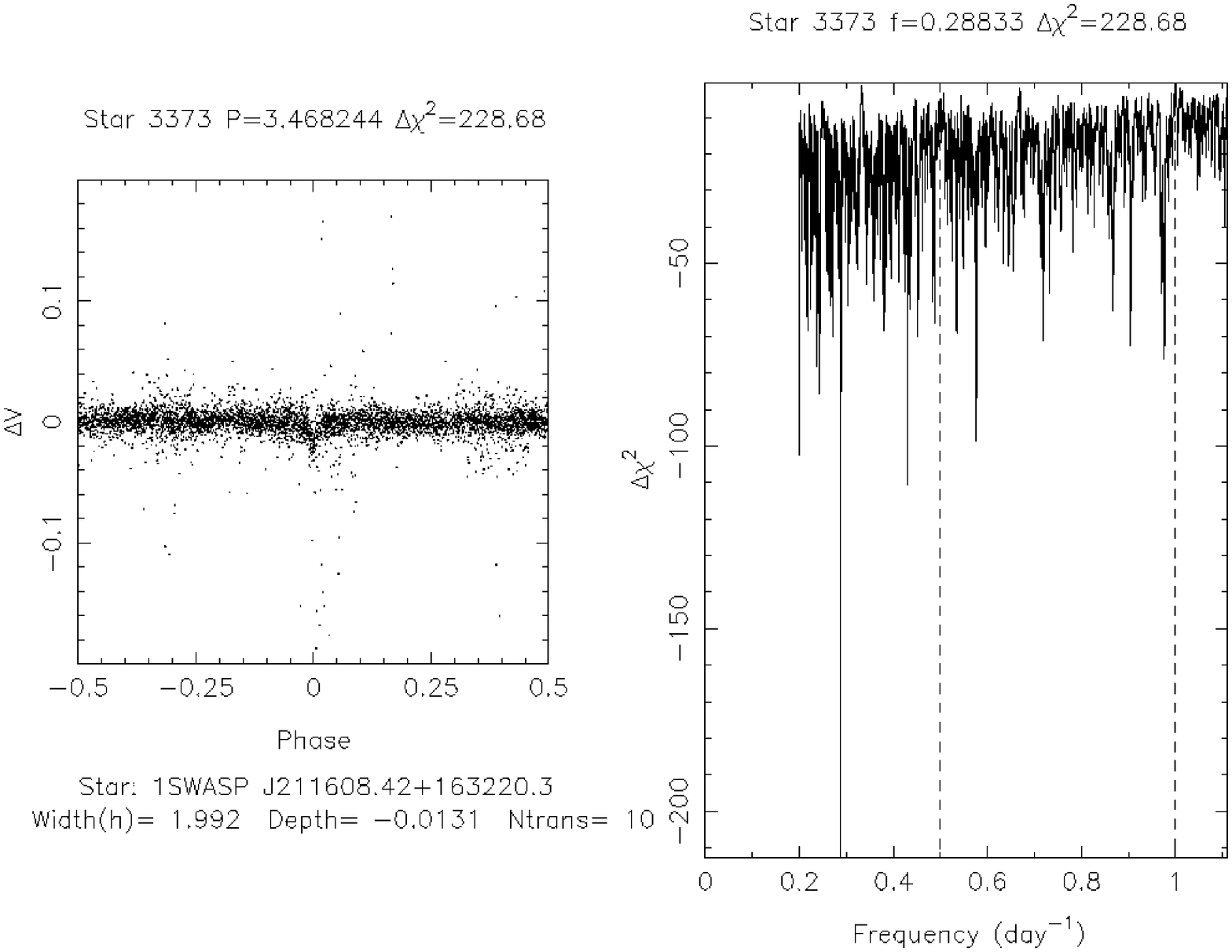}} 
&
\subfigure[1SWASP J211645.22+192136.8]{\label{fig:J211645.22+192136.8}
\includegraphics[angle=270,width=6.5cm]{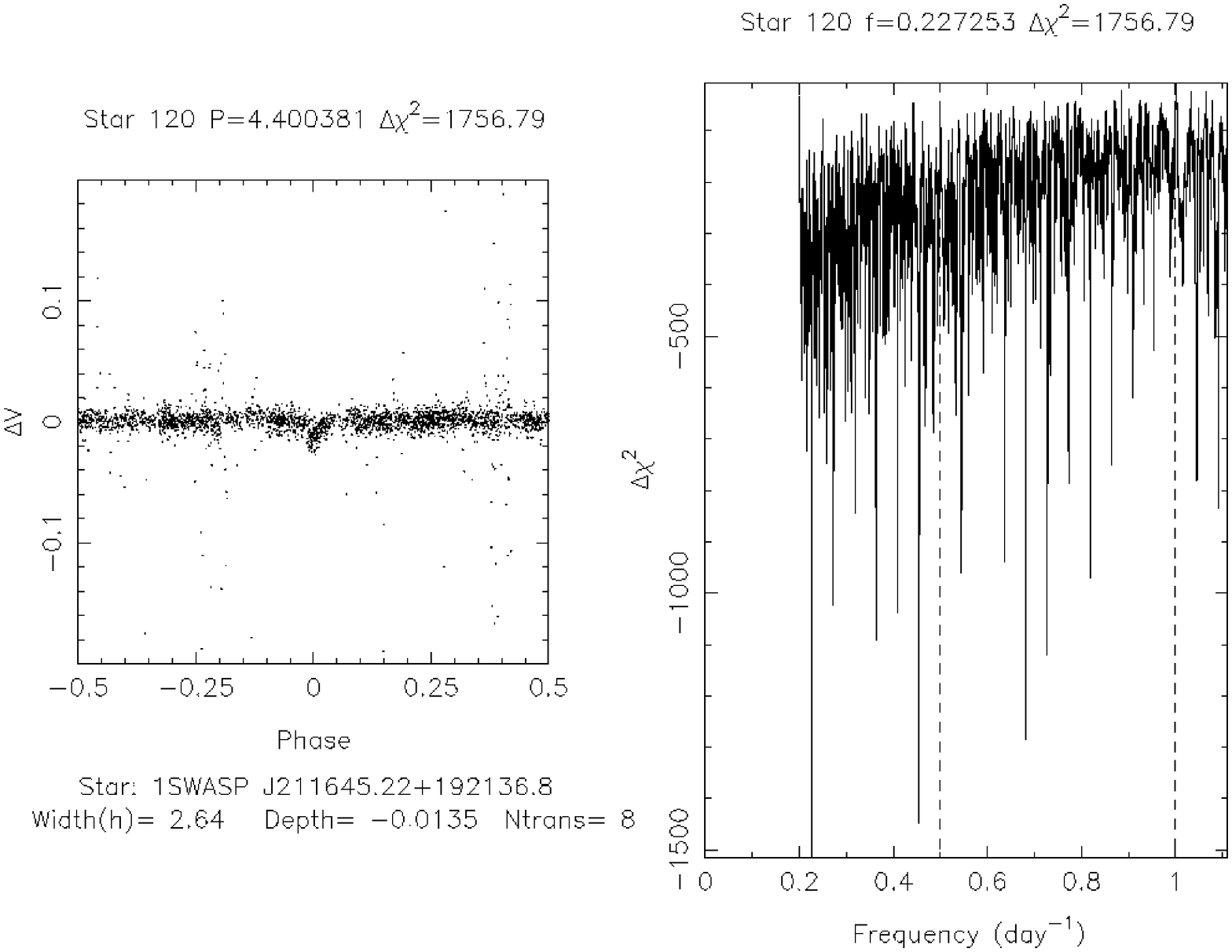}} \\

\end{tabular}
\caption{The lightcurves of the selected transit candidates, folded on the measured
period.  }
\protect\label{fig:candlcs4}
\end{figure*}

\begin{figure*}
\def\subfigtopskip{4pt}
\def\subfigbottomskip{8pt}
\def\subfigcapskip{4pt}
\centering
\begin{tabular}{cc}

\subfigure[1SWASP J212532.55+082904.4]{\label{fig:J212532.55+082904.4}
\includegraphics[angle=270,width=6.5cm]{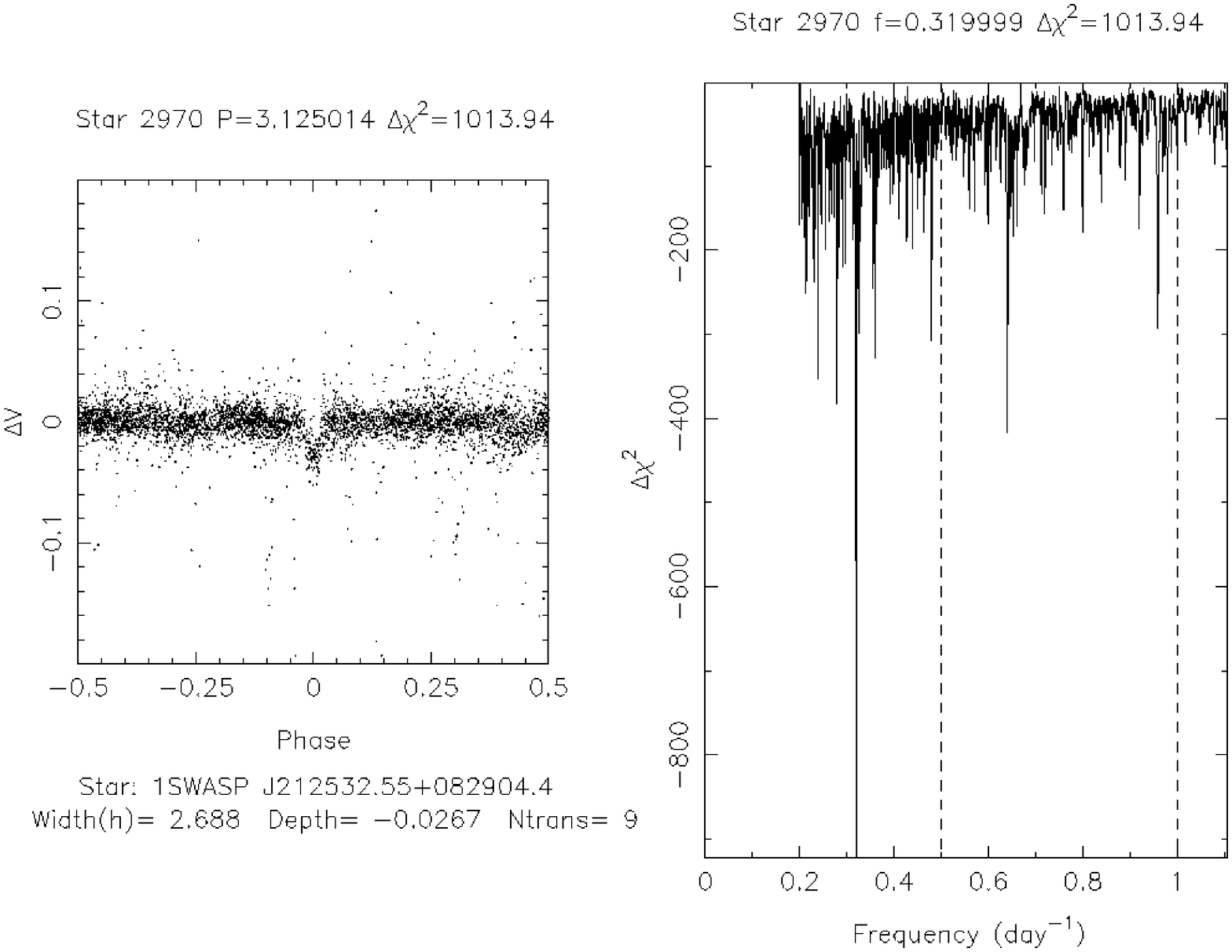}} 
& 
\subfigure[1SWASP J212843.62+160806.2]{\label{fig:J212843.62+160806.2}
\includegraphics[angle=270,width=6.5cm]{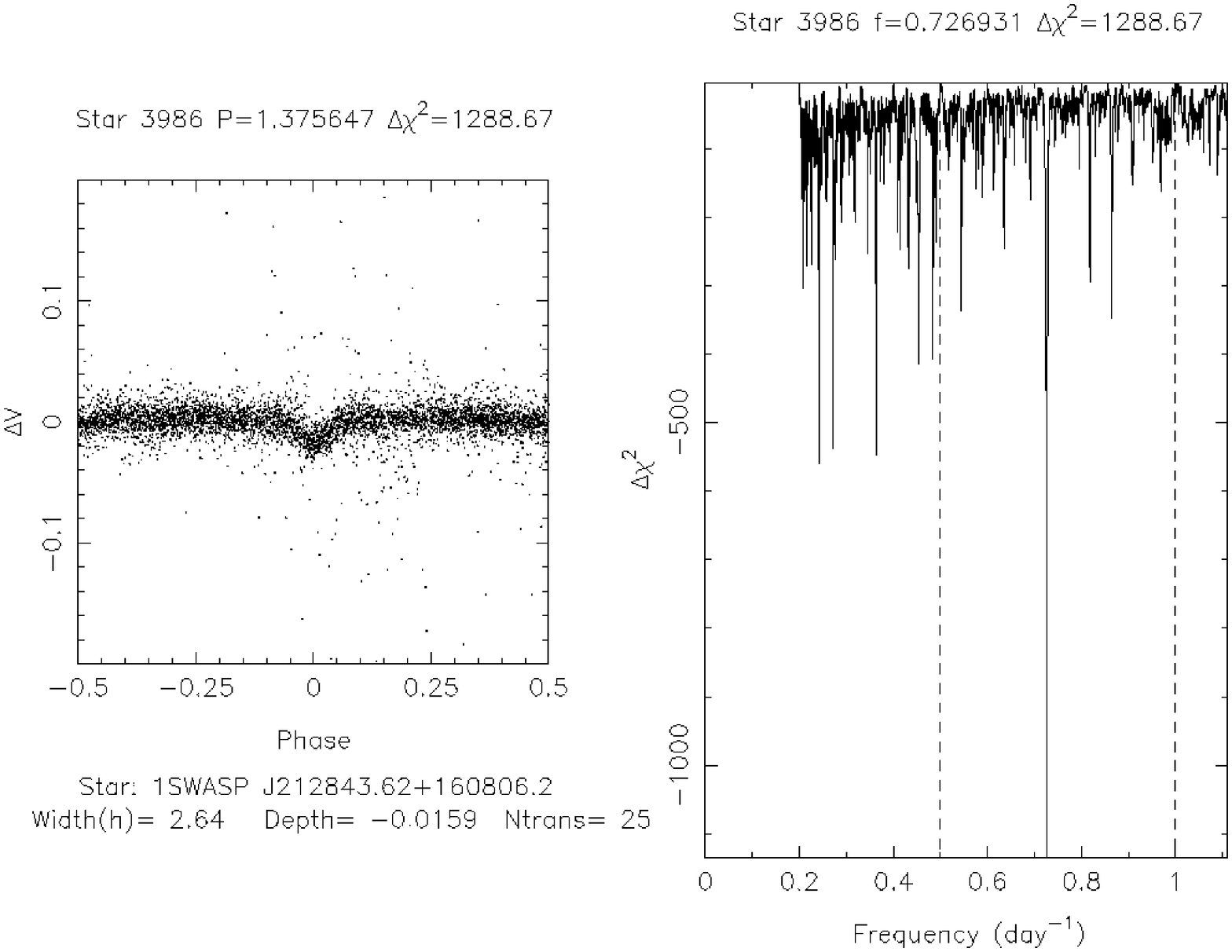}} \\

\subfigure[1SWASP J212855.03+075753.5]{\label{fig:J212855.03+075753.5}
\includegraphics[angle=270,width=6.5cm]{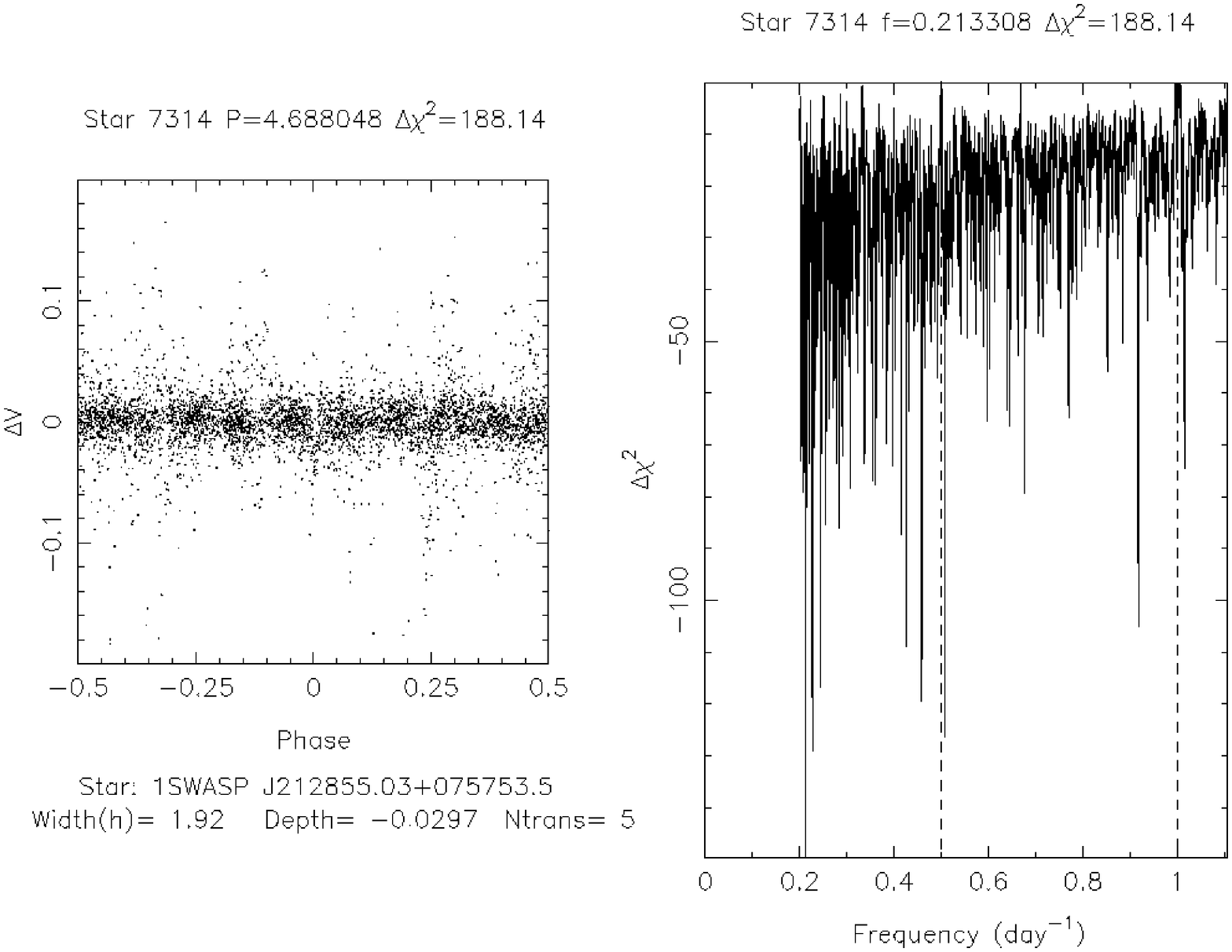}} \\

\end{tabular}
\caption{The lightcurves of the selected transit candidates, folded on the measured
period.  }
\protect\label{fig:candlcs5}
\end{figure*}

Evaluating all the information available for all candidates highlighted 35
objects of particular interest at the stage 4; the remaining objects being
rejected as likely stellar binaries, some blended.  These are printed in bold in
Tables~\ref{tab:stage3list} \& \ref{tab:stage4list} and their folded lightcurves
and $\Delta \chi^{2}$ periodograms are presented in
Figures~\ref{fig:candlcs1}--\ref{fig:candlcs5}.  We discuss these objects
individually below, and indicate particularly strong planetary candidates. 
However, all of these objects deserve follow-up observations as `false alarms'
from a transit survey include interesting low-mass binaries.  

\subsubsection{1SWASP J181317.03+305356.0} This object displayed a distinct, if
noisy, dip when folded on its original period of 4.499\,days but this resulted
in gaps in the phase coverage.   The transit is still visible when the data is
folded on a period of 2.248\,days but this time the lightcurve is more smoothly
sampled and flat out of transit to visual inspection.  The new parameters imply a Jovian-sized companion object
($R_{p}$=1.05\,\Rj) supported by a reasonable $\eta_{p}$=0.71, but while the
target is the brightest object in its field it has sufficient nearby faint stars
for blending to be a possibility.  More observations are required for this
object.  

\subsubsection{1SWASP J181454.99+391146.0} The faintness of this object
(12.796\,mag) accounts for the degree of noise in the lightcurve, but the transit is
still visible.  The noise makes it difficult to judge the flatness out of transit,
though the $S/N_{ellip}$ is 0.659.  The period is close to the 1-day alias at 1.10
days, but this is derived from a clear strong peak in $\Delta \chi^{2}$.  Otherwise,
the amplitude and the transit duration are reasonable, supported by an
$\eta_{p}$=0.92.  The primary star appears to be late type, implying a relatively
small companion (0.89\Rj).  However, this object lies in a fairly crowded field,
so it may be a blended stellar binary.  

\subsubsection{1SWASP J181958.25+492329.9} The brightness of this 10.6\,mag object
allows us to detect transits only $\sim$6\,mmag deep in this flat lightcurve.  The
period was confirmed independently with the S-C algorithm and transit signatures
identified by visual inspection of the unfolded lightcurve.  The host star has a
solar spectral type so the estimated companion radius is very low: 0.69\Rj,
supported by an $\eta_{p}$ close to 1.  This makes it an exciting candidate for
follow-up despite the serious crowding in this field.  However, further
observations are required to eliminate the possibility of a blended eclipsing
binary.   

\subsubsection{1SWASP J182620.36+475902.8} The folded lightcurve clearly shows a
fairly deep, wide, `V'-shaped dip (which might indicate a stellar binary) but no
obvious ellipsoidal variations.  The period is 3.04\,days, close to a multiple of
the 1-day alias, but the signal is clear with a credible number of transits
observed.  The object is unblended and has an estimated companion radius of 1.6\Rj;
however the $\eta_{p}$ of 1.49 would support the stellar binary hypothesis.  

\subsubsection{1SWASP J182924.67+232200.2} We handle this object with caution
because the transit signature is unclear for the partially owing to its period
(3.68\,days) and also to the intrinsic scatter in the
lightcurve.  Nevertheless, transit-like dips were identified from visual inspection
of the unfolded lightcurve.  No other variability is evident.  The companion radius
is credible for a planet at 1.26\Rj supported by $\eta_{p}$=0.88.  This star is
significantly brighter than any other object within $\sim$3\arcmin although
blending cannot be ruled out.  We recommend obtaining more data on this object, to
confirm the transit-like signal.  

\subsubsection{1SWASP J183104.01+323942.7} The low amplitude (0.0089\,mag) and
short duration (1.8\,hrs) of this event would have made it difficult to detect
in a fainter star.  Our lightcurve shows little out-of-transit variation and a
clear, credible period.  The predicted radius of 0.97\Rj is supported by a
slightly low but acceptable value of $\eta_{p}$=0.61.  As this candidate lies in an
uncrowded field it is a strong planetary candidate.  

\subsubsection{1SWASP J183431.62+353941.4} The classic, flat-bottomed transit
signature is clear in the folded lightcurve of this bright (10.5\,mag) star,
which shows no other signs of variability and a reasonable if quite short
period.  The companion radius of 1.3\Rj is within the expected range for a hot
Jupiter, and an $\eta_{p}$ of 0.78 makes it believable.  The high degree of
blending around this candidate raises a warning flag for an otherwise strong
candidate.  

\subsubsection{1SWASP J183805.57+423432.3} This folded lightcurve shows a degree of
clumping because the period of $\sim$3.5\,days requires a longer timebase of
observations to cover the full phase range.  Dips are clearly visible in the
unfolded data although the $V\sim$12.6\,mag means there is a high degree of
intrinsic scatter in the data.  However, the star lies in a relatively uncrowded
field and the nearest companions are $\gtrsim$10\,arcmins away.  The late-type host star
leads us to infer a small companion radius of 0.86\Rj.  Although this is tempered
by an $\eta_{p}$ of 1.6, this object remains a candidate.  

\subsubsection{1SWASP J184119.02+403008.4} The transit signature in this folded
lightcurve is unclear for the same reasons given for 1SWASP J183805.57+423432.3. 
As above, the validity of the measured signal was confirmed by visual inspection
of the unfolded data.  No other variation is evident in the lightcurve.  The 
predicted companion radius of 0.92\Rj is tempered by a slightly elevated
$\eta_{p}$=1.45, but is the brightest object in an uncrowded field.  

\subsubsection{1SWASP J184303.62+462656.4} 
\begin{figure*}
\def\subfigtopskip{4pt}
\def\subfigbottomskip{8pt}
\def\subfigcapskip{4pt}
\centering
\begin{tabular}{cc}

\subfigure[1SWASP J184119.02+403008.4]{\label{fig:J184119.02+403008.4}
\includegraphics[angle=270,width=6.5cm]{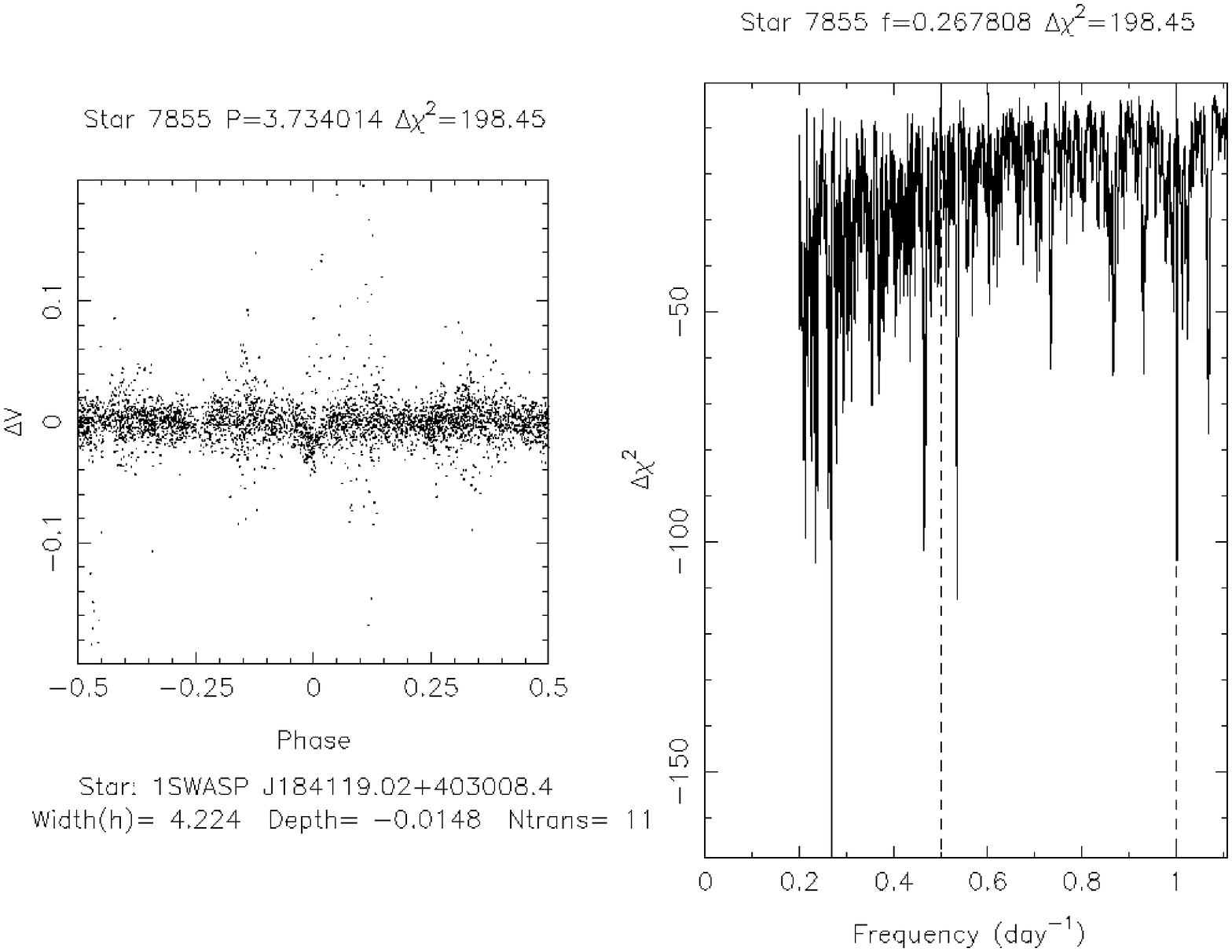}}
&
\subfigure[1SWASP J184303.62+462656.4]{\label{fig:J184303.62+462656.4}
\includegraphics[angle=270,width=6.5cm]{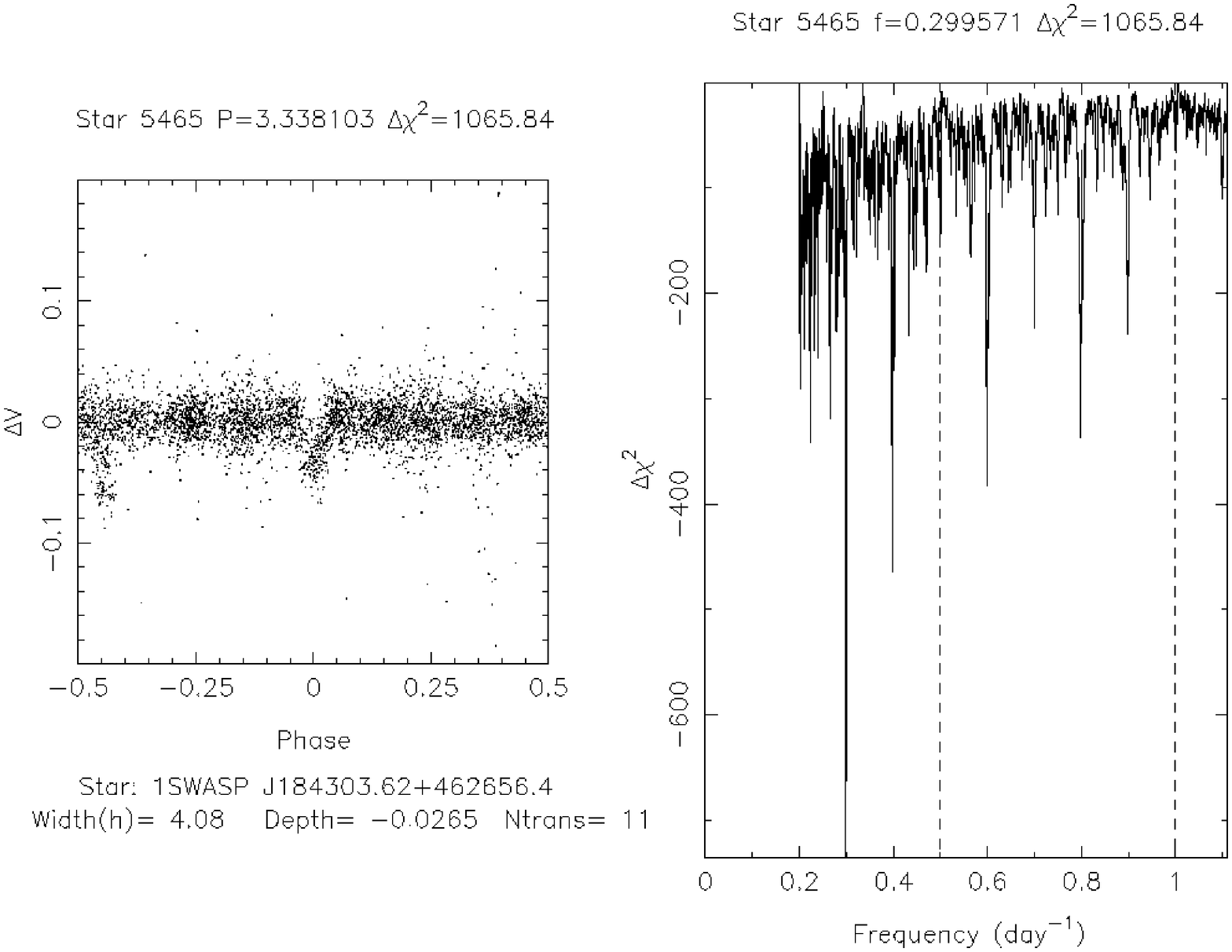}} \\

\subfigure[1SWASP J202824.02+192310.2]{\label{fig:J202824.02+192310.2}
\includegraphics[angle=270,width=6.5cm]{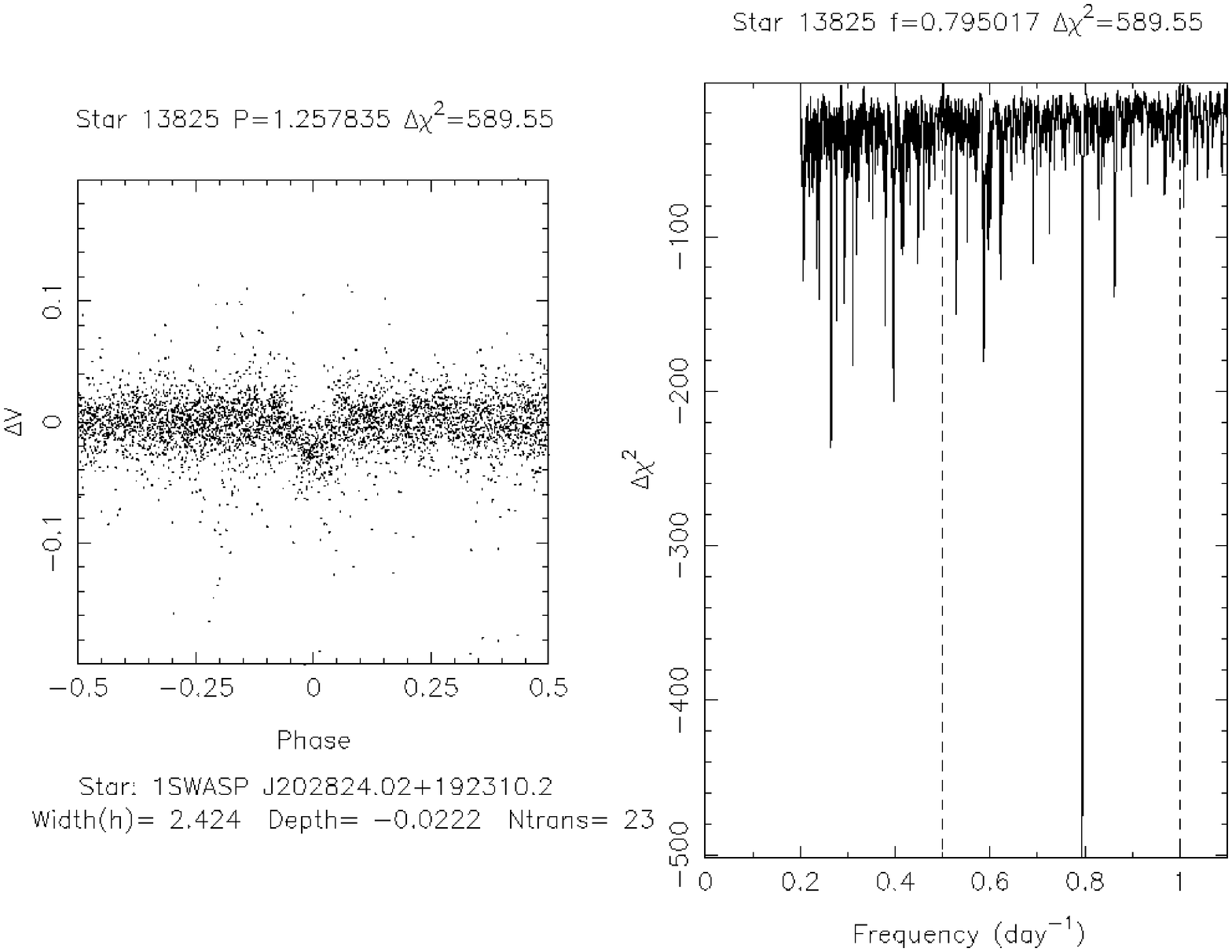}}
&
\subfigure[1SWASP J203054.12+062546.4]{\label{fig:J203054.12+062546.4}
\includegraphics[angle=270,width=6.5cm]{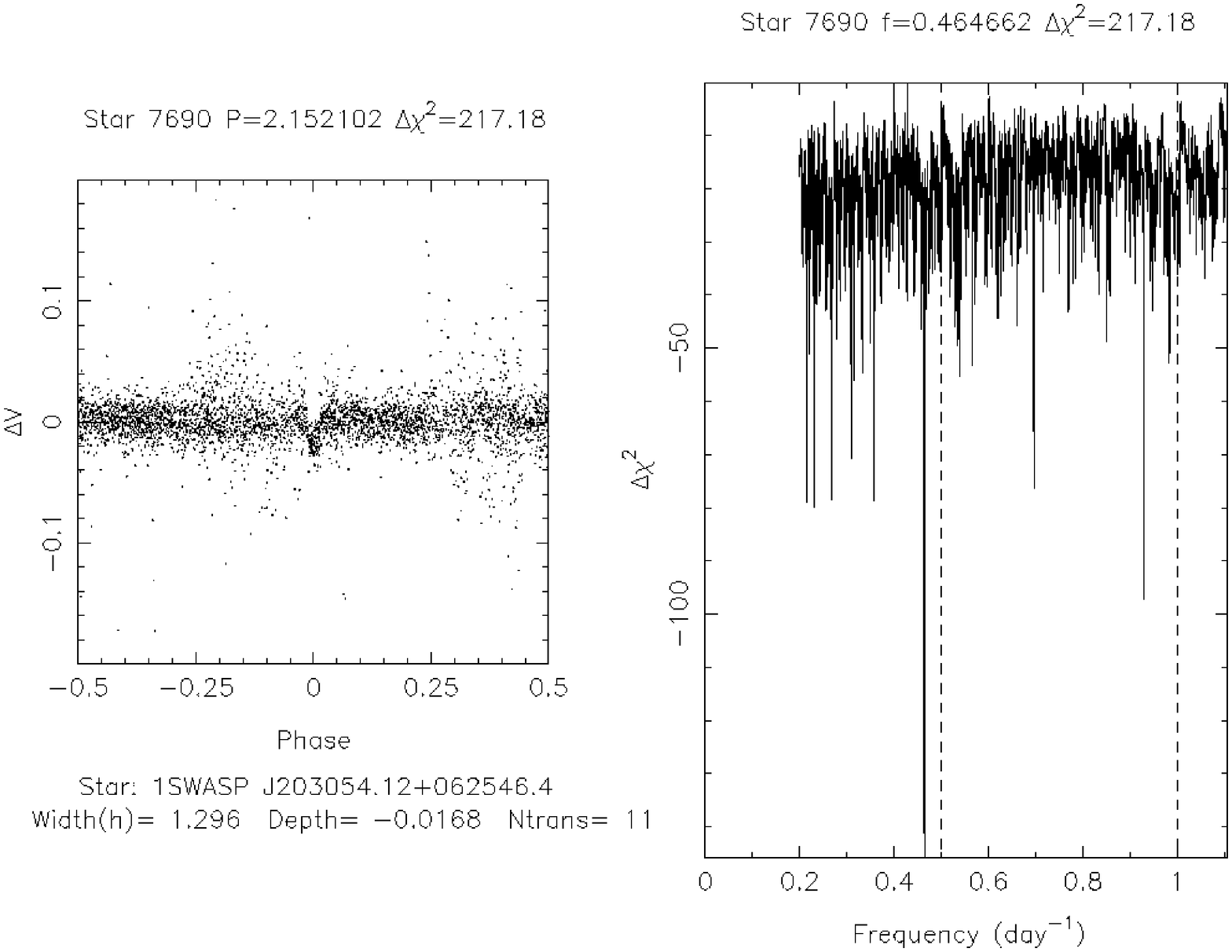}} \\

\subfigure[1SWASP J203314.77+092823.4]{\label{fig:J203314.77+092823.4}
\includegraphics[angle=270,width=6.5cm]{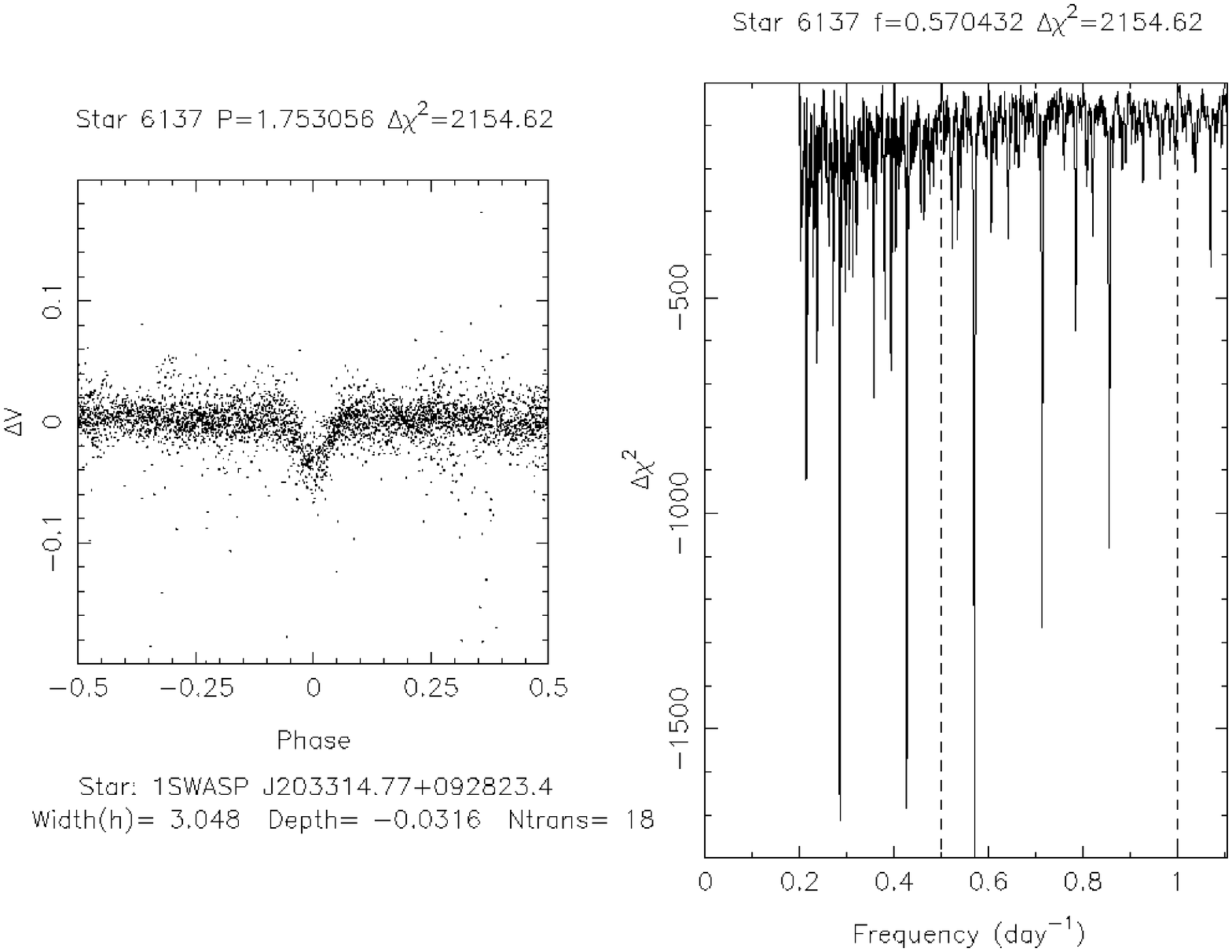}}
&
\subfigure[1SWASP J203315.84+092854.2]{\label{fig:J203315.84+092854.2}
\includegraphics[angle=270,width=6.5cm]{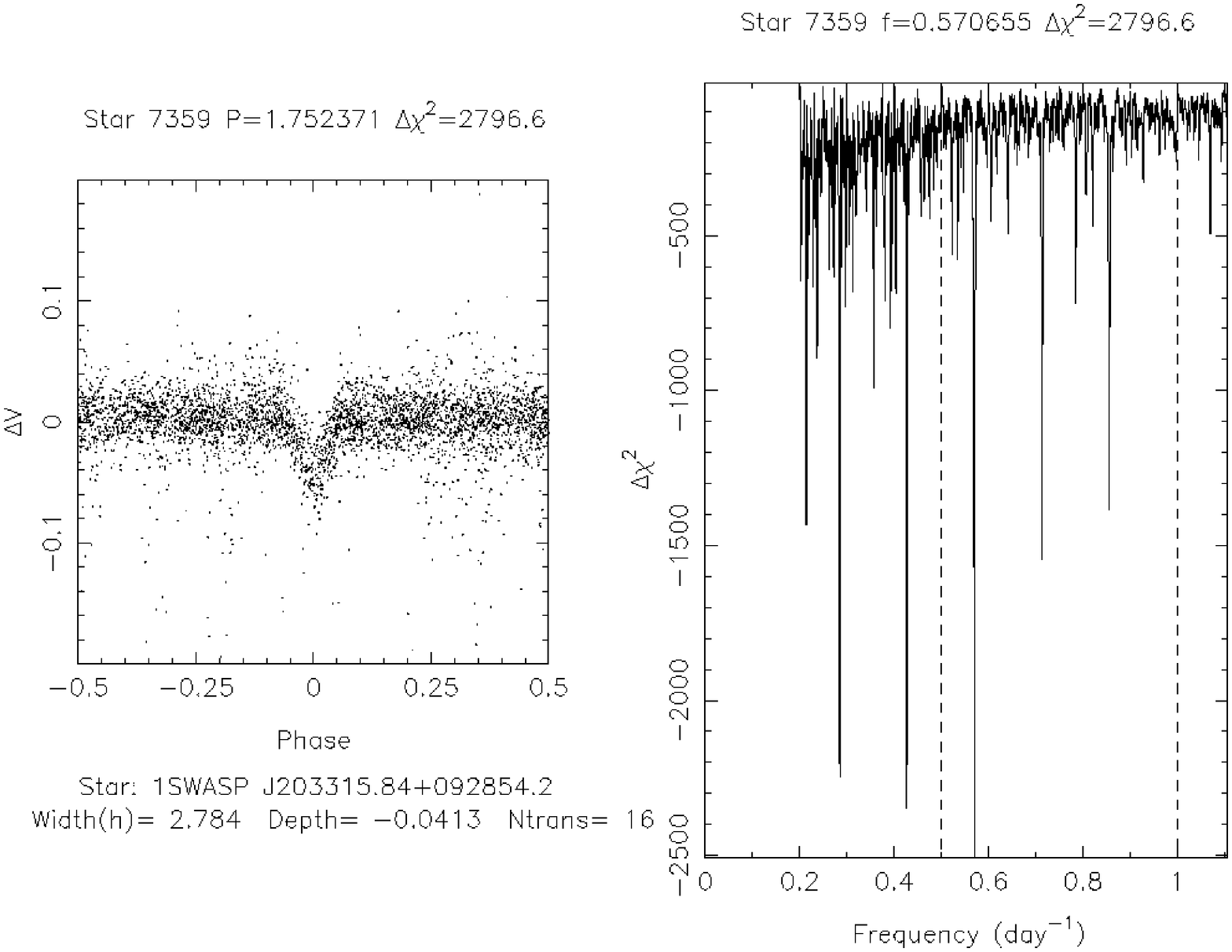}} \\

\subfigure[1SWASP J203704.92+191525.1]{\label{fig:J203704.92+191525.1}
\includegraphics[angle=270,width=6.5cm]{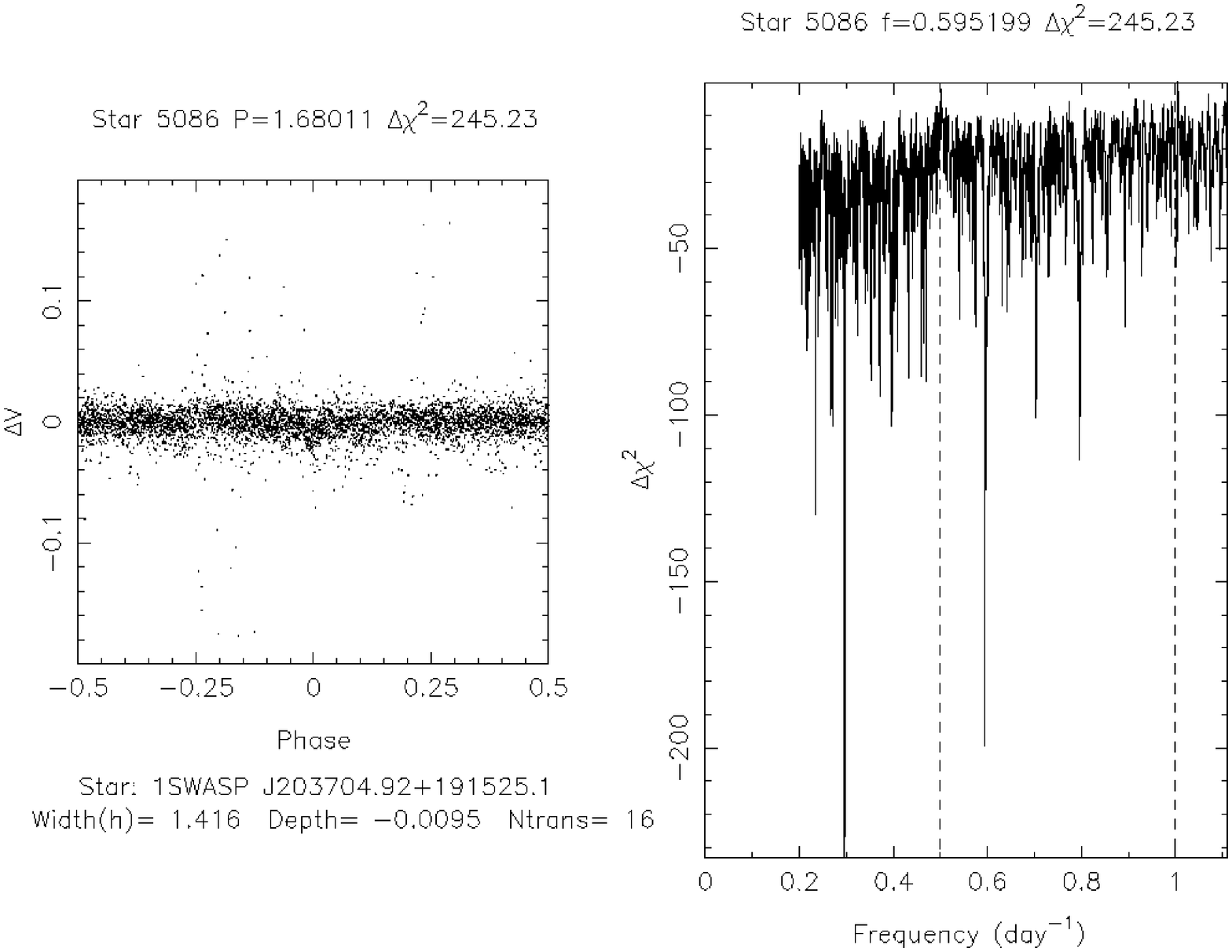}} 
&
\subfigure[1SWASP J203906.39+171345.9]{\label{fig:J203906.39+171345.9}
\includegraphics[angle=270,width=6.5cm]{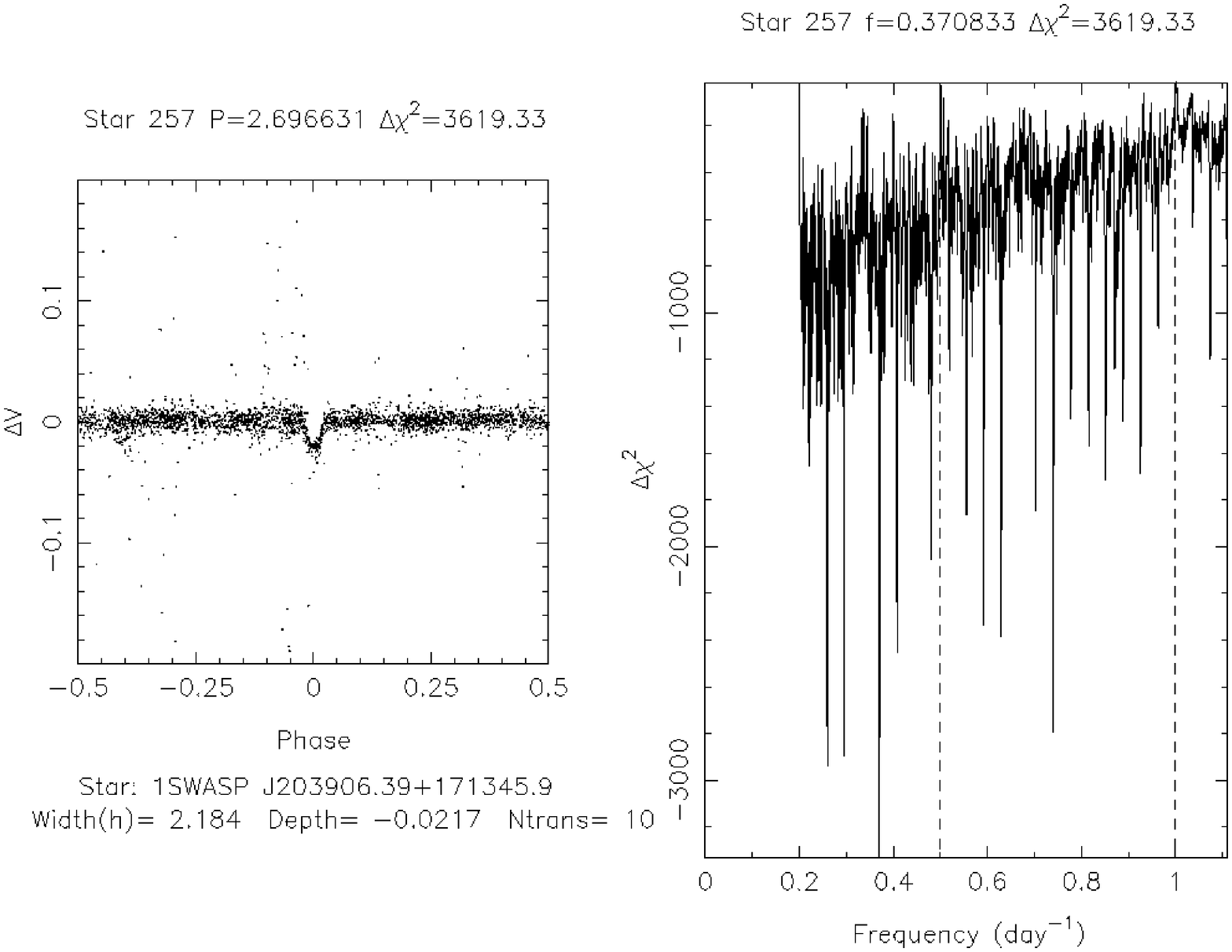}} \\

\end{tabular}
\caption{The lightcurves of the selected transit candidates, folded on the measured
period.  }
\protect\label{fig:candlcs2}
\end{figure*}

The original lightcurve showed a
`V'-shaped dip at phase 0.0 with additional points around phase -0.45, which gave
the appearance that the correct period was not identified.  The gaps in the
lightcurve indicate that the true period lies close to an alias making it difficult
to determine.  This is supported by investigation with the S-C algorithm, which
suggested a period around 10\,days; the lightcurve in
Figure~\ref{fig:J184303.62+462656.4} is shown folded on the strongest peak found by
{\sc huntsman}.  The predicted companion radius given these parameters is only
1.25\Rj, although the eclipse durations are longer than those expected for a
planetary transit ($\eta_{p}$=1.86).  This object could be a low-mass binary and
although it suffers from blending, we recommend that it continue to be observed.  

\subsubsection{1SWASP J202824.02+192310.2} This object displays transits of credible
width and depth in an otherwise flat, if noisy, lightcurve.  The host star colour
implies a radius of 1.29\Rsol and a fairly large companion object at 1.64\Rj
($\eta_{p}$=0.94).  However, light from a number of nearby stars will have contaminated
the photometry, so this could be a stellar binary. 

\subsubsection{1SWASP J203054.12+062546.4} The data for this target show a brief
but quite well defined signal in an otherwise flat, if noisy, lightcurve.  The
period and amplitude are believable for a planetary companion of 0.83\Rj with a
low but acceptable $\eta_{p}$ of 0.59.  The few nearby objects raise the
possibility of contaminating light but this remains a candidate.  

\subsubsection{1SWASP J203314.77+092823.4 \& J203315.84+092854.2} These objects
both display a similar periodicity at $P$$\sim$1.75\,days and are blended.   It should
be noted that J203315.84+092854.2 was actually eliminated at Stage 3 since it has
$S/N_{ellip}$ = 9.663.  This object was only retained because
J203314.77+092823.4 passed the automatic criteria, but could not be considered in
isolation.   Both lightcurves are a little noisy and the transit has quite shallow
in/egress slopes, but no other activity is apparent.  The late spectral type of
the former star makes this system interesting, implying a 0.94\Rj companion radius
but the $\eta_{p}$=1.59 suggest the observed dips are longer than expected for a
planetary object.  The eclipses are more likely to be due to the latter object, an
F2-F5 type, with a companion of radius 2.53\Rj ($\eta$=0.87).  

\subsubsection{1SWASP J203704.92+191525.1} The very low amplitude (9.5\,mmag)
and short (1.4\,hr) duration of this candidate makes the transit dips difficult
to detect, but the signal is seen in the unfolded lightcurve and S-C
periodogram. Obtaining follow-up photometry with a large telescope is therefore
recommended.  The predicted companion radius is close to that of Jupiter but the
value of $\eta_{p}$ is quite low, 0.55, implying that the observed transit duration
is short compared with theoretical predictions.  The target does have 2 other
stars nearby so blending is a consideration.  

\subsubsection{1SWASP J203906.39+171345.9} Datapoints overlapping the clear
transit-like dip indicated that the true period for this object was twice that
found by {\sc hunter}, i.e. 2.697\,days.  The `V'-shape morphology then becomes
clear in a flat, if noisy, lightcurve, and the predicted planet radius is only
1.35\Rj with $\eta_{p}$=0.79.  This object is the brightest in a crowded field, and
suffers from significant blending.  

\subsubsection{1SWASP J204125.28+163911.8} Despite the low amplitude of this
candidate, visual inspection of the unfolded data confirms the occurrences of
transit-like dips, and the S-C algorithm produces a strong spike at a frequency
of $1/1.221$\,days.  The predicted companion radius is extraordinarily low at
0.53\Rj owing to the very red colour of the host star, which is classified as a
mid-K type.  The high value of $\eta_{p}$ though, warns that the eclipse
duration is longer than expected, and the star, while by far the brightest
object in its field, does have nearby companions.  Overall, we recommend this
object for further investigation. 

\subsubsection{1SWASP J204142.49+075051.5} The low amplitude (10.2\,mmags) and
faint magnitude ($V$$\sim$12\,mag) of this object conspire to produce a very
shallow transits of $\sim$2.3\,hrs duration.  Their existence was confirmed by
visual inspection however, and the strongest peak in the S-C periodogram
corresponds to 2.763\,days.  Once folded on this period, the lightcurve shows no
other form of variation from the mid- to late-K type host star.  The low predicted
companion radius, 0.59\Rj, makes this an exciting candidate, particularly in the
light of the $\eta_{p}$=1.04.  Some nearby stars raise a caution of potential
blending.  
  
\subsubsection{1SWASP J204323.83+263818.7} This star displays a clear `V'-shaped dip
when the lightcurve is folded on the period of one of the top five peaks,
$P$=1.421\,days.  Transits were observed of reasonable amplitude (0.04\,mag) and
fairly short (1.32\,hr) duration, however there are hints of ellipsoidal variation
and faint signs of secondary eclipses.  The estimate radius for the companion object
is a promising 1.32\Rj, though with a comparatively low $\eta_{p}$=0.63, but this
star has a very close companion and is certainly affected by blending.  High
resolution imaging and/or spectroscopic observations are required to confirm or
dismiss this candidate.  

\subsubsection{1SWASP J204617.02+085412.0} Another case where faint magnitude
($V$$\sim$12.3\,mag) and low amplitude (9.5\,mmag) mask the transit signal, but
close inspection reveals a series of shallow dips.  The S-C periodogram is
unclear, the period being so close to 2\,days, but the folded lightcurve shows a
transit-like dip in an otherwise flat dataset.  The G-type host star has one
very close blended star, albeit a much fainter one as well as a group of other
stars within the aperture, meaning the true companion radius could well be
greater than the predicted 0.91\Rj.  Nevertheless, we recommend this object for
follow-up observations.  

\subsubsection{1SWASP J204712.42+202544.5} The faintness of this star (12.386\,mag)
leads to a noisy but apparently flat lightcurve except for a clear and credible
transit dip.  The host star's IR colour suggests a mid-K spectral type and a
companion radius of 0.95\Rj, supported by the $\eta_{p}$=0.91.  Despite a number of
much fainter companions, the level of blending is low in this field, strengthening
the case for a planetary explanation in this case. 

\subsubsection{1SWASP J204745.08+103347.9} This is another case of a clear
`V'-shaped dip implying a stellar companion in spite of a low ($\sim$0.03\,mag)
amplitude in a lightcurve which shows no signs of ellipsoidal variation.  The
estimated companion radius of 1\Rj is belied by a long transit duration
($\eta_{p}$=1.47).  The likelihood of blending in this case points to a stellar binary.

\subsubsection{1SWASP J205027.33+064022.9} This bright ($V$$\sim$10.2\,mag) star
displays a very shallow (9.6\,mmag) but clear `U'-shaped dip with an out of transit
lightcurve that shows slight signs of ellipsoidal variation.  The photometric
precision is such that the transits are immediately obvious in the unfolded data. 
This is a good candidate, with a prediction companion radius of 0.92\Rj though the
transits are slightly longer than expected ($\eta_{p}$ = 1.43).  The star has two
nearby companions of similar magnitude, so we have flagged it `B' for a potential
blend.  

\subsubsection{1SWASP J205308.03+192152.7} The very low amplitude (0.0068\,mag)
transit signal is just visible over the noise in this otherwise flat lightcurve but
appears to exhibit a flat-bottomed dip.  The amplitude means that despite a host
star radius of 1.24\Rsol the estimated companion radius is only 0.87\Rj, supported
by an $\eta_{p}$=1.04.  The 5 nearby stars means that contamination of the
photometry cannot be ruled out without further observations.  

\subsubsection{1SWASP J210009.75+193107.1} This folded lightcurve displays a shallow
but clear `U'-shaped dip which can also been seen in the unfolded data.  The
periodogram exhibits a strong peak on the frequency $1/3.054875$ although the period
is close to an integer multiple of 1\,day.  The predicted radius implies a
Jovian-sized companion, supported by an $\eta_{p}$ of 0.71, but this object does
suffer from blending.  

\subsubsection{1SWASP J210151.43+072326.7} At $V$=12.476\,mag, this is one of our
faintest candidates, and the lightcurve has a commensurate level of noise, but
transit-like dips can be seen in the unfolded data also and no other variability is
visible in the lightcurve.  The estimated companion radius of 0.92\Rj is supported
by $\eta_{p}$=0.99.  This star does have three nearby objects of similar brightness,
and a much fainter object within $\sim$4\arcsec, so blending is a possibility
here.  

\subsubsection{1SWASP J210318.01+080117.8}  This lightcurve shows a clear transit
dip with believable width, depth and period and although the intrinsic noise makes
the true morphology unclear there is no sign of any other variability.  The
measured duration is a close match for that predicted, and the companion radius of
1.01\Rj makes this a strong candidate.  A single nearby star raises a 
possibility of blending.  

\subsubsection{1SWASP J210352.56+083258.9} While noisy, this lightcurve clearly
exhibits a $\sim$0.02\,mag dip and is flat out of transit though the relatively long
period (close to the 4$\times$ multiple of the 1\,day alias) results in a certain
amount of `clumping' of datapoints.  The 1.61\Rj companion radius is on the
borderline for a planetary companion, but is supported by an $\eta_{p}$=0.95.  Three
nearby stars mean that the photometry for this object could be contaminated and that
follow-up observations are necessary.  

\subsubsection{1SWASP J210909.05+184950.9} This bright ($V\sim$9.9\,mag) object
shows a remarkably smooth lightcurve out of transit, allowing {\sc huntsman} to
detect the very shallow (8.2\,mmag), `U'-shaped transit dip.  Closer inspection
however, reveals a marked ellipsoidal variation, flagging this object as a probable
stellar binary.  The host star is found to be of mid F-type yet the predicted
companion radius is only 1.07\Rj, supported with an $\eta_{p}$ of 0.71.  While this
object is certainly the brightest in its field, it is likely that nearby, fainter
stars will have affected the SW-N photometry.  We encourage follow-up observations
of this target.  

\subsubsection{1SWASP J210912.02+073843.3} This star was included despite a high
$S/N_{ellip}$=12.508 because the folded lightcurve appeared fairly flat to visual
inspection, and showed clear, flat bottomed transits with a duration of 2.28\,hrs \&
$\delta$=0.0213\,mag appropriate for an exoplanet.  The F-type host star implies a
$R_{p}$=1.52\Rj \& $\eta_{p}$=0.89.  Further inspection reveals the object to be
severely blended, so the true eclipses will be deeper.  As they are flat bottomed,
the orbit must be edge-on.  The companion could therefore be a low mass star or
brown dwarf and higher resolution photometry is recommended.  

\subsubsection{1SWASP J211608.42+163220.3} The brief dip in this flat lightcurve
appears to be `V'-shaped, although intrinsic noise makes the morphology difficult
to judge.  The strong $\Delta \chi^{2}$ peak implies a credible period of
3.47\,days.  The estimated companion radius is Jovian at 1.18\Rj though the low
$\eta_{p}$ of 0.59 implies the observed duration is shorter than predicted.  As
this star does not suffer from any blending it is a strong candidate for
follow-up.  

\subsubsection{1SWASP J211645.22+192136.8} This object has a relatively long period
of $\sim$4.4\,days which means that a single observing station will normally only
observe roughly one transit in two, weather permitting.  For this reason there are
some gaps in the lightcurve and, although a reasonable number of transits were
detected, there is a higher degree of uncertainty on the period.  This may explain
the somewhat unclear transit curve.  Nevertheless, this is a promising candidate: it
is an isolated bright object, and the predicted companion radius is 1.23\Rj with
$\eta_{p}$=0.71.  

\subsubsection{1SWASP J212532.55+082904.4} The transit signal is clearly visible in
this slightly noisy lightcurve though the shape is not well defined.  The companion
radius is large but still within the planetary range at 1.58\Rj, backed up by an
$\eta_{p}$=0.82.  There are no other stars close by this object, so it too is a
target for further observations.  

\subsubsection{1SWASP J212843.62+160806.2} The folded lightcurve clearly shows a
shallow dip of $\sim$0.02\,mag with a believable period of 1.376\,days.  Closer
inspection is needed to spot faint signs of a secondary eclipse and possible
ellipsoidal variation ($S/N_{ellip}$=8.841).  The target has three objects nearby
though all are $\geq$4\,mags fainter.  The late spectral type, derived from IR
colours, leads us to suggest that this could be a low-mass binary.  

\subsubsection{1SWASP J212855.03+075753.5} The faintness ($V$$\sim$12.2\,mag) of
this host star and the long period result in a low number of transits detected,
and an under sampled, sharply `V'-shaped signal in a noisy, but apparently flat,
lightcurve.  The nearby presence of a star of similar magnitude will also have
contributed to the photometric uncertainty.  The colour indicates a late G-type
host star with a companion of radius 1.35\Rj, though the measured transit
duration is shorter than expected for a planet ($\eta_{p}$=0.58).  

\begin{landscape}
\begin{table} 
\centering
\caption{Candidate list after Stage 4.  $N_{bri,faint}$ gives the number of USNO-B1.0 objects listed within 48\arcsec of the target that are brighter or 
$<$5\,mag fainter respectively. Spectral types marked with an asterisk were estimated from the 2MASS $J-H$ index in cases where the $V_{SW}-K$ index was at the extremity of 
the range, and unreliable.  Borderline candidates are marked with $^{\dagger}$ and are listed 
for information.} 
\label{tab:stage4list} 
\begin{tabular}{lccccccccccccccc} 
\hline 
 Identifier   	      	  & Period    	    & Duration    & $\delta$      & $V_{SW}-K$  & $J-H$       & Spectral    &$R_{*}$	& $R_{p}$   & $\eta_{p}$       & $N_{bri}$ & $N_{faint}$  &  & Codes & \\
1SWASP...             	  & (days)    	    & (hrs)   	  & (mag)     	  &     	&	      & Type	    &(\Rsol)	& (\Rj)     &		       &      &	      	& R   & Eta  & Blend \\
\hline 
$^{\dagger}$J175919.79+353935.1 & 4.846186  & 4.272 	  & 0.026 	  & 3.51 	& 0.61 	      & M0    	    & 0.64    	& 0.88      & 1.58    	       & 0    & 2     	& A   & C    & B \\
$^{\dagger}$J180103.13+511557.1 & 4.785081  & 3.672 	  & 0.0145 	  & 2.46 	& 0.48        & K4    	    & 0.73    	& 0.75      & 1.3     	       & 0    & 1     	& B   & A    & B \\
 J180304.96+264805.4  	  & 2.364723  	    & 5.136   	  & 0.0254    	  & 2.52	&  0.53       & K4	    & 0.72	& 0.98      & 2.26	       & 0    & 6     	& A   & C    & C \\ 
 J180726.64+224227.9  	  & 4.246971  	    & 4.752   	  & 0.0205    	  & 1.91	&  0.29       & G9	    & 0.87	& 1.06      & 1.56	       & 0    & 2     	& A   & C    & B \\ 
$^{\dagger}$J181129.19+235412.4 & 4.234895  & 8.568 	  & 0.0578 	  & 1.91 	& 0.48        & G9    	    & 0.87    	& 1.78      & 2.61    	       & 0    & 2     	& C   & C    & B \\
{\bf J181317.03+305356.0} & {\bf 2.248420}  & {\bf 1.896} & {\bf 0.0145}  & {\bf 1.6}	& {\bf 0.28}  & {\bf G3}    & {\bf 1.02}& {\bf 1.05}& {\bf 0.71}       & {\bf 0} & {\bf 2} & {\bf A} & {\bf A} & {\bf B}\\ 
{\bf J181454.99+391146.0} & {\bf 1.102625}  & {\bf 1.56}  & {\bf 0.0235}  & {\bf 2.89}  & {\bf 0.74}  & {\bf K5}    & {\bf 0.68}& {\bf 0.89}& {\bf 0.92}       & {\bf 0} & {\bf 10}& {\bf A} & {\bf A} & {\bf C}\\ 
{\bf J181958.25+492329.9} & {\bf 2.368548}  & {\bf 2.424} & {\bf 0.0061}  & {\bf 1.57}  & {\bf 0.26}  & {\bf G2}    & {\bf 1.04}& {\bf 0.69}& {\bf 0.92}       & {\bf 0} & {\bf 2} & {\bf A} & {\bf A} & {\bf B}\\ 
 J182127.09+200011.7  	  & 2.647752  	    & 4.248   	  & 0.0366    	  & 1.26	& 0.18        & F7	    & 1.25	& 2.04      & 1.27	       & 0    & 6     	& C   & A    & C \\ 
 J182131.07+483735.5  	  & 1.809191  	    & 2.832   	  & 0.0167    	  & 0.56	& 0.26        & A7-F0	    & 1.79	& 1.97      & 0.82	       & 0    & 3     	& C   & C    & C \\ 
 J182333.22+222801.2  	  & 1.821008  	    & 3.432   	  & 0.0421    	  & 1.59	& 0.21        & G3	    & 1.03	& 1.8	    & 1.29	       & 0    & 2     	& C   & A    & B \\
$^{\dagger}$J182339.64+210805.5 & 1.585846  & 2.088 	  & 0.0245 	  & 1.16 	& 0.28        & F6    	    & 1.32    	& 1.76      & 0.74    	       & 0    & 13    	& C   & A    & C \\
 J182346.12+434241.3  	  & 11.87746  	    & 6.841   	  & 0.065     	  & 1.26	&  0.19       & F7	    & 1.25	& 2.72      & 1.19	       & 0    & 1     	& C   & A    & B \\
{\bf J182620.36+475902.8} & {\bf 3.04365}   & {\bf 4.032} & {\bf 0.0628}  & {\bf 2.35}  & {\bf 0.45}  & {\bf K3}    & {\bf 0.75}& {\bf 1.6} & {\bf 1.49}       & {\bf 0} & {\bf 0}& B & A    & A \\ 
$^{\dagger}$J182626.38+374954.8 & 4.698312  & 4.944 	  & 0.0157 	  & 1.26 	& 0.25        & F7    	    & 1.25    	& 1.34      & 1.29    	       & 0    & 6     	& A   & A    & C \\
 J182916.00+235724.8  	  & 8.901122  	    & 4.168   	  & 0.038     	  & 1.48	&  0.34       & G0	    & 1.1	& 1.83      & 0.9	       & 0    & 6     	& C   & A    & C \\
{\bf J182924.67+232200.2} & {\bf 3.678186}  & {\bf 2.952} & {\bf 0.0173}  & {\bf 1.45}  & {\bf 0.21}  & {\bf G0}    & {\bf 1.12}& {\bf 1.26}& {\bf 0.88}       & {\bf 0} &{\bf  3}& A & A    & C \\ 
$^{\dagger}$J182927.04+233217.1 & 4.903747  & 4.704 	  & 0.0063 	  & 2.5 	& 0.52        & K4    	    & 0.73    	& 0.49      & 1.71    	       & 0    & 5     	& A   & C    & C \\
$^{\dagger}$J183043.97+230526.1 & 3.680977  & 4.296 	  & 0.0098 	  & 1.68 	& 0.26        & G5    	    & 0.98    	& 0.83      & 1.43    	       & 0    & 2     	& A   & A    & B \\
{\bf J183104.01+323942.7} & {\bf 2.378781}  & {\bf 1.776} & {\bf 0.0089}  & {\bf 1.33}  & {\bf 0.21}  & {\bf F8}    & {\bf 1.2} & {\bf 0.97}& {\bf 0.61}       & {\bf 0} & {\bf 2}& A & A    & B\\ 
 J183104.12+243739.3  	  & 0.746192  	    & 3.836   	  & 0.0197    	  & 1.47	&  0.23       & G0	    & 1.1	& 1.32      & 1.96	       & 0    & 6     	& A   & C    & C \\
{\bf J183431.62+353941.4} & {\bf 1.846796}  & {\bf 2.28}  & {\bf 0.0127}  & {\bf 1.12}  & {\bf 0.2}   & {\bf F5}    & {\bf 1.35}& {\bf 1.3} & {\bf 0.78}       & {\bf 0} & {\bf 3}& A & A    & C\\ 
$^{\dagger}$J183517.51+390316.2 & 4.073428  & 5.16 	  & 0.012 	  & 2.71 	& 0.49        & K5    	    & 0.7     	& 0.65      & 2       	       & 0    & 7     	& A   & C    & C \\
 J183723.62+373721.9  	  & 3.300887  	    & 4.32    	  & 0.0251    	  & 2.69	& 0.51        & K5	    & 0.7	& 0.95      & 1.73	       & 0    & 4     	& A   & C    & C \\
{\bf J183805.57+423432.3} & {\bf 3.515957}  & {\bf 4.104} & {\bf 0.0197}  & {\bf 2.51}  & {\bf 0.55}  & {\bf K4}    & {\bf 0.72}& {\bf 0.86}& {\bf 1.6}        & {\bf 0} & {\bf 4}& A & C    & C\\ 
{\bf J184119.02+403008.4} & {\bf 3.734014}  & {\bf 4.224} & {\bf 0.0148}  & {\bf 1.86}  & {\bf 0.29}  & {\bf G8}    & {\bf 0.89}& {\bf 0.92}& {\bf 1.45}       & {\bf 0} & {\bf 1}& A & A    & B\\ 
{\bf J184303.62+462656.4} & {\bf 10.07384}  & {\bf 7.253} & {\bf 0.037}   & {\bf 2.3}	& {\bf 0.55}  & {\bf K3}    & {\bf 0.76}& {\bf 1.25}& {\bf 1.86}       & {\bf 0} & {\bf 3}& {\bf A}  & {\bf C}  & {\bf C} \\ 
 J202820.25+094651.0  	  & 2.146933  	    & 4.776   	  & 0.0085    	  & 2.37	& 0.48        & K3	    & 0.75	& 0.59      & 2.23	       & 0    & 2     	& A   & C    & B\\ 
{\bf J202824.02+192310.2} & {\bf 1.257835}  & {\bf 2.424} & {\bf 0.0222}  & {\bf 1.2}	& {\bf 0.2}   & {\bf F6}    & {\bf 1.29}& {\bf 1.64}& {\bf 0.94}       & {\bf 0} & {\bf 8}& B & A    & C\\ 
$^{\dagger}$J202907.09+171631.7 & 4.117398  & 4.968 	  & 0.0309 	  & 1.61 	& 0.37        & G3    	    & 1.02    	& 1.53      & 1.46    	       & 0    & 13    	& A   & A    & C \\
{\bf J203054.12+062546.4} & {\bf 2.152102}  & {\bf 1.296} & {\bf 0.0168}  & {\bf 2.35}  & {\bf 0.41}  & {\bf K3}    & {\bf 0.75}& {\bf 0.83}& {\bf 0.59}       & {\bf 0 } & {\bf 3}& A & A   & C \\ 
$^{\dagger}$J203229.10+132820.9 & 4.632829  & 4.608 	  & 0.047 	  & 2.35 	& 0.61        & K3    	    & 0.75    	& 1.39      & 1.51    	       & 0    & 15    	& A   & C    & C \\
 J203247.55+182805.3  	  & 2.522688  	    & 7.776   	  & 0.0118    	  & 1.42	&  0.27       & F9	    & 1.14	& 1.06      & 2.66	       & 0    & 10    	& A   & C    & C\\ 
 {\bf J203314.77+092823.4} & {\bf 1.753056}  & {\bf 3.048} & {\bf 0.0316}  & {\bf 3.87}  & {\bf 0.77}  & {\bf M0}    & {\bf 0.62}& {\bf 0.94}& {\bf 1.59}      & {\bf 0} & {\bf 2}& A & C    & B\\ 
{\bf J203315.84+092854.2} & {\bf 1.752371}  & {\bf 2.784} & {\bf 0.0413}  & {\bf -0.46} & {\bf 0.198} & {\bf F2-F5$^{*}$} & {\bf 1.46}& {\bf 2.53}& {\bf 0.87} & {\bf 1} & {\bf 12}& {\bf C}   & {\bf A}  & {\bf C} \\ 
 J203543.98+072641.1  	  & 1.85463   	    & 2.76	  & 0.0195    	  & 0.99	& 0.24        & F3	    & 1.43	& 1.7	    & 0.9	       & 0    & 3     	& B   & A    & C\\  	      
{\bf J203704.92+191525.1} & {\bf 1.68011}   & {\bf 1.416} & {\bf 0.0095}  & {\bf 1.37}  & {\bf 0.27}  & {\bf F9}    & {\bf 1.17}& {\bf 0.97}& {\bf 0.55}       & {\bf 0} & {\bf 2} & A   & A    & B\\		  
 J203717.02+114253.5  	  & 3.118049  	    & 2.496   	  & 0.0274    	  & 1.31	& 0.25        & F8	    & 1.21	& 1.71      & 0.74	       & 0    & 1     	& B   & A    & B\\  	      
{\bf J203906.39+171345.9} & {\bf 2.696631}  & {\bf 2.184} & {\bf 0.0217}  & {\bf 1.33}	& {\bf 0.22}  & {\bf F8}    & {\bf 1.2}& {\bf 1.35}& {\bf 0.79}       & {\bf 0} & {\bf 2} & {\bf A}   & {\bf A}  & {\bf B} \\  
$^{\dagger}$J203932.30+162451.1 & 1.520504  & 8.976 	  & 0.02 	  & 3.35 	& 0.62        & K7    	    & 0.65    	& 0.78      & 4.92    	       & 0    & 2     	& A   & C    & B \\				     
{\bf J204125.28+163911.8} & {\bf 1.221506}  & {\bf 2.88}  & {\bf 0.008}   & {\bf 2.77}  & {\bf 0.54}  & {\bf K5}    & {\bf 0.69}& {\bf 0.53}& {\bf 1.71}       & {\bf 0} & {\bf 4} & A   & C & C \\ 	     
$^{\dagger}$J204142.31+052007.5 & 3.216912  & 4.776 	  & 0.0279 	  & 2.05 	& 0.38        & K0    	    & 0.82    	& 1.17      & 1.75    	       & 0    & 5     	& A   & C    & C \\				     
{\bf J204142.49+075051.5} & {\bf 2.763125}  & {\bf 2.328} & {\bf 0.0102}  & {\bf 2.86}  & {\bf 0.59}  & {\bf K5}    & {\bf 0.69}& {\bf 0.59}& {\bf 1.04}       & {\bf 0} & {\bf 4} & {\bf A}   & {\bf A}  & {\bf C} \\  
\hline								  	        									       
\end{tabular}																		       
\end{table}																		       
\end{landscape}

\begin{landscape}
\begin{table} 
\centering
\contcaption{Candidate list after Stage 4.  $N_{bri,faint}$ gives the number of USNO-B1.0 objects listed within 48\arcsec of the target that are
brighter or $<$5\,mag fainter, respectively. 1SWASP J211448.98+203557.1 was identified in two fields, SW2114+1628 \& SW2115+2351 and independent results are given for each.  
Spectral types marked with an asterisk were estimated from the 2MASS $J-H$ index in cases where the $V_{SW}-K$ index was at the extremity of the range, and 
unreliable.  Borderline candidates are marked with $^{\dagger}$ and are listed for information. Parenthesis around an object indicates that spectroscopic data are discussed in Section~\ref{sec:specdata}.} 
\label{tab:secstagecand2} 
\begin{tabular}{lccccccccccccccc} 
\hline 
 Identifier   	      	  & Period	    & Duration	  & $\delta$	  & $V_{SW}-K$  & $J-H$       & Spectral    & $R_{*}$	& $R_{p}$   & $\eta_{p}$  & $N_{bri}$  & $N_{faint}$   &	& Code & \\
1SWASP...         	  & (days)	    & (hrs)	  & (mag)	  &            &	      & Type	    & (\Rsol)	& (\Rj)     &		  &	  &   	     &	&     & \\
\hline 
 J204211.19+240145.1  	  & 3.362228  	    & 2.664   	  & 0.0544    	  & 1.65	&  0.19       & G4	    & 1.0	& 1.99      & 0.81	       & 0    & 2     	& C   & C    & B \\ 		  
{\bf J204323.83+263818.7} & {\bf 1.421123}  & {\bf 1.32}  & {\bf 0.0366}  & {\bf 2.1}	& {\bf 0.32}  & {\bf K1}    & {\bf 0.81}& {\bf 1.32}& {\bf 0.63}       & {\bf 0} & {\bf 2} & {\bf A}   & {\bf A}  & {\bf B} \\  
$^{\dagger}$J204328.95+054823.1 & 3.939179  & 2.328 	  & 0.0617 	  & 1.69 	& 0.25        & G5    	    & 0.97    	& 2.06      & 0.68    	       & 0    & 2     	& C   & A    & B \\					 
 (J204456.57+182136.0  	  & 8.147196  	    & 7.821   	  & 0.044     	  & 1.61	&  0.24       & G3	    & 1.02	& 1.83      & 1.79	       & 0    & 2     	& C   & C    & B )\\		 
{\bf J204617.02+085412.0} & {\bf 1.947141}  & {\bf 2.184} & {\bf 0.0095}  & {\bf 1.48}  & {\bf 0.25}  & {\bf G0}    & {\bf 1.1} & {\bf 0.91}& {\bf 0.84}       & {\bf 0} & {\bf 5} & A   & A   & C\\	      
{\bf J204712.42+202544.5} & {\bf 2.61264}   & {\bf 2.064} & {\bf 0.0275}  & {\bf 3.01}  & {\bf 0.58}  & {\bf K5$^{*}$}    & {\bf 0.67}& {\bf 0.95}& {\bf 0.91} & {\bf 0} & {\bf 8} & A   & A   & C\\ 
{\bf J204745.08+103347.9} & {\bf 3.235407}  & {\bf 3.648} & {\bf 0.0289}  & {\bf 2.79}  & {\bf 0.63}  & {\bf K7$^{*}$}    & {\bf 0.69}& {\bf 1.00}& {\bf 1.47} & {\bf 0} & {\bf 2} & A   & A   & B\\ 
$^{\dagger}$J204905.55+110000.4 & 1.371571  & 1.584 	  & 0.023 	  & 1.07 	& 0.24        & F5    	    & 1.38    	& 1.79      & 0.57    	       & 0    & 7     	& C   & A    & C \\
{\bf J205027.33+064022.9} & {\bf 1.229345}  & {\bf 3.192} & {\bf 0.0096}  & {\bf 1.47}  & {\bf 0.24}  & {\bf G0}    & {\bf 1.1} & {\bf 0.92}& {\bf 1.43}       & {\bf 0} & {\bf 2} & A   & A   & B\\ 
$^{\dagger}$J205218.75+182330.0 & 2.197814  & 3.48 	  & 0.0441 	  & 1.45 	& 0.2 	      & G0    	    & 1.12    	& 2.01      & 1.17    	       & 0    & 4     	& C   & A    & C \\
 J205223.03+151046.8  	  & 2.910170  	    & 2.400   	  & 0.0409    	  & 1.5 	&  0.2        & G1	    & 1.08	& 1.86      & 0.75	       & 0    & 0       & C   & A    & A \\
 J205302.40+201748.3  	  & 4.931719  	    & 8.88	  & 0.0084    	  & 1.58	&  0.33       & G2	    & 1.03	& 0.81      & 2.61	       & 0    & 1       & A   & C    & B \\ 
{\bf J205308.03+192152.7} & {\bf 1.676449}  & {\bf 2.736} & {\bf 0.0068}  & {\bf 1.27} & {\bf 0.21}   & {\bf F7}    & {\bf 1.24}& {\bf 0.87}& {\bf 1.04}  & 0 	  & {\bf 5} & A & A   & C\\ 
 J205438.05+105040.7  	  & 4.198031  	    & 3.048   	  & 0.0468    	  & 1.42       &  0.19        & F9	    & 1.14	& 2.10      & 0.81	  & 0 	  & 0	    & C & A   & A \\
{\bf J210009.75+193107.1} & {\bf 3.054875}  & {\bf 2.424} & {\bf 0.0082}  & {\bf 1.08} & {\bf  0.11}  & {\bf F5}    & {\bf 1.38}& {\bf 1.07}& {\bf 0.71}  & 0 	  & {\bf 1} & A & A   & B\\ 
$^{\dagger}$J210130.24+190021.7 & 2.683466  & 1.608 	  & 0.0709    	  & 1.9        & 0.34 	      & G9    	    & 0.88    	& 2   	    & 0.56    	  & 0 	  & 3 	    & C & A   & C \\
{\bf J210151.43+072326.7} & {\bf 2.220785}  & {\bf 2.472} & {\bf 0.0138}  & {\bf 1.79} &  {\bf 0.33}  & {\bf G7}    & {\bf 0.92}& {\bf 0.92}& {\bf 0.99}  & 0 	  & {\bf 3} & A & A   & C\\ 
$^{\dagger}$J210231.79+101014.5 & 1.506187  & 1.608 	  & 0.0296    	  & 1.79       & 0.35 	      & G7    	    & 0.92    	& 1.35      & 0.71    	  & 0 	  & 6 	    & A & A   & C\\
{\bf J210318.01+080117.8} & {\bf 1.223824}  & {\bf 1.92}  & {\bf 0.0167}  & {\bf 1.79} & {\bf 0.31}   & {\bf G7}    & {\bf 0.92}& {\bf 1.01}& {\bf 0.93}  & 0 	  & {\bf 1} & A & A   & B\\ 
$^{\dagger}$J210335.82+125637.6 & 1.447543  & 2.856 	  & 0.0146    	  & 1.27       & 0.18 	      & F7    	    & 1.24    	& 1.28      & 1.11    	  & 0 	  & 2 	    & A & A   & B \\
{\bf J210352.56+083258.9} & {\bf 3.89368}   & {\bf 3.504} & {\bf 0.0227}  & {\bf 1.25} & {\bf 0.2 }   & {\bf F7}    & {\bf 1.25}& {\bf 1.61}& {\bf 0.95}  & 0 	  & {\bf 3} & B & A   & C\\ 
{\bf J210909.05+184950.9} & {\bf 2.91879}   & {\bf 2.664} & {\bf 0.0083}  & {\bf 0.86} & {\bf  0.07}  & {\bf F1}    & {\bf 1.51}& {\bf 1.17}& {\bf 0.75}  & 0 	  & {\bf 3} & A & A   & C\\ 
{\bf J210912.02+073843.3} & {\bf 1.36983}   & {\bf 2.28}  & {\bf 0.0213}  & {\bf 1.3}  & {\bf 0.21}   & {\bf F8}    & {\bf 1.22}& {\bf 1.52}& {\bf 0.89}  & 0 	  & {\bf 2} & {\bf A} & {\bf A}  & {\bf B} \\ 
 J211127.41+182653.3  	  & 4.216933  	    & 3.168   	  & 0.0464    	  & 1.44       & 0.19	      & G0	    & 1.12	& 2.06      & 0.85	  & 0 	  & 5	    & C & A   & C\\ 
 J211417.15+112741.0  	  & 6.579094  	    & 8.23    	  & 0.0336    	  & 1.6        &  0.35        & G3	    & 1.02	& 1.6	    & 2.06	  & 0 	  & 3	    & B & C   & C \\
 J211448.98+203557.1  	  & 4.864623  	    & 4.656   	  & 0.0530    	  & 1.63       & 0.25	      & G3	    & 1.01	& 1.98      & 1.25	  & 0 	  & 4	    & C & A   & C \\
 J211448.98+203557.1  	  & 4.864666  	    & 4.632   	  & 0.0525    	  & 1.63       & 0.25	      & G3	    & 1.01	& 1.98      & 1.25	  & 0 	  & 4	    & C & A   & C \\
{\bf J211608.42+163220.3} & {\bf 3.468244}  & {\bf 1.992} & {\bf 0.0131}  & {\bf 1.31} & {\bf 0.21}   & {\bf F8}    & {\bf 1.21}& {\bf 1.18}& {\bf 0.59}  & 0 	  & {\bf 0} & A & A   & A\\ 
{\bf J211645.22+192136.8} & {\bf 4.400381}  & {\bf 2.640} & {\bf 0.0135}  & {\bf 1.27} &  {\bf 0.16}  & {\bf F7}    & {\bf 1.24}& {\bf 1.23}& {\bf 0.71}  & 0 	  & {\bf 0} & {\bf A} & {\bf A} & {\bf A} \\ 
 J211817.92+182659.9  	  & 7.715382  	    & 8.888	  & 0.0357    	  & 1.04       &  0.3	      & F4	    & 1.4	& 2.45      & 0.4	  & 0 	  & 7	    & C & B   & C \\
{\bf J212532.55+082904.4} & {\bf 3.125014}  & {\bf 2.688} & {\bf 0.0267}  & {\bf 1.43} & {\bf 0.23}   & {\bf F9}    & {\bf 1.13}& {\bf 1.58}& {\bf 0.82}  & 0 	  & {\bf 0} & A & A   & A\\ 
 J212749.35+190246.0  	  & 7.810082  	    & 8.4	  & 0.10      	  & 2.08       &  0.34        & K1	    & 0.82	& 2.21      & 2.05	  & 0 	  & 1	    & C & C   & B \\
$^{\dagger}$J212815.28+082933.7 & 4.91815   & 5.592 	  & 0.0083    	  & 1.27       & 0.23 	      & F7    	    & 1.24    	& 0.96      & 1.48    	  & 0 	  & 3 	    & A & A   & C \\
{\bf J212843.62+160806.2} & {\bf 1.375647}  & {\bf 2.64}  & {\bf 0.0159}  & {\bf 2.59} & {\bf 0.53}   & {\bf K5$^{*}$}& {\bf 0.71}& {\bf 0.76}& {\bf 1.44}& 0 	  & {\bf 3} & {\bf A} & {\bf A} & {\bf C} \\ 
{\bf J212855.03+075753.5} & {\bf 4.688048}  & {\bf 1.92}  & {\bf 0.0297}  & {\bf 1.8}  &  {\bf 0.36}  & {\bf G7}	    & {\bf 0.92}& {\bf 1.35}& {\bf 0.58}  & 0 	  & {\bf 2} & A & A   & B\\ 
\hline								  	       
\end{tabular}
\end{table}
\end{landscape}

\section{Spectroscopic Confirmation of Candidate Assessment}
\protect\label{sec:specdata}

While the analysis discussed above was at a preliminary stage, the opportunity
arose to obtain echelle spectra of 7 objects using the ESPaDOnS
spectrograph on the Canada-France-Hawai'i Telescope, Hawai'i (CFHT) between 2005
September 23--24.  These targets were taken from a preliminary selection list,
according to the visibility from the telescope.  The spectrograph was configured
in spectropolarimetric mode during these observations, with a 79 rules/mm
grating achieving a resolution of $R\sim$68,000 and spanning over 40 orders in
wavelength between 370--1050\,nm.  The filter in place was Stokes I band, and
the exposure time was set between 300-600s, depending on the magnitude of the
target.  These data were reduced at the telescope using the {\em
Libre-ESpRIT}
\footnote{http://www.cfht.hawaii.edu/Instruments/Spectroscopy/\\Espadons/Espadons\_esprit.html}
online data reduction facility to perform the usual bias subtraction,
flat-fielding and wavelength calibrations, followed by the order extraction of
the polarisation information.  

Echelle spectroscopy provides a wavelength range of several thousand Angstroms and
hence a large number ($n_{l}$=4688) of images of a large sample (3507) of
photospheric lines.  These were used to boost the signal-to-noise of the spectra by
a factor of $\sim \sqrt(n_{l})$ by applying the Least Squares Deconvolution method
(see e.g. \citealt{donati97}) in conjunction with a G2 line list.  This analysis
increased the S/N from $\sim$30 to $\sim$323.  The telluric water lines within the
echellogram were used to obtain a velocity calibration accurate to $\sim$few m/s of
the heliocentric reference frame.  

\begin{figure}
\centering
\includegraphics[angle=270,width=6cm]{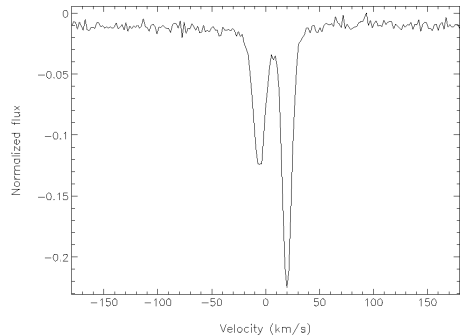}
\caption{Deconvolved spectrum of 1SWASP~J204456.57+182136.0 taken with ESPaDOnS at CFHT}
\protect\label{fig:cfhtspec}
\end{figure}

One of these stars, 1SWASP~J204456.57+182136.0, falls within this dataset.  This
object survived the selection procedure as far as the final stage, where it
received a grade of `CCB' because of a high companion radius estimate (1.83\Rj)
and $\eta_{p}$=1.79.  The low number of transits is a consequence of the long period
($\sim$8.15\,days).  Under our current selection procedure, this object was
judged to be a blended stellar binary independently of the spectroscopic data. 
This assessment was confirmed by the CFHT spectra, shown in
Figure~\ref{fig:cfhtspec}, which clearly shows a double-lined binary.  These
data give us confidence that our candidate selection procedure eliminates
many astrophysical false positives, and prioritises strong
exoplanet candidates for follow-up observations.  

\section{Discussion}  
\protect\label{sec:concs} 

We have whittled down the original {\sc huntsman} list of 11,626 stars observed
between RA=18\,hr--21\,hr and have identified 35 objects of particular interest which
we recommend for follow-up observations.  We find 6 objects for which all the data
currently at our disposal supports the hypothesis that the companion object is
planetary.  These are summarised in Table~\ref{tab:priority1}.  However, we encourage
investigation of all these objects, since some narrowly missed the priority list,
chiefly due to blending.  In the tabulated data and discussions above we have noted
any causes of uncertainty on a case-by-case basis.  Furthermore, `false alarms' from
transit surveys provide a new sample of low-mass binaries, which are of interest in
their own right.  

The SW-N instrument has proven to be an excellent way of finding transiting
candidates among millions of bright field stars but it cannot conclusively
determine the nature of these systems alone.  As experience from a number of
earlier transit surveys has shown (e.g. OGLE \citealt{udalski04}, Vulcan
\citealt{borucki01}), a large ($\sim$90\%) percentage of the candidates will turn
out to be stellar binaries.  \citet{lister07} estimate that $\sim$20--30 genuine
exoplanets will be discovered in the 2004 season data as a whole, so we
anticipate 2--4 to lie within this sample.  This is an inescapable part of the
nature of transit surveys: there are many astrophysical phenomena which mimic
the signal of a transiting exoplanet (for a discussion, see \citealt{brown03},
\citealt{charbonneau04}).  Some of our candidates will be binary stars eclipsing
at grazing incidence as seen from Earth, others are likely to be binaries whose
eclipses appear shallower than they are in reality because of light from a third
object contaminating our photometry.  

\begin{table} 
\centering
\caption{Priority exoplanet candidates.} 
\label{tab:priority1} 
\begin{tabular}{lcccc} 
\hline 
 Identifier	     & Period	& Duration  & Depth  & $R_{p}$ \\
1SWASP...	     & (days)	& (hrs)     & (mag)  & (\Rj)   \\
\hline 
J183104.01+323942.7  & 2.378781 & 1.776     & 0.0089 & 0.97 \\
J184119.02+403008.4  & 3.734014 & 4.224     & 0.0148 & 0.92 \\
J204712.42+202544.5  & 2.61264  & 2.064     & 0.0275 & 0.95 \\
J210318.01+080117.8  & 1.223824 & 1.92      & 0.0167 & 1.01 \\
J211608.42+163220.3  & 3.468244 & 1.992     & 0.0131 & 1.18 \\
J211645.22+192136.8  & 4.400381 & 2.640     & 0.0135 & 1.23 \\
\hline  									     
\end{tabular}
\end{table}

It is therefore necessary to execute a systematic and careful series of follow-up
observations to finally establish the true nature of these objects, and in the
process, accurately determine their physical and orbital parameters.  We have an
extensive program of photometric and spectroscopic follow-up on-going.  We 
initially obtain 1--2 medium-resolution spectra of all priority candidates to
confirm the spectral typing and hence the estimate of the minimum companion
radius.  These data will also eliminate single- and double-lined binaries and
line-of-sight blends from the asymmetries in the line profiles.  An imaging
campaign running in parallel provides high-precision two-colour photometry at
higher resolution around the times of transit of the best candidates.  This can
identify stellar companions from a detectable ($\lesssim$0.01\,mag) difference in
transit depth.  Finally, the best candidates are subject to full radial
velocity observations.  

Note: Shortly after this paper was submitted, our follow-up program confirmed
the planetary nature of the companion to the shortlisted star
1SWASP~J203054.12+062546.4, henceforth dubbed WASP-2b.  For a detailed
discussion of this discovery, see \cite{cameron07}.

\section{Acknowledgements} 
 
The WASP consortium consists of representatives from the Queen's University
Belfast, University of Cambridge (Wide Field Astronomy Unit),  Instituto de
Astrof\'{i}sica de Canarias,  Isaac Newton Group of Telescopes (La Palma), 
University of Keele, University of Leicester, Open University, and the
University of St Andrews.  The SuperWASP-N \& -S instruments were constructed
and operated  with funds made available from Consortium Universities and the
Particle Physics and Astronomy Research Council.  SuperWASP-N is located in the
Spanish Roque de Los Muchachos Observatory on La Palma,   Canary Islands which
is operated by the Instituto de Astrof\'{i}sica de Canarias (IAC). 
This publication makes use of data products from the Two Micron All Sky Survey
which is a joint project of the University of Massachusetts and the Infrared
Processing and Analysis Center/California Institute of Technology, funded by the
National Aeronautics and Space Administration and the National Science
Foundation.

RAS was funded by a PPARC Post-Doctoral Fellowship during the course of this
work. 

\clearpage

\bibliographystyle{mn2e}
\bibliography{iau_journals,references}

\bsp\label{lastpage}
\end{document}